\documentclass[twocolumn,showpacs,superscriptaddress,preprintnumbers,amsmath,amssymb,prc]{revtex4}
\usepackage{graphicx}
\usepackage{dcolumn}
\usepackage{bm}
\usepackage{color}

\begin{document}

\title{Astrophysical S factor of {$^{12}$C($\alpha,\gamma$)$^{16}$O} Calculated with the Reduced R-matrix Theory}

\author{Zhen-Dong An}
\affiliation{Shanghai Institute of Applied Physics, Chinese Academy of Sciences, Shanghai 201800, China}
\affiliation{University of Chinese Academy of Sciences, Beijing 100049, China}

\author{Zhen-Peng Chen} \thanks{Email: zhpchen@tsinghua.edu.cn}
\affiliation{Department of Physics, Tsinghua University, Beijing 100084, China}

\author{Yu-Gang Ma}     \thanks{Email: ygma@sinap.ac.cn}
\affiliation{Shanghai Institute of Applied Physics, Chinese Academy of Sciences, Shanghai 201800, China}
\affiliation{ShanghaiTech University, Shanghai 200031, China}

\author{Jian-Kai Yu}
\affiliation{Department of Engineering Physics, Tsinghua University, Beijing 100084, China}

\author{Ye-Ying Sun}
\affiliation{Department of Materials, Tsinghua University, Beijing 100084, China}

\author{Gong-Tao Fan}
\author{Yong-Jiang Li}
\affiliation{Shanghai Institute of Applied Physics, Chinese Academy of Sciences, Shanghai 201800, China}

\author{Hang-Hua Xu}
\author{Bo-Song Huang}
\affiliation{Shanghai Institute of Applied Physics, Chinese Academy of Sciences, Shanghai 201800, China}
\affiliation{University of Chinese Academy of Sciences, Beijing 100049, China}

\author{Kan Wang}
\affiliation{Department of Engineering Physics, Tsinghua University, Beijing 100084, China}

\begin{abstract}
Determination of the accurate astrophysical S factor of {$^{12}$C($\alpha,\gamma$)$^{16}$O} reaction has been regarded
as a holy grail of nuclear astrophysics for decades. In current stellar models, a knowledge of that value to better than
10\% is desirable. Due to the practical issues, tremendous experimental and theoretical efforts over nearly 50 years
are not able to reach this goal, and the published values contradicted with each other strongly and their uncertainties
are 2 times larger than the required precision. To this end we have developed a Reduced R-matrix Theory, based on
the classical R-matrix theory of Lane and Thomas, which treats primary transitions to ground state and
four bound states as the independent reaction channels in the channel spin representation.
With the coordination of covariance statistics and error propagation theory, a global fitting for almost all available
experimental data of $^{16}$O system has been multi-iteratively analyzed by our powerful code. A reliable, accurate and
self-consistent astrophysical S factor of {$^{12}$C($\alpha,\gamma$)$^{16}$O} was obtained with a recommended value
$S_{tot}$ (300) = 162.7 $\pm$ 7.3 keV b (4.5\%) which could meet the required precision.

\end{abstract}


\maketitle

\section{Introduction}
During core-He burning 3$\alpha$ and {$^{12}$C($\alpha,\gamma$)$^{16}$O} reactions compete to determine the helium burning time scale,
together with the convection mechanism, and the relative abundances of oxygen and carbon prior to core-C burning. W. A. Fowler,
a Nobel laureate in physics in 1983, definitely held a view that the abundance ratio of $^{12}$C to $^{16}$O and the solar neutrino
problem are serious difficulties in the most basic concepts of nuclear astrophysics~\cite{Rolf88}. The work of many investigators
has resulted in a knowledge of the predicted reaction rate of 3$\alpha$ process within about 10\%~\cite{Fynb05} accuracy at the usual
helium-burning temperatures. Unfortunately, the same of the {$^{12}$C($\alpha,\gamma$)$^{16}$O} reaction is much less well determined,
and it is highly desirable to know this rate with comparable accuracy~\cite{Woos07} to that of the 3$\alpha$ process in order to
provide adequate constraints on stellar evolution and the synthesis of elements, e.g. the yield of the neutrino-process isotopes
$^{7}$Li, $^{11}$B, $^{19}$F, $^{138}$La and $^{180}$Ta in core-collapse supernovae~\cite{Aust11,Aust14}, and the production of
the important radioactive nuclei $^{26}$Al, $^{44}$Ti, and $^{60}$Fe~\cite{Tur10}.

\begin{table}[b]
\caption{\label{tab:table1}
Comparison of the astrophysical S factor (keV b) at 0.3 MeV obtained in various fits, including this work, for the E10, E20,
cascade transitions components, as well as the total.}
\begin{ruledtabular}
\begin{tabular}{cccccccc}
$Reference$               & $S_{tot}$     & $S_{E10}$     & $S_{E20}$    & $S_{cas}$   \\
\hline
This work                 & 162.7$\pm$7.3 & 98.0$\pm$7.0 & 56.0$\pm$4.1    & 8.7$\pm$1.8 \\
Sch\"urmann~\cite{Schu12} & 161$\pm$19    & 83.4          & 73.4            & 4.4         \\
Oulebsir~\cite{Oule12}    & 175$\pm$63    & 100$\pm$28    & 50$\pm$19       & ---         \\	
Sayre~\cite{Sayr12}       & ---           & ---           & $62_{-6}^{+9}$  & ---         \\
Tang~\cite{Tang10}        & ---           & 84$\pm$21     & ---             & ---         \\
Matei~\cite{Mate08}       & ---           & ---           & ---             & $S_{6.92}$=7.0$\pm$1.6 \\
Matei~\cite{Mate06}       & ---           & ---           & ---   & $S_{6.05}$=$25_{-16}^{+25}$ \\
Hammer~\cite{Hamm05}      & 162$\pm$39    & 77$\pm$17     & 80$\pm$20       & 4$\pm$4     \\
Tischhauser~\cite{Tisc02} & 149$\pm$29    & 80$\pm$20     & $53_{-18}^{+13}$    & 16$\pm$16   \\
Kunz~\cite{Kunz01}        & 165$\pm$50    & 76$\pm$20     & 85$\pm$30       & 4$\pm$4     \\
Brune~\cite{Brun99}       & 159           & 101$\pm$17    & $42_{-23}^{+16}$       & 16   \\
Ouellet~\cite{Ouel96}     & 120$\pm$40    & 79$\pm$16     & 36$\pm$6        & ---         \\
Buchmann~\cite{Buch96}    & 165$\pm$75    & 79$\pm$21     & 70$\pm$70        & 16$\pm$16  \\
\end{tabular}
\end{ruledtabular}
\end{table}

The most direct and trustworthy way to obtain the astrophysical S factor of the {$^{12}$C($\alpha,\gamma$)$^{16}$O}
reaction is to measure the cross section for that reaction till as low energy as possible, and to extrapolate to energies
of astrophysical interest. The astrophysical S factor for {$^{12}$C($\alpha,\gamma$)$^{16}$O} reaction reaction is given by
\begin{equation}\label{SF}
  S(E)= \sigma(E)\cdot E\cdot \exp(2\pi\eta),
\end{equation}
where $\eta$ = $Z_{\alpha}Z_{C}e^2/ \hbar \nu$ is the Sommerfeld parameter for the interaction particles.
To investigate the specific role of the $^{16}$O nucleus for the S factor, a wealth of experimental data have been accumulated
over the past few decades, including the precise measurements of total cross-section of {$^{12}$C($\alpha,\gamma$)$^{16}$O}~\cite{Schu05,Schu11,Plag12,Fuji13}, $\gamma$-ray angular distributions of ground state
transition~\cite{Dyer74,Redd87,Ouel96,Kunz97,Kunz01,Assu06,Fey04,Maki09,Lars64,Ophe76,Kern71,Mitc64},
cascade transitions~\cite{Schu11,Redd87,Kett82,Mate06}, $\beta$-delayed $\alpha$ spectra for $^{16}$N~\cite{Tang07,Tang10,Azum94,Zhao93},
transfer reaction~\cite{Brun99,Belh07,Oule12}, elastic scattering {$^{12}$C($\alpha,\alpha$)$^{12}$C}~\cite{Plag87,Tisc02,Tisc09,Morr68,Brun75}
and additional particle reaction pathways {$^{12}$C($\alpha,\alpha${$_{1}$})$^{12}$C}~\cite{Mitc65,Debo12a}
and {$^{12}$C($\alpha,p$)$^{15}$N}~\cite{Mitc65,Debo12b} at high energies. However, complexity of reaction mechanisms of
{$^{12}$C}+$\alpha$ makes it extremely difficult to determine the S factor, despite five decades of experimental investigations yet,
the desired accuracy and precision associated with the {$^{12}$C($\alpha,\gamma$)$^{16}$O} reaction continues to be an obstacle
(See the Table~\ref{tab:table1}). Recent two collaborations~\cite{Gai12,Rehm12} and our team~\cite{Xu07} have pursued
complementary approaches to obtain the inverse {$^{16}$O($\gamma,\alpha$)$^{12}$C} reaction to energies lower than the currently achieved,
which could offer significant advantages over traditional approaches. But the expected outputs of these proposals
in low energy are far from what's required in the stellar models. R-matrix analysis is the most effective method
for the fitting and extrapolation of existing data of $^{16}$O system, which are main content of this work.

The classical R-Matrix theory of Lane and Thomas~\cite{Lane58} deduced the standard R-matrix formulae to describe two body nuclear reaction.
However, these formulae were thought not justified for $\gamma$ radiative capture, because of the possibilities of particle production
and annihilation, and it is hard to select suitable channel radius for the long-range electromagnetic transition. After that, another paper
of Lane~\cite{Lane60}  expanded the collision matrix to radiative capture for a sum of three parts, viz. an internal resonant,
an external resonant and a non-resonant part corresponding to the channel integral from hard sphere scattering. Based on this conclusion,
the angle-integrated cross section formulae are derived from perturbation theory in Refs.~\cite{Holt78,Bark91},
and the adjustable parameter of photon reduced-width amplitude can be split into internal and asymptotic channel contributions
in practical applications~\cite{Bark91,Schu12}. Recently, a vital progress in R-matrix code, AZURE was presented in Refs.~\cite{Azum10,Debo13},
and it allows simultaneous analysis of the integrated and differential data for the electromagnetic transition.

In the most widely used theory~\cite{Bark91}, since the S factor of the $E$1 and $E$2 multipoles has different energy dependence,
one must have an independent and precise information on each multipole cross section for an extrapolation to 0.3 MeV.
So the secondary data of $E$1 and $E$2 multipoles were in general used for the R- matrix analysis in the previous publications.
The primary data most often consist of angular distributions measured at many discrete energies.
Each primary distribution was then analyzed independently in terms of the appropriate set of Legendre polynomials~\cite{Dyer74},
a non R-matrix analysis neglecting the energy- and angle-dependence, to yield the secondary data, $\sigma_{E10}$ and
$\sigma_{E10}$/$\sigma_{E20}$ at this discrete energy. And these data and their error values, being derived quantities,
are no longer proportional to the experimentally measured quantities, i.e., the angular distribution yields,
and thus lead to complications and discrepancy for the extrapolation of the {$^{12}$C($\alpha,\gamma$)$^{16}$O} S factor~\cite{Buch96}.

A R-matrix code for a nuclear system is in principle an exact model as long as a complete set of quantum states of a nuclear system
and all the corresponding experimental data can be accurately described simultaneously (called Global fitting).
Any inconformity that does not meet the principle will induce inestimable uncertainty, because each channel,
each level and each data are intimately correlated and strong interference in the nuclear system.
By now the R-matrix approach mentioned above~\cite{Bark91},
has got popular application with the procedure, but their results have very larger difference (See Table~\ref{tab:table1}).
And no one used it to do a global fitting for $^{16}$O system in the astrophysical energy yet. For these reasons,
based on the theory of Lane and Thomas~\cite{Lane58}, we develop a Reduced R-matrix Theory to make the global fitting
for the special problem to search for the S factor of {$^{12}$C($\alpha,\gamma$)$^{16}$O}.

Section II summarizes the construction of the Reduced R-matrix Theory and general aspects of the R-matrix approach in the global analysis.
Section III presents the construction of reaction channel, evaluation and fits of experimental data. Results and discussion of the global analysis for each reaction channel is presented in Sec. IV. Finally, conclusions are given in Sec. VII.

\section{Reduced R-matrix Theory}

\subsection{The representations~\cite{Devo57,Ferg65} for decay channel of {$\gamma$}$_{n}$+$^{16}$O$_{n}$}

For the {$^{12}$C($\alpha,\gamma$)$^{16}$O} reaction, the decay channel of {$\gamma$}$_{n}$+$^{16}$O$_{n}$ (n=0, 1, 2, 3, 4)
has the same total angular momentum and parity {\bf J}$^{\pi}$ as the entrance channel {$^{12}$C+$\alpha$}.
Three angular momenta are involved at this stage for the state of compound nucleus, the spins of the target (or residual nucleus)
and of the incident (or emitting) particle and the orbital angular momentum of the incident (or emitting) particle.
There are two conventional ways of combining these for the decay channel {$\gamma$}$_{n}$+$^{16}$O$_{n}$.
One is the channel spin scheme in which the vector sum of the spins of photon ({\bf I}$_{\gamma}$={\bf 1})
and residual nucleus $^{16}$O$_{n}$ ($\mathbf {J_f}$) is firstly formed giving the channel spin ${\mathbf s}$
\begin{equation}\label{EQ_s}
    {\mathbf I}_{\mathbf \gamma }+{\mathbf J}_{\mathbf f}={\mathbf s}.
\end{equation}
Channel spin ${\mathbf s}$ and the orbital angular momentum ${\mathbf l}$ of the photon are then combined to form
the spin of the compound nucleus $^{16}$O,
\begin{equation} \label{EQ_Ji c}
    {\mathbf s}+{\mathbf l}={\mathbf J}_{\mathbf i}.
\end{equation}
Finally a state in the representation is labeled by the set of quantities
\{${\mathbf \alpha }\left({{\mathbf I}}_{{\mathbf \gamma }}{{\mathbf J}}_{{\mathbf f}}\right){\mathbf sl}{{\mathbf J}}_{{\mathbf i}}{\mathbf M}$\}.

The alternative coupling scheme is called Devons and Goldfarb the ``${\mathcal L}$-representation''~\cite{Devo57},
which is used widely in the R-matrix theory mentioned above~\cite{Bark91,Azum10}. Here the spin of photon is combined with
the corresponding orbital angular momentum to form total angular momentum for the photon states. It gives
\begin{equation} \label{EQ_L}
    {\mathbf I}_{\mathbf \gamma }+{\mathbf l}={\mathbf L}.
\end{equation}
The compound state's spin is then given by
\begin{equation} \label{EQ_Ji_L}
    {{\mathbf L}{\rm +}{\mathbf J}}_{{\mathbf f}}{\rm =}{{\mathbf J}}_{{\mathbf i}}.
\end{equation}
And a state of compound nucleus is labeled by the set of quantities \{${\mathbf \alpha }\left({{\mathbf I}}_{{\mathbf \gamma }}{{\mathbf J}}_{{\mathbf f}}\right){{\mathbf L}{\mathbf J}}_{{\mathbf f}}{{\mathbf J}}_{{\mathbf i}}{\mathbf M}$\} in the scheme.

These two coupling schemes represent the same physical situation in two different representations. In the ${\mathcal L}$-representation,
parity conservation implies that
\begin{equation} \label{EQ_Pi_L}
    {{\pi _i} = {\pi _f}{\pi _\gamma }{\left( { - 1} \right)^{L + P}}},
\end{equation}
with \emph{L} the multipolarity and the P mode (1 = electric, 0 = magnetic) of the gamma ray, and ${\pi _i}$, ${\pi _f}$,  ${\pi _\gamma}$
are the parity of the initial state ${{\mathbf J}}_{{\mathbf i}}$ , final state ${{\mathbf J}}_{{\mathbf f}}$ and the intrinsic parity
of photon, respectively. Electric multiploes correspond to parity (-1)$^L$ and \emph{l} = \emph{L}$\pm $1, while magnetic multiploes correspond
to parity (-1)$^{L+1}$ and \emph{l} = \emph{L}~\cite{Devo57,Ferg65}. Based on the expression of Eq.~\eqref{EQ_Pi_L},
we can deduce the parity conservation in the channel spin scheme
\begin{equation} \label{EQ_Pi_C}
{\pi _i} = {\pi _f}{\pi _\gamma }{\pi _l} = \left\{ {\begin{array}{*{20}{c}}
{{\pi _f}{\pi _\gamma }{{\left( { - 1} \right)}^l},  l = L \pm 1}\\
{{\pi _f}{\pi _\gamma }{{\left( { - 1} \right)}^{l{\rm{}}}},  l = L},
\end{array}} \right.
\end{equation}
where ${\pi _l}$ is parity of orbit angular momentum for the exit channels.

\begin{figure}
\center
\includegraphics[scale=0.28]{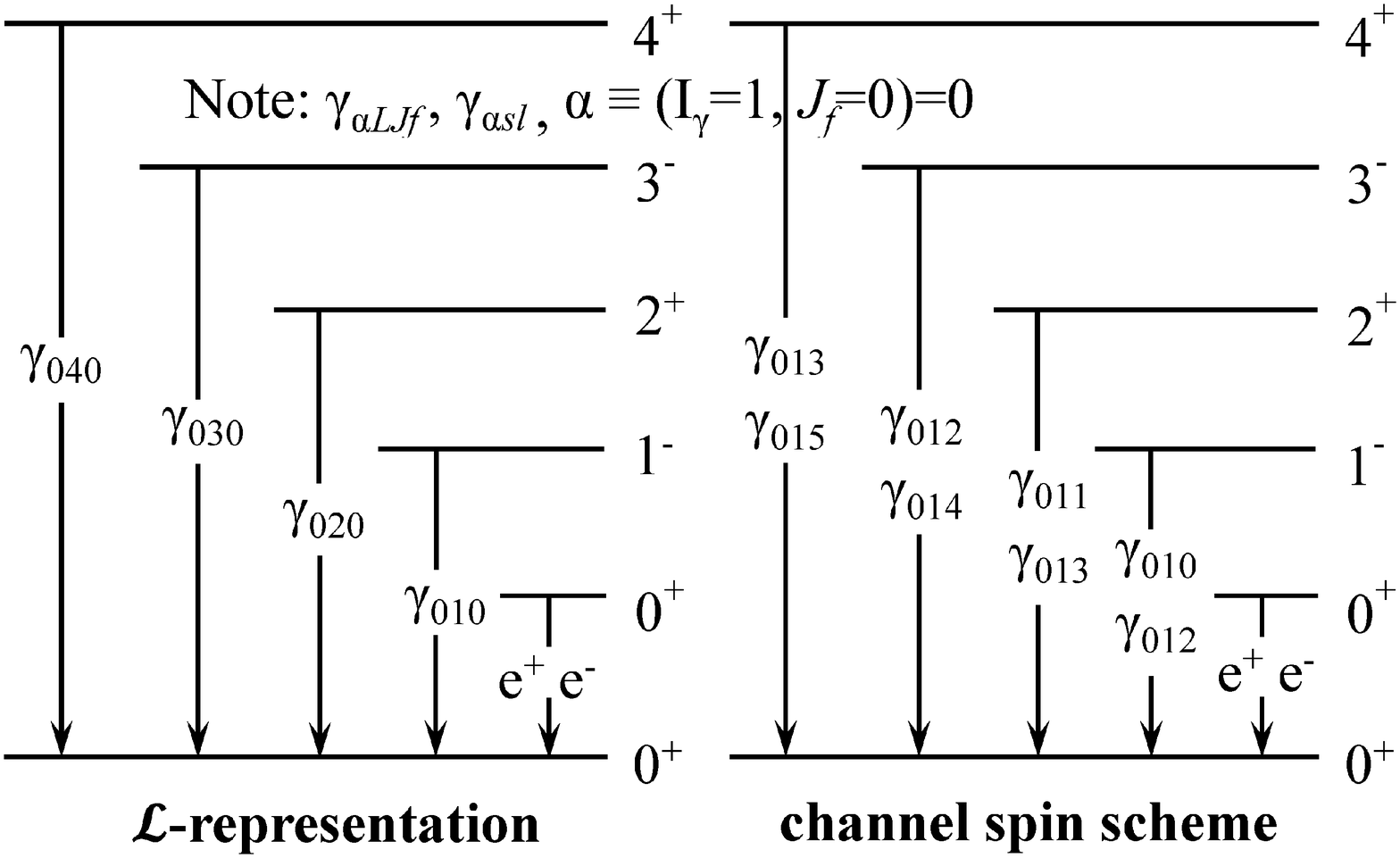}
\caption{
    The electric multiploes transition processes to ground state (0${}^{+}$) described by channel spin scheme and
    ${\mathcal L}$-representation for ${}^{16}$O system.
}
\label{fig:com_1}
\end{figure}

For the comparison of two coupling schemes, Fig.~\ref{fig:com_1} displays the all possible transition processes to the ground-state
(J$^\pi$ = $0^+$) of ${}^{16}$O system. The ground-state transition of ${\mathcal L}$-representation, set of components has one value
for each decay, which is less than the two components in the channel spin scheme of the Fig.~\ref{fig:com_1}, and the relevant photon
reduced-width amplitude of this processes only exists one parameter in the collision matrix. Only when the channel capture is considered
in ${\mathcal L}$-representation, this parameter can be split into internal and asymptotic channel contributions~\cite{Bark91,Azum10}.
So the channel spin scheme provides more subsets to denote the ground-state transition and cascade transition, which is extremely
beneficial to the interpretation of the observed experimental data more reliable.

\subsection{Wave function of compound nucleus $^{16}$O}
For the {$^{12}$C($\alpha,\gamma$)$^{16}$O} reaction, the transition of compound nucleus $^{16}$O, initial radioactive decay
to ground state and four bound states are regarded as two body particle reaction channels, denoted by {$\gamma$}$_{n}$+$^{16}$O$_{n}$
(n=0, 1, 2, 3, 4), and the reduced masses of these channels are represented by relativistic energy.
It is not necessary to consider how to decay to the ground state finally, so the problem of `particles are created or destroyed'
is avoided. The sum of integral cross sections of each reaction channel {$\gamma$}$_{n}$+$^{16}$O$_{n}$ is equal to the cross section
of $^{16}$O production.

Owing to the advantage of the channel spin scheme, all the channels of the $^{16}$O system are represented as c = ${\alpha sl}$,
where \emph{s} is the channel spin, \textit{l} is the relative angular momentum of the interacting particle of entrance or exit channels,
and $\alpha $ identifies the interacting particle pair. The primary wave function ${\Psi }$ can be unfolded with different exit channels
${{\mathbf \psi }}_{c}$. Furthermore ${{\mathbf \psi }}_{c}$ can be expended with level wave functions ${X_\lambda}$,
which have different total angular momentum, parity ${J}^{\pi}$ and ${E}_{\lambda}$. Finally ${\Psi }$ is expressed with Eq. 2.6
(page 283) in Ref.~\cite{Lane58}:
\begin{equation} \label{EQ_Func}
 \Psi=\mathop \sum \limits_c \left[ {\mathop \sum \limits_\lambda  \frac{{{X_\lambda }{\gamma _{\lambda c}}}}{{{E_\lambda } - E}}} \right]D_c^0.
\end{equation}
So the total wave function for the initial state of ${}^{16}$O can be expanded by the complete orthogonal set,
the coefficient of the expanded formula represent the probability of different reaction channel of all sorts of resonance energy state.
Eq.~\eqref{EQ_Func} demonstrates that, if the primary gamma decay $\gamma $${}_{n}$+${}^{16}$O${}_{n}$ as the independent
two body reaction channel, the set of level wave function ${X_\lambda }$, of the theoretical model contains all types of $\gamma $- transition,
whether direct decay to the ground state transition or cascade transition. Using the Eq.~\eqref{EQ_Func},
one can obtain the fundamental R-matrix relation, the collision matrix, the cross sections and so on.

For the channel $\gamma $${}_{n}$+${}^{16}$O${}_{n}$, the electromagnetic interaction is long range, therefore contributions to
the collision matrix for radiative capture reactions can come from large distances. Thus, in addition to the internal contribution
to the collision matrix, there should also be channel contributions~\cite{Bark91}. However, for the chief ground state transition,
owing to the large binding energy (7.16 MeV) of the ${}^{16}$O with respect to the $\alpha $+${}^{12}$C threshold,
its wave function decreases rapidly when the radius is larger than a certain value. So the internal contribution is strongly dominant,
and the external part can be neglected. The results in Refs.~\cite{Bark91,Kett82} prove to be reasonable and effective for this approximation,
that the external contribution accounts for lower than 3\% at 0.3 MeV. For the cascade transitions, a parameter for the final state
can be used to characterize the direct capture process of these transitions(see the parameters table).
So a global fitting for whole ${}^{16}$O system can be done using the standard R-matrix formulae of Ref.~\cite{Lane58}
with a suitable channel radius.

\subsection{Mathematical formalism of RAC code}

The practical formulae of our RAC code are introduced from the literatures~\cite{Chen03,Carl08,Carl09}. On the R-matrix and
the reaction cross sections, the codes are strictly compiled in accordance with the formulae of classic literatures~\cite{Lane58},
without any approximation.

Explicitly, the \textit{R}-matrix, which represents all the internal information concerning the structure of the compound system,
is defined as
\begin{equation}\label{EQ_Rmatrix}
  {\bf{R}}^J _{\alpha 's'l',\alpha sl} = \mathop \sum \limits_{\lambda \mu }^N \gamma _{\alpha 's'l'}^J\gamma _{\alpha sl}^J{A_{\lambda \mu }}{\delta _{J{J_0}}},
\end{equation}
where $\gamma _{\alpha 's'l'}^J$ and $\gamma _{\alpha sl}^J$ are the reduced-width amplitude of entrance and exit channel,
respectively. The matrix ${{\mathbf A}}_{{\mathbf \lambda }{\mathbf \mu }}$ is defined by its inverse
\begin{equation}\label{EQ_Ainverse}
  {\left[ {{A^{ - 1}}} \right]_{\lambda \mu }} = \left( {E_\lambda - E} \right){\delta _{\lambda \mu }} - {\Delta }_{\lambda \mu }- \frac{i{\Gamma}_{\lambda \mu}}{2},
\end{equation}
where $E_\lambda$ is the position of resonance level, ${\Delta }_{\lambda \mu }$ is the energy shift, ${\Gamma }_{\lambda \mu }$ is
the total reduced channel width, which can represent the contribution of all un-considered channels, such as the
$^{12}$C($\alpha,\alpha_2$)$^{12}$C in our fit.
The additional quantity appearing in Eqs.~\eqref{EQ_Ainverse} is
\begin{equation}\label{EQ_Delta1}
  {{\Delta }_{\lambda \mu }} =  - \mathop \sum \limits_{\alpha sl}^N \left( {{{S}_{\lambda \mu }} - {{B}_{\lambda \mu } }} \right)\gamma _{\alpha 's'l'}^{}\gamma _{\alpha sl}^{},
\end{equation}
where $S_{\lambda\mu}$ is the shift factor calculated at the channel radius, and $B_{\lambda\mu}$ is the constant boundary parameter.

The literature of of Lane and Thomas (Page 273)~\cite{Lane58}, gives the correction formula on the level width $\Gamma_{\lambda c}$
and level shift $\Delta_{\lambda c}$:

\begin{equation}\label{EQ_Gamma}
   {\Gamma _{\lambda c}} = 2{P_c}\gamma _{\lambda c}^2/{d_c},
\end{equation}

\begin{equation}\label{EQ_Delta2}
  {\Delta _{\lambda c}} = \frac{{{P_c}\left( {R_{cc}^0{P_c}} \right) - S_c^0\left( {1 - R_{cc}^0S_c^0} \right)}}{{{d_c}}}\gamma _{\lambda c}^2.
\end{equation}
where
\begin{equation}\label{EQ_dc}
  {d_c} = {\left( {1 - R_{cc}^0S_c^0} \right)^2} + {\left( {R_{cc}^0{P_c}} \right)^2}.
\end{equation}
Here, $\lambda$ is the level of c reaction channel, $P_c$ is the penetration factor and notation zero is the constant background.
These formulae are workable only based upon an approximation of single level. RAC is the multi-channel and multi-level \emph{R}-matrix
formula without constant background. When calculating the width and shift of some levels, the calculated values of R-Matrix with remaining levels are taken as the constant background of the level. The observed width can be related to the physical reduced width amplitudes with a formula as follow:
\begin{equation}\label{EQ_Gamma_2}
    {\Gamma }_{\lambda c}^{obs} = {\Gamma}_{\lambda c}\left( {1 + \sum\nolimits_k {\gamma _{\lambda k}^2} \frac{{d{S_k}}}{{dE}}} \right)_{{E_\lambda }}^{ - 1}.
\end{equation}
The total width for a state $\lambda$ is then the sum
\begin{equation}\label{TotGamma}
    \Gamma_{\lambda}^{obs}=\sum_c {\Gamma }_{\lambda c}^{obs}.
\end{equation}

With the relation between T-matrix and U-matrix of formula
\begin{equation}\label{EQ_T-matrix}
  {T_{\alpha 's'l',asl}^J = {e^{2i{\omega _{\alpha l}}}}{\delta _{\alpha 's'l',\alpha sl}} - U_{\alpha 's'l',\alpha sl}^J},
\end{equation}
for a reaction going through $\alpha \to \alpha '$ the angle-integrated cross section is given as,
\begin{equation}\label{EQ_sigma}
  {\sigma _{\alpha ',a}} = \frac{\pi }{{k_\alpha ^2}}\mathop \sum \limits_{sl's'lJ} \frac{{\left( {2J + 1} \right)}}{{\left( {2{I_1} + 1} \right)\left( {2{I_2} + 1} \right)}}{\left| {T_{\alpha 's'l',asl}^J} \right|^2},
\end{equation}
and ${I}_{1}$ and ${I}_{2}$ are the projectile and target spins, respectively. It should be noted that the above equation does not hold for charged particle elastic scattering.

For the corresponding differential cross section formula, a more rigorous calculation is involved in Ref.~\cite{Lane58}
\begin{equation}\label{EQ_Diff}
\frac{d\sigma_{\alpha,\alpha{'}}}{d\Omega_{\alpha{'}}}=\frac{1}{(2I_{1}+1)(2I_{2}+1)}
\sum_{ss{'}\nu\nu{'}}^{ }|A_{\alpha{'}s{'}\nu{'},\alpha s\nu}(\Omega_{\alpha{'}})|^{2},
\end{equation}
where $A_{\alpha{'}s{'}\nu{'},\alpha s\nu}$ is the amplitudes of the outgoing waves.
\begin{widetext}
\begin{equation}\label{EQ_A}
  A_{\alpha{'}s{'}\nu{'},\alpha s\nu}(\Omega_{\alpha^{'}})=\frac{\sqrt{\pi}}{k_{\alpha}}
  [-C_{\alpha{'}}(\theta_{\alpha{'}})\delta_{\alpha{'}s{'}\nu{'},\alpha s\nu}
  +i\sum_{JMll{'}m{'}}^{ }\sqrt{2l+1}(sl\nu0|JM)(s{'}l{'}\nu{'}m{'}|JM)
  T^{J}_{\alpha{'}s{'}l{'},\alpha sl}Y_{m{'}}^{(l{'})}(\Omega_{\alpha{'}})].
\end{equation}
\end{widetext}
Several new quantities have been introduced in Eq.~\eqref{EQ_A} to define the angular dependence of the cross section.
The term $-C_{\alpha{'}}(\theta_{\alpha{'}})$ represents the Coulomb amplitudes, while $Y_{m'}^{(l')}$ is the spheric harmonics function.

The transverse character of electromagnetic wave requires that the projections of intrinsic spins of photon can not be zero.
For the transition with $\emph{\textbf{J}}_{f}$ = 0, when $\nu{'}$ = 0, the $\gamma$ spin projection is zero. So when using the
Eq.~\eqref{EQ_Diff} to calculate the angular distribution of $\gamma$ decay, as long as ignoring the loop for $\nu{'}$ = 0,
the calculation will not include the contribution of $\gamma$ spin projection component zero. When this method is adopted,
the angle-integrated cross section can be described effectively in our fit, but the corresponding differential cross section
is not fitted accurately. So in the actual work, the longitudinal contribution of photon is employed to give a precise description
for the all available data.

\subsection{Covariance statistic and error propagation law}

The uncertainty determination of the extrapolated S factor requires an error propagation of all relevant fit parameters through
the fit function taking into account the covariances. The theoretical formula about error propagation~\cite{Smit91} for our
\emph{R}-matrix model fitting is as following:
\begin{equation}\label{Eq_ERR}
    {\mathbf y}-{\mathbf y} _{0}= {\mathbf D}({\mathbf P}-{\mathbf P} _{0}),
\end{equation}
\begin{equation}\label{Eq_Sens}
    {D}_{ki}=(\partial y_{k}/\partial P_{i})_{0}.
\end{equation}
Here ${\mathbf y}$ refers to vector of calculated values, ${\mathbf D}$ to sensitivity matrix, ${\mathbf P}$ to vector of \emph{R}-matrix
parameters. Subscript 0 means optimized original value, k and i stand for fitted data and \emph{R}-matrix parameter, respectively.
The covariance matrix of parameter ${\mathbf P}$ is
\begin{equation}\label{EQ_Cov_para}
  {\textbf{V}}_{\textbf{P}}=(\textbf{D}^{+}\textbf{V}^{-1}\textbf{D})^{-1}.
\end{equation}
Here \textbf{V} refers to covariance matrix of the data to be fitted, and its inversion matrix can be expressed as following:
\begin{equation}\label{Eq_cov_matrix}
    \mathbf{\textbf{V}^{-1}} =
    \left( \begin{array}{cccc}
    \textbf{V}^{-1}_{1} & {}                  & {}     & {0}     \\
    {}                  & \textbf{V}^{-1}_{2} & {}     & {}      \\
    {}                  & {}                  & \ddots & {}      \\
    {0}                 & {}                  & {}     &\textbf{V}^{-1}_{k}
    \end{array} \right),
\end{equation}
where \emph{V}${}_{1}$, \emph{V}${}_{2}$$\cdots $\emph{V}${}_{k }$ refer to the covariance matrixes of the sub-set data,
which are independent with each other. The covariance matrix of calculated values is
\begin{equation}\label{Eq_cov_cal}
    {\textbf{V}}_{\textbf{y}}=\textbf{D}\textbf{V}_{\textbf{P}}\textbf{D}^{+}.
\end{equation}
The sensitivity matrix is quite useful in eliminating redundant fit parameters and in understanding which fit parameters
are the most effective on the low-energy extrapolation of the \textit{S }factor as discussed below.

Formula adopted for optimizing with \emph{R}-matrix fitting is
\begin{equation}\label{CHI}
\chi^{2}=({\mathbf \eta} - {\mathbf y})^{+} {\mathbf V}^{-1} ({\mathbf \eta} - {\mathbf y})\Longrightarrow minimum.
\end{equation}
Here ${\mathbf \eta }$ refers to the vector of experimental data, ${\mathbf y}$ refers to the vector of calculated values.
Using covariance statistics and error propagation law, it enables us to get accurate expected values and standard deviation of S factor.

In addition the Peelle Pertinent Puzzle (PPP) was corrected by the method used in~\cite{Carl09}. RAC was used to produce the accurate
(error= 1 \%) ${}^{6}$Li(n,$\alpha $) and ${}^{10}$B(n,$\alpha $) cross sections, for International Evaluation of Neutron
Cross section Standards~\cite{Carl08,Carl09}. And RAC was comprehensively compared with R-Matrix code EDA and SAMMY of
USA~\cite{Carl08,Carl09}, the results were highly identical when the same parameters are used. To verify the performance of
the \textit{R}-matrix code of $^{16}$N $\alpha $ spectrum, we repeated the analysis of Ref.~\cite{Tang10} using their input data,
and the same results were obtained. In a word, it is proved that the code RAC is reliable.

\section{Evaluation and Fit of experimental data}
\subsection{The construction of reaction channel}

\begin{table}[b]
\caption{\label{tab:tablechannle}
The reaction channels in our fit including Q-values, radii, the maximum of the orbital angular momentum,
and the data, respectively, in the R-matrix calculation. For the capture reaction, Magnetic \emph{L}-pole radiation is weaker
than the corresponding electric \emph{L}-pole radiation significantly, so the Magnetic transitions are not considered in our fit. }
\begin{ruledtabular}
\begin{tabular}{cccccccc}
$Channel$     & $Q(MeV)$  & $R(fm)$  & $\emph{l}_{max}$  & $Data$  \\
\hline

$\alpha+{}^{12}$C          & $0.000$  & $6.5$ & $6$  &  $AD\footnotemark[1],^{16}$N    \\
$\gamma_{0}+{}^{16}$O$_{0}$  & $7.162$  & $6.5$ & $3$  &  $AD,S_{g.s.}$  \\
$\gamma_{1}+{}^{16}$O$_{1}$  & $1.113$  & $6.5$ & $1$  &  $S_{6.05}$  \\
$\gamma_{2}+{}^{16}$O$_{2}$  & $1.032$  & $6.5$ & $1$  &  $S_{6.13}$  \\
$\gamma_{3}+{}^{16}$O$_{3}$  & $0.245$  & $6.5$ & $1$  &  $S_{6.92}$  \\
$\gamma_{4}+{}^{16}$O$_{4}$  & $0.045$  & $6.5$ & $1$  &  $S_{7.12}$  \\
$\alpha_{1}+{}^{12}$C      & $-4.438$ & $6.5$ & $2$  &  $AD,\sigma$  \\
$p+{}^{15}$N               & $-4.968$ & $6.5$ & $2$  &  $AD$  \\
\end{tabular}
\end{ruledtabular}
\footnotetext[1]{Here AD is the abbreviation of angular distribution.}
\end{table}

\begin{figure*}
\centering
\includegraphics[scale=0.90]{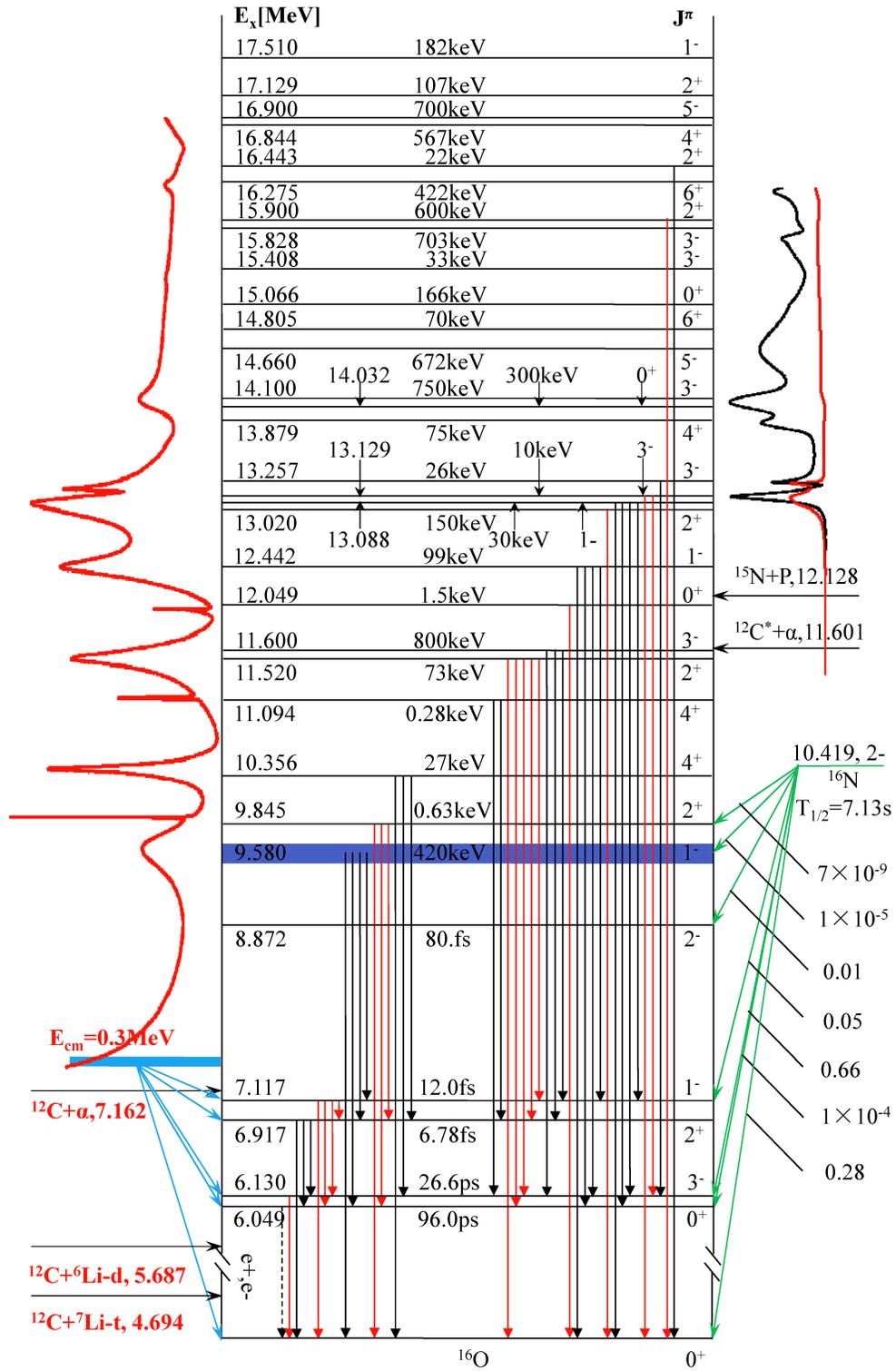}
\caption{(Color online)
Level scheme of the ${}^{16}$O nucleus~\cite{Till93}. All states relevant for the analysis are indicated.}
\label{fig:com_3}
\end{figure*}

The observed position, width or life time of ${}^{16}$O levels are displayed in Fig.~\ref{fig:com_3} up to the 17.5 MeV energy region
from Ref.~\cite{Till93}. The cascade transition data in Ref.~\cite{Schu11} and Ref.~\cite{Mate08} reveal several $\gamma$-ray
cascade transitions from the $1^{-}$, $3^{-}$ and $2^{+}$ states at $E_{x}$ = 7.12, 11.49 and 11.51 MeV, respectively,
which are considered in our fit. Except the level $J^{\pi}$ = $2^{-}$ ($E_{x}$ = 8.872 MeV), the other 31 levels contain
$\alpha$+${}^{12}$C reaction channel. Based on the level scheme of the ${}^{16}$O nucleus, the \emph{R}-matrix analysis of
the ${}^{16}$O compound nucleus considers one particle entrance channel ${}^{12}$C+$\alpha$, eight particle exit channels
as shown in the Table~\ref{tab:tablechannle}. The \emph{R}-matrix calculations were performed for $J^{\pi}$ = $0^{+}$
(four real levels, one background level), $J^{\pi}$ = $1^{-}$ (five real levels, one background level), $J^{\pi}$ = $2^{+}$
and $3^{-}$ (seven real levels, and one background level), $J^{\pi}$ = $4^{+}$ (four real levels, one background level),
$J^{\pi}$ = $5^{-}$ (two levels, one background level) and $J^{\pi}$ = $6^{+}$ (two real levels). The available data sets
cover the energy from $E_{c.m.}$ = 0.9 MeV to $E_{c.m.}$ = 7.5 MeV in this fit, so the parameters above this region are fixed
at values determined from the Ref.~\cite{Till93}.

\subsection{Evaluation of experimental data}
The basic principle of the data evaluation is that the database can reflect the information of nuclear structure and nuclear reaction
accurately and objectively, no matter which is to use the original data or the appropriate amendment. \emph{R}-matrix fitting
requires the experimental data covering full energy region with complete energy points and continuous values, especially
in the resonance peak area with the different types of data. Reliable experimental data subset should satisfy the following requirements:
In the resonance peak area, the sum of \emph{S} factor in different reaction channel should be equal to the total S factor;
The peak position of the different types of data should be consistent within the range of error; The principal value of different groups should be consistent within the range of uncertainty; The width data of resonance peaks are matched to the implied width information of the other data; The integral value of the differential data should be equal to the corresponding integral data; The integral data of different groups should span a broad energy range with a number of data points and have a good match with each other.

According to the principle of maximum likelihood, a fit to a dataset with many types and large amount of points needs to meet
the approximate statistical distribution, so the revisions of some dataset are reasonable. If one experimental point deviates
from the expectations obviously, such as the residual error larger than three times of uncertainty, the error of this point
can be enlarged with the Letts' criteria (3$\sigma$ criteria); In the same type of data, if the difference of principal value is
far greater than their uncertainties, the error of corresponding data should be amplified in the fitting; If the principal value
in one group data deviates from the expected value wholly, the normalization to this dataset is needed in the fitting;
If one high precision dataset is selected as the standard data in the evaluation, then some data with systematical deviation
should be normalized to the standard data.

\subsection{Iterative fit}

\begin{figure}
\center
\includegraphics[scale=0.36]{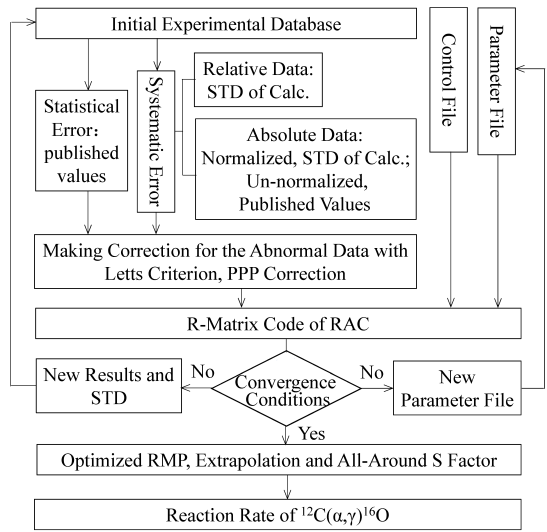}
\caption{
    The flow chart for the procedure of iterative fit.}
\label{fig:Iterat}
\end{figure}

The fits to the data of ${}^{16}$O system are iterated, to achieve internal consistency. A file is a fixed record of the original data, which is to provide the original statistical error for the fit. Another file is a dynamic data file recording the evaluation process,
which role is to provide the actually used data in fitting and is updated in the iterative process. In the file the original
relative data values are replaced with the new normalized value, and the systematic error values are updated by the
standard deviation (STD) of the new calculation. And the statistical errors are renewed with the original one at the beginning,
but some of them are corrected according to the Letts' criteria. The ratio of the corresponding data in there two files
is the new scaling factor or normalization coefficient. The scaling factor is adjustable in RAC, which is recorded in the parameter file together with the new R-matrix parameters. Fig.~\ref{fig:Iterat} shows the flow chart of R-matrix iterative fit procedure.

With the continuity of iterative fit, the variation of scaling factor becomes smaller and smaller, and the principal values
of relative experimental data are closer to their expectations. Similarly, the R-matrix parameters (RMP), fitted values and
their standard deviations become more accurate. At last, all calculated values tend to  very slight fluctuations,
and the $\chi^{2}$ approaches the minimum.

\section{Results and discussion}

The following subsections give details of the different reaction channels included in this analysis. Although they are described individually, the fits to the different reaction channel data sets have been performed simultaneously.

\subsection{The reduced $\alpha$-width amplitude for the bound states}

At energies of astrophysical interest, direct cross section measurement of capture reaction, such as
$^{11}$B(p,$\gamma$)$^{12}$C, is very difficult because of the Coulomb barrier, but it can be derived by
the proton spectroscopic factor and asymptotic normalization coefficients (ANC) from the transfer reaction
$^{12}$C($^{11}$B,$^{12}$C)$^{11}$B~\cite{Li14} based on distorted wave Born approximation (DWBA) analysis~\cite{Du15,Canb15,Wu14}.
The S factor of ${}^{12}$C($\alpha $, $\gamma $)${}^{16}$O at astrophysical energies arises largely
from the high-energy tails of subthreshold states $2^+_1$(Ex = 6.92 MeV) and $1^-_1$(Ex = 7.12 MeV) of ${}^{16}$O,
but the properties of these states are only weakly constrained by cross-section measurements
at higher energies. The cross section of transfer reactions $({}^{6}Li, d)$ and $ ({}^{7}Li, t)$ provides an alternative way
for extracting the reduced $\alpha$-widths for these states of ${}^{16}$O. In this fit, the $\gamma_{\alpha}$ of $1^-_1$ and $2^+_1$
bound states are fixed to the weighted average of two new measurements~\cite{Belh07,Oule12}, and the other subthreshold states,
the $\gamma_{\alpha}$ of $0^+_1$ and $3^-_1$ are adopted by the literature value of Ref.~\cite{Oule12}. While the $\gamma_{\gamma}$ of
the four states could vary within their uncertainties of literature ~\cite{Till93}.

\subsection{Total S factor}

\begin{figure}
\center
\includegraphics[scale=0.29]{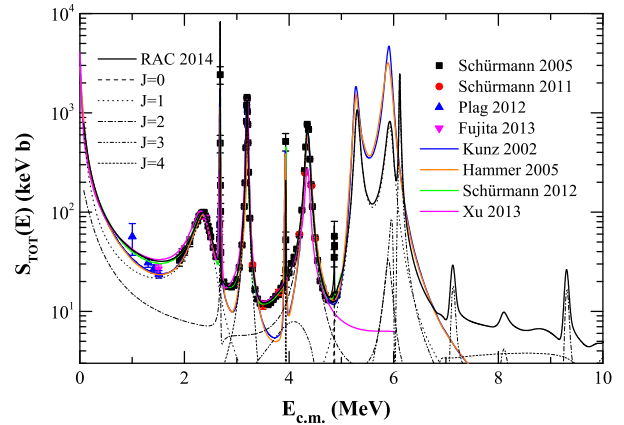}
\caption{
    (Color online) Results of the best R-matrix fit for the $S_{tot}$ data from Sch\"urmann 2005:~\cite{Schu05}, Sch\"urmann 2011:~\cite{Schu11}, Plag 2012:~\cite{Plag12}, and Fujita 2013:~\cite{Fuji13}, together with the decomposition into
    different energy level contributions. For comparison, the results of Kunz 2002:~\cite{Kunz02}, Hammer 2005:~\cite{Hamm05},
    Sch\"urmann 2012:~\cite{Schu12}, and Xu 2013:~\cite{Xu13} are shown in this figure, respectively.}
\label{fig:Stot}
\end{figure}

The available total S factor of ${}^{12}$C($\alpha $, $\gamma $)${}^{16}$O have been obtained in four independent experiments
~\cite{Schu05,Schu11,Plag12,Fuji13}. Fig.~\ref{fig:Stot} illustrates the corresponding fitted values. In general,
the fits are perfect where all the energy levels are accurately described. The measurement of Sch\"urmann \emph{et al}.~\cite{Schu05,Schu11}
in inverse kinematics using the recoil mass separator ERNA allowed to collect data with high precision in a wide energy range,
which would make a good restriction to the extrapolation of ground transition, cascade transitions and the total S factor.
The data of Ref.~\cite{Schu05} have not given definite numerical value of three narrow peaks 2$^{+}_{2}$, 4$^{+}_{1}$ and 0$^{+}_{2}$,
and the author's personal communication considers that the relative numerical value is difficult to be determined, so the excitation energies and partial widths are fixed by including in the dataset of pseudo cross section points which were assigned by 50\% errors around the resonance peaks.

In the peak region of the 1$^{-}_{2}$ energy level ($E_{x}$ = 9.58 MeV), the evaluation of data is as follow. We can learn
the $S_{tot}$ $\approx$ 97.0 keV b from the measurement of Sch\"urmann \emph{et al}.~\cite{Schu05}. The evaluated S${}_{g.s} \approx$ 76.0
is from six groups of S${}_{g.s}$. Estimating the S${}_{6.05}$ $\approx$ 1.0 keV b, S${}_{6.92}$ $\approx$ 7.0 keV b and
S${}_{7.12}$ $\approx$ 20.0 keV b from Ref.~\cite{Mate06} and Ref.~\cite{Kunz01} and making assumption for S${}_{6.13}$ $\approx$ 1.0 keV b.
In view of the above, the sum of the partial S factor is 105 keV b , which is 1.08 times of the results $S_{tot}$ = 97 keV b
from the measurement of Sch\"urmann \emph{et al}.~\cite{Schu05}, that is to say, the experimental value of the total S-factor
is significantly less than the sum of the experimental partial S factor. Relevant Ref.~\cite{Schu05} accounts for that
the maximal systemical error is 6.5\%, so we choose 1.03 as the normalization coefficient of the data~\cite{Schu11} in the careful exploration,
then the data become S${}_{tot}$ = 100 keV b, which is lower than the sum of partial S factors. The systematical study shows
that the S${}_{6.92 }$ of Ref.~\cite{Kunz01} has an increasing trend, and the S${}_{7.12 }$ of this paper has a decreasing trend.
When taking the normalization coefficient of S${}_{6.92}$ and S${}_{7.12}$ as 1.00 and 0.95, respectively, then the sum of partial S-factor
is approximately equal to 100 keV b. Theresore we can get a satisfied dataset which has complete types and numerical self-consistency
for the main resonance peak 1$^{-}_{2}$. These constitute the skeleton of the whole database for the fits.

Another skeleton of the dataset is the data on the peak region of 2$^{-}_{3}$ at $E_{c.m.}$ = 4.358 MeV (See Fig.~\ref{fig:Stot}).
The data of Sch\"urmann \emph{et al}.~\cite{Schu11} is obtained by the adding of their components S${}_{g.s}$, S${}_{6.05}$,
S${}_{6.13}$, S${}_{6.92 }$ and S${}_{7.12}$, and it is consistent with the total S factor of Sch\"urmann \emph{et al}.~\cite{Schu05}
very well. All kinds of the data of Ref.~\cite{Schu11} are used as the standard data, and the normalization coefficient is 1.03.

Recently, the ${}^{12}$C($\alpha $, $\gamma $)${}^{16}$O cross sections of Plag \emph{et al}.~\cite{Plag12} have been measured
at four energy points, $E_{c.m.}$ between 1.00 and 1.51 MeV, and the $E10$ and $E20$ components were derived with
an accuracy comparable to the previous best data obtained with HPGe detectors. This data are first employed in the $S_{tot}$ fit,
which have great influences on the $S_{tot}$(0.3 MeV). In Ref.~\cite{Fuji13}, total cross section measurements for
$E_{c.m.}$= 2.4 and 1.5 MeV were performed at KUTL by using a tandem accelerator. And our fit results are relatively
close to the principal values.

In current research on S factor of Ref.~\cite{Kunz02,Hamm05} at higher energies, i.e., at $E_{c.m.}$ $>$ 2.8 MeV, resonance parameters
taken from Ref.~\cite{Till93} were used in their R-matrix fit, in which the published data at high energy, such as $\alpha$
capture measurements of Ref.~\cite{Broc73} were neglected. So the high-energy resonances from $E_{c.m.}$ = 5 MeV to $E_{c.m.}$ = 6 MeV
are overestimated apparently (please see the fit of $S_{g.s.}$). In addition, one should note that in the analysis of
Ref.~\cite{Kunz02,Hamm05} there is a clear disagreement at energies around $E_{c.m.}$ = 3 and 4 MeV, where the calculation
underestimates total cross-section. The latest results of Ref.~\cite{Schu12} are consistent with the available experimental data,
but the high energy data are not analysed in a similar way of the R-matrix fit. In Ref.~\cite{Xu13}(NACREII) the total
and partial S factors are analyzed with the potential model, where the S factor at $2^{+}_3$ ($E_{x}$ = 11.52 MeV)
is underestimated  by the calculation.

\subsection{${}^{{\mathbf 16}}{{\mathbf N}}$ $\alpha $ spectrum}

\begin{figure}
\center
\includegraphics[scale=0.29]{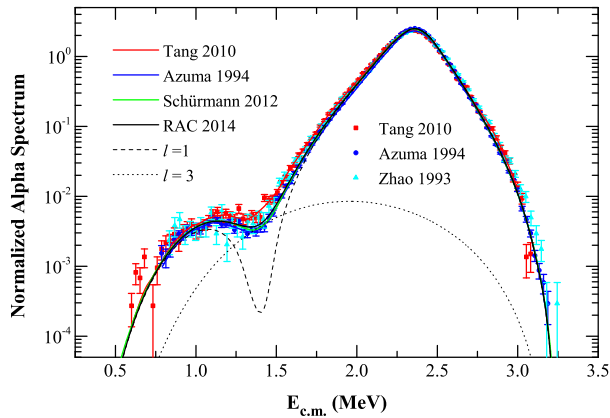}
\caption{
 (Color online) Results of the best R-matrix fits (Black line) for the ${}^{{\mathbf 16}}{{\mathbf N}}$ $\alpha $ spectra
 from Tang 2010:~\cite{Tang10}, Azuma 1994:~\cite{Azum94}, and Zhao 1993:~\cite{Zhao93}, together with the decomposition into
 p- (dashed line) and f- wave (dotted line) contributions. For comparison, the best fits of Tang 2010:~\cite{Tang10},
 Azuma 1994:~\cite{Azum94}, and Sch\"urmann 2012:~\cite{Schu12} scaled by the corresponding coefficients are shown.}
\label{fig:16N}
\end{figure}

\begin{figure}
\center
\includegraphics[scale=0.29]{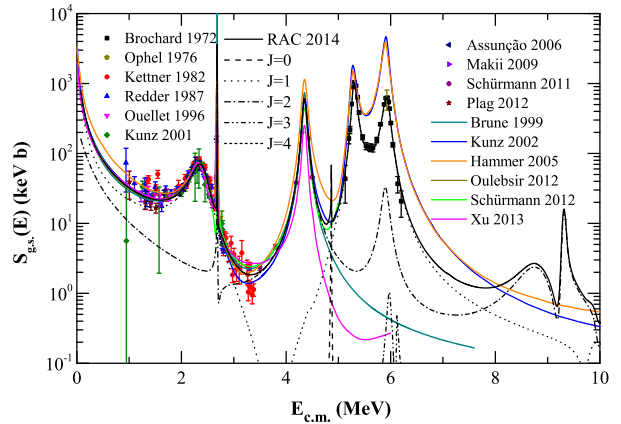}
\caption{
    (Color online) Results of the best R-matrix fits for the ground state transitions $S_{g.s.}$ data from Brochard 1972:~\cite{Broc73},
    Ophel 1976:~\cite{Ophe76}, Kettner 1982:~\cite{Kett82}, Redder 1987:~\cite{Redd87}, Ouellet 1996:~\cite{Ouel96},
    Kunz 2001:~\cite{Kunz01}, Assun\c c\~{a}o 2006:~\cite{Assu06}, Makii 2009:~\cite{Maki09}, Sch\"urmann 2011:~\cite{Schu11},
    Plag 2012:~\cite{Plag12}, together with the decomposition into different energy level contributions. For comparison,
    the results of Brune 1999:~\cite{Brun99}, Kunz 2002:~\cite{Kunz02}, Hammer 2005:~\cite{Hamm05},
    Sch\"urmann 2012:~\cite{Schu12}, Oulebsir 2012:~\cite{Oule12}, and Xu 2013:~\cite{Xu13} are shown.}
\label{fig:SCG0}
\end{figure}

The shape of the low-energy part of the $\beta $-delayed $\alpha $ spectrum of ${}^{16}$N is very sensitive to the
$\alpha $ + ${}^{12}$C reduced width of the $1^{-}_1$ subthreshold state and $1^{-}_2$ state of ${}^{16}$O, in turn,
which dominates the low-energy p-wave capture $S_{E10}$(0.3MeV) of $^{12}$C($\alpha,\gamma_0$)$^{16}$O${}_0$.
In this energy region the reduced $\alpha$ widths are determined by the $\alpha $ spectra and the angular distributions
of {$^{12}$C($\alpha,\alpha$)$^{12}$C}, which results in the competition with each other in the fit.
As shown in Ref.~\cite{Buch09}, there exists a limitation by
the use of {$^{12}$C($\alpha,\alpha$)$^{12}$C} data in Ref.~\cite{Plag87} for obtaining reliable values of
S${}_{E10}$(0.3 MeV). So the ${}^{16}$N $\alpha $ spectrum may help to give a better confirmation of
the reduced $\alpha $ width amplitude of 1$^{-}_{1}$ and 1$^{-}_{2}$. Included in this analysis are the three independent
$\alpha $ spectra data of Refs.~\cite{Tang07,Tang10,Azum94,Zhao93}, and the normalization for probability spectrum is used
in the practice to reduce the influence of systematical errors. Fig.~\ref{fig:16N} shows the fit to the normalized
${}^{16}$N $\alpha $ spectrum together with the decomposition into p- and f-wave contributions, which suggests a significantly
negative interference of the bound ${{\rm 1}}^-_1$ state with the broad ${{\rm 1}}^-_2$ state that leads to a second peak
at $E_{c.m.}$ = 1.1 MeV and a minimum in the vicinity of 1.4 MeV. The dotted line denotes the contribution of 3$^{-}$ state,
which perfectly compensates this negative interference. The fit concluded that the measurement of Azmua \emph{et al}~\cite{Azum94}
most likely represents the currently closest approximation to the true $\alpha $ spectrum.

\subsection{$^{12}$C($\alpha,\gamma_0$)$^{16}$O${}_0$}

For the ground state transitions, the secondary data of E${}_{10}$ and E${}_{20}$ multipoles were used in the previous
R-matrix analysis independently~\citep{Bark91,Schu12}. In general, these secondary data were obtained from the Legendre
polynomials fit~\citep{Dyer74} (page 510) to the angular distributions of $^{12}$C($\alpha,\gamma_0$)$^{16}$O${}_0$
measured at many discrete energies. Two methods of analysis (phase fixed or free) are often applied,
however, the derived S factors S${}_{E10}$ and S${}_{E20}$ are significantly different for the same $\gamma$
angular distributions, see the Fig.12 of Ref.~\cite{Assu06}, especially for the S${}_{E20}$.

In our fit, the S${}_{g.s.}$=S${}_{E10}$+S${}_{E20}$ is used for the ground state transition, which the proportion of
S${}_{E10}$ and S${}_{E20}$ are determined by the R-matrix fit to the relevant $\gamma $ angular distributions.
Fig.~\ref{fig:SCG0} shows the fit to ${}^{{\rm 12}}{{\rm C}}\left(\alpha {\rm ,}{\gamma }_0\right){{}^{{\rm 16}}{{\rm O}}}_0$
data of ten independent measurements and the calculations of previous works. It is worth mentioning that after the experiments
by Dyer and Barnes~\cite{Dyer74}, Kettner \emph{et al}~\cite{Kett82}, and Redder \emph{et al}~\cite{Redd87},
a weighted average value $\sigma $ of 47 $\pm$ 3 nb at resonance peak of ${{\rm 1}}^-_2$ was used to derive a cross section
at low energy, this data plays a vital role for the determination of S${}_{g.s.}$, and can be regarded as a criterion
for normalizing the experimental data. Even though the S${}_{g.s.}$ of Kettner \emph{et al}~\cite{Kett82} deviates
systematically from the other data in this resonant region, it is dispensable since it is the only one that
has the data points at above 3.0 MeV. It is noteworthy that if the normalization coefficient is fixed by a factor of 0.87,
the data of ${{\rm 1}}^-_2$ peak region is consistent with the other data, meanwhile the data above 3.0 MeV is well
consistent with the S${}_{g.s.}$ of Sch\"urmann \emph{et al}.~\cite{Schu11}.
\begin{figure}
\center
\includegraphics[scale=0.29]{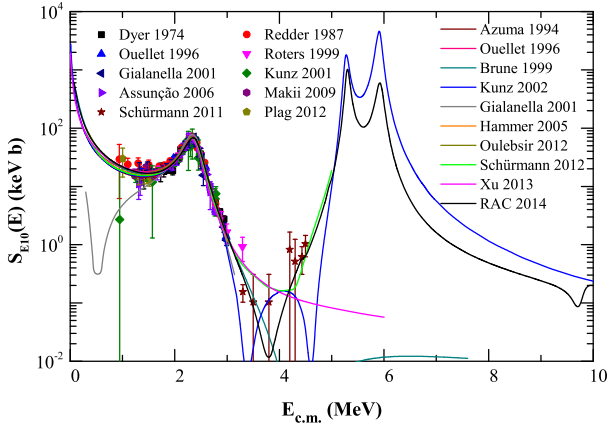}
\caption{
    (Color online) The calculation of the $S_{E10}$ with the best R-matrix fit. For comparison,
    previous fits of Azuma 1994:~\cite{Azum94}, Ouellet 1996:~\cite{Ouel96}, Brune 1999:~\cite{Brun99},
    Gialanella 2001:~\cite{Gial01}, Kunz 2002:~\cite{Kunz02}, Hammer 2005:~\cite{Hamm05},
    Sch\"urmann 2012:~\cite{Schu12}, Oulebsir 2012:~\cite{Oule12}, and Xu 2013:~\cite{Xu13} are shown in this figure.
    Data points shown are taken from Dyer 1974:~\cite{Dyer74}, Redder 1987:~\cite{Redd87}, Ouellet 1996:~\cite{Ouel96},
    Roters 1999:~\cite{Rote99}, Kunz 2001:~\cite{Kunz01}, Gialanella 2001:~\cite{Gial01}, Assun\c c\~{a}o 2006:~\cite{Assu06},
    Makii 2009:~\cite{Maki09}, Sch\"urmann 2011:~\cite{Schu11}, and Plag 2012:~\cite{Plag12}.}
\label{fig:SE10}
\end{figure}

\begin{figure}.
\center
\includegraphics[scale=0.29]{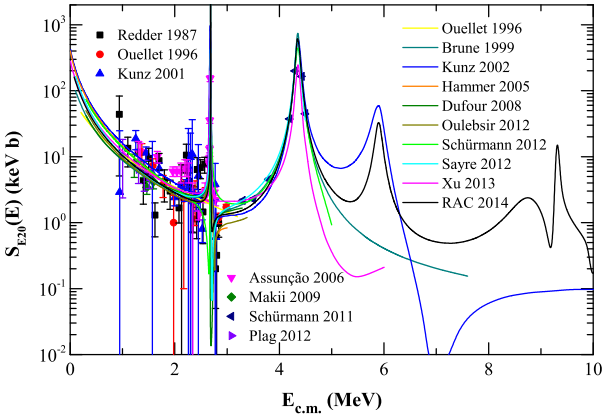}
\caption{
    (Color online) The calculation of the $S_{E20}$ with the best R-matrix fit. For comparison,
    previous fits of Ouellet 1996:~\cite{Ouel96}, Brune 1999:~\cite{Brun99}, Kunz 2002:~\cite{Kunz02},
    Hammer 2005:~\cite{Hamm05}, Dufour 2008:~\cite{Dufo08}, Oulebsir 2012:~\cite{Oule12},
    Sch\"urmann 2012:~\cite{Schu12}, Sayre 2012:~\cite{Sayr12} and Xu 2013:~\cite{Xu13} are shown in this figure.
    Data points shown are taken from Redder 1987:~\cite{Redd87}, Ouellet 1996:~\cite{Ouel96},
    Kunz 2001:~\cite{Kunz01}, Assun\c c\~{a}o 2006:~\cite{Assu06}, Makii 2009:~\cite{Maki09},
    Sch\"urmann 2011:~\cite{Schu11}, and Plag 2012:~\cite{Plag12}.}
\label{fig:SE20}
\end{figure}

\begin{figure}
\center
\includegraphics[scale=0.28]{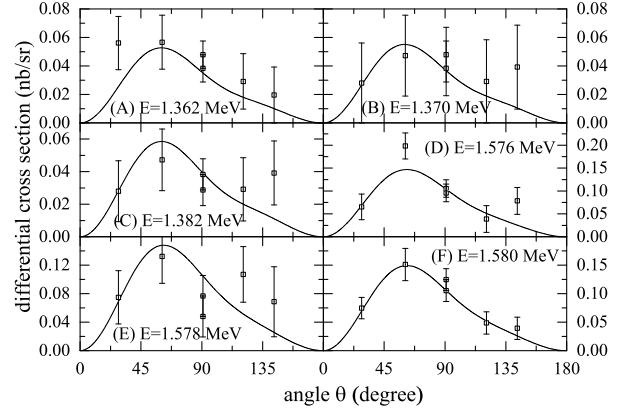}
\caption{
 Fits to the $^{12}$C($\alpha,\gamma_0$)$^{16}$O${}_0$ angular distributions of Ouellet 1996:~\cite{Ouel96}
 at $E_{c.m.}$ = 1.362 (A), 1.370 (B), 1.382 (C), 1.576 (D), 1.578 (E) and 1.580 (F) MeV.}
\label{fig:Ouellet_1}
\end{figure}

\begin{figure}
\center
\includegraphics[scale=0.28]{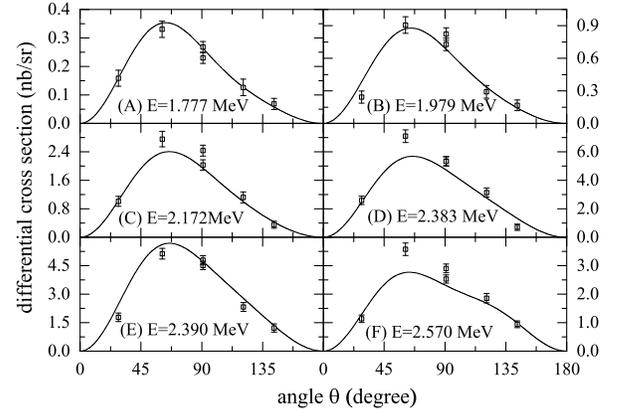}
\caption{
 Fits to the $^{12}$C($\alpha,\gamma_0$)$^{16}$O${}_0$ angular distributions of Ouellet 1996:~\cite{Ouel96}
 at $E_{c.m.}$ = 1.777 (A), 1.979 (B), 2.172 (C), 2.383 (D), 2.390 (E) and 2.570 (F) MeV.}
\label{fig:Ouellet_2}
\end{figure}

\begin{figure}
\center
\includegraphics[scale=0.28]{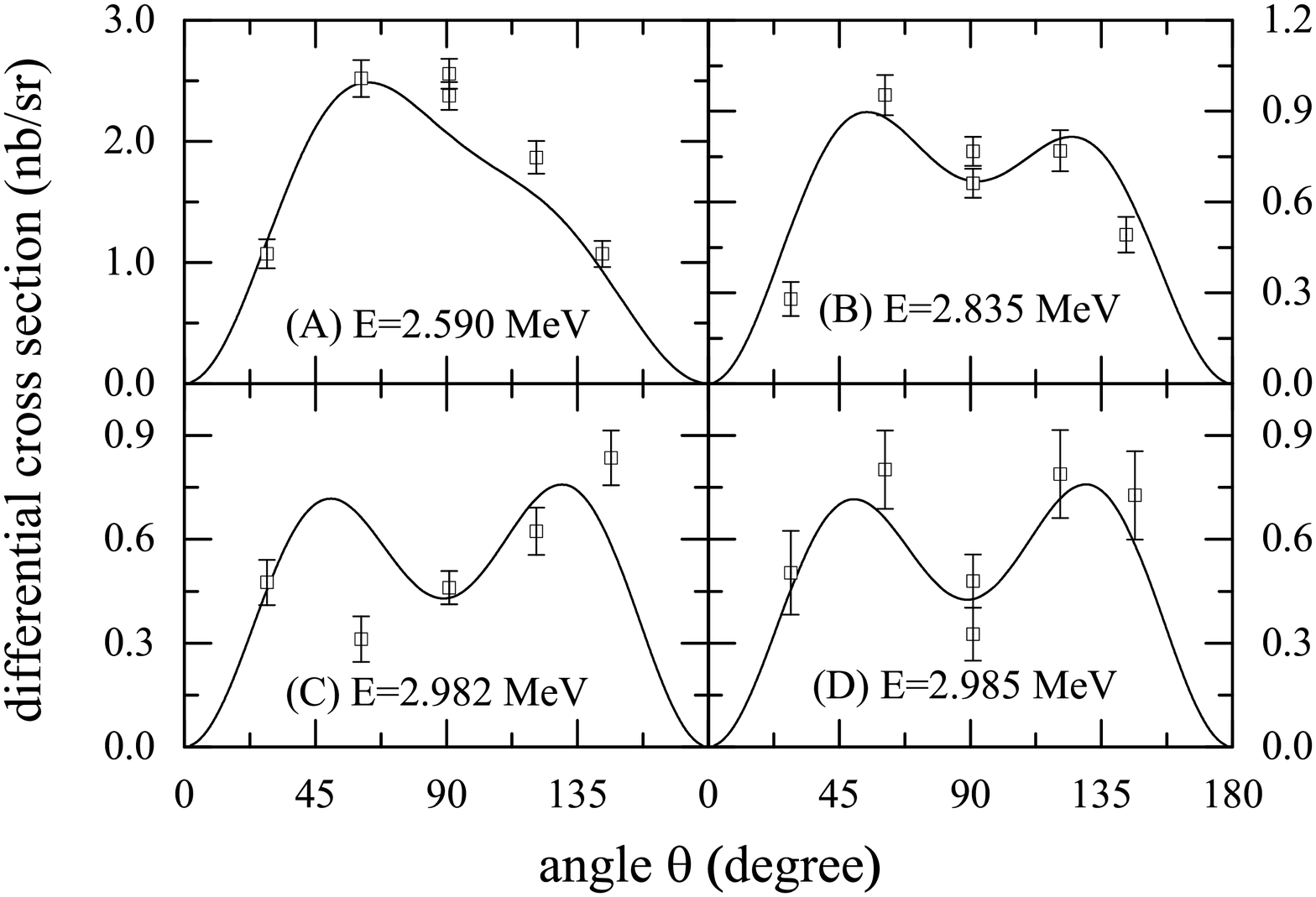}
\caption{
 Fits to the $^{12}$C($\alpha,\gamma_0$)$^{16}$O${}_0$ angular distributions of Ouellet 1996:~\cite{Ouel96}
 at $E_{c.m.}$ = 2.590 (A), 2.835 (B), 2.982 (C) and 2.985 (D) MeV.]}
\label{fig:Ouellet_3}
\end{figure}

\begin{figure}
\center
\includegraphics[scale=0.28]{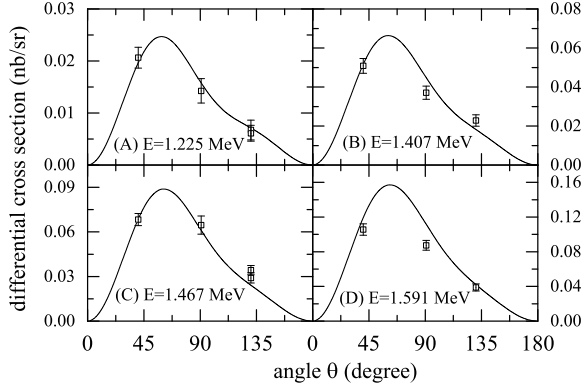}
\caption{
 Fits to the $^{12}$C($\alpha,\gamma_0$)$^{16}$O${}_0$ angular distributions of Makii 2009:~\cite{Maki09}
 at $E_{c.m.}$ = 1.225 (A), 1.470 (B), 1.467 (C), and 1.591 (D) MeV.}
\label{fig:Maki}
\end{figure}

In higher energy region, there are five independent experiments~\cite{Lars64, Mitc64, Kern71, Broc73,Ophe76}
covering the $1^-_3$ ($E_{x}$ = 12.44 MeV) and $1^-_4$ ($E_{x}$ = 13.09 MeV).
All the available measurements of the relative ratio of peak cross sections $\sigma$($1^-_3$)
/$\sigma$($1^-_4$), are tabulated in table 3 of Ref.~\cite{Ophe76}, in which the largest deviation of
Ref.~\cite{Broc73} data from the other three data, lower about 20\%, are evident. So in our fit, the data of
Ref.~\cite{Broc73} near $1^-_4$ resonance are corrected to the data at the peak of
$1^-_3$ with a factor 0.81. Then the cross sections are found to be in a good agreement
with the absolute data from Ref.~\cite{Ophe76} if normalization corrections (maximum of $\pm$20\%) are applied.
The remaining data from Ref.~\cite{Lars64,Mitc64,Kern71} show good agreement in the shape of the excitation curves,
and the normalization factors are given in Table.~\ref{tab:Normalization}.

\begin{figure}
\center
\includegraphics[scale=0.28]{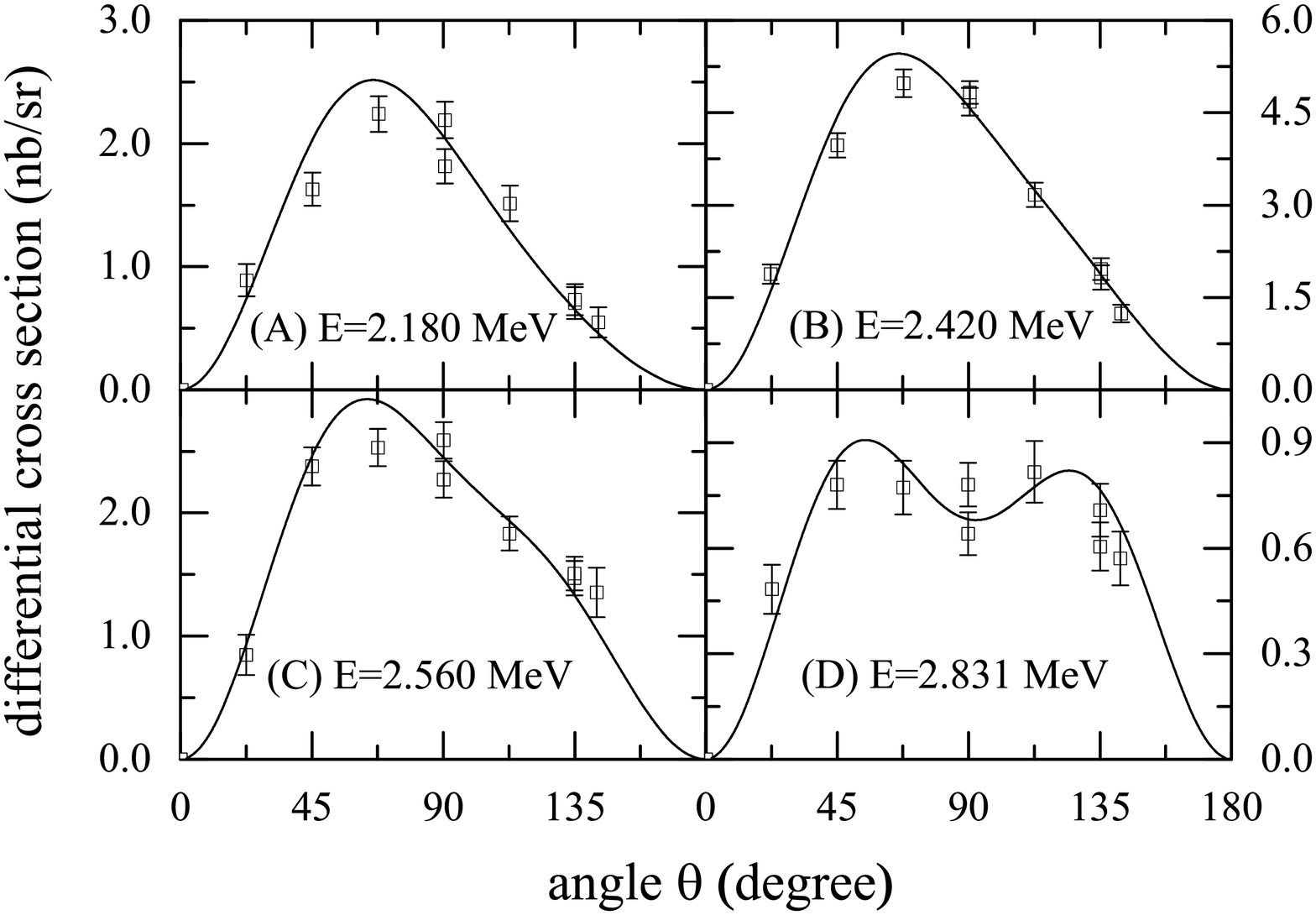}
\caption{Fits to the $^{12}$C($\alpha,\gamma_0$)$^{16}$O${}_0$ angular distributions of Dyer 1974:~\cite{Dyer74}
         at $E_{c.m.}$ =  2.180 (A), 2.420 (B), 2.560 (C), and 2.831 (D) MeV.}
\label{fig:Dyer}
\end{figure}

\begin{figure}
\center
\includegraphics[scale=0.28]{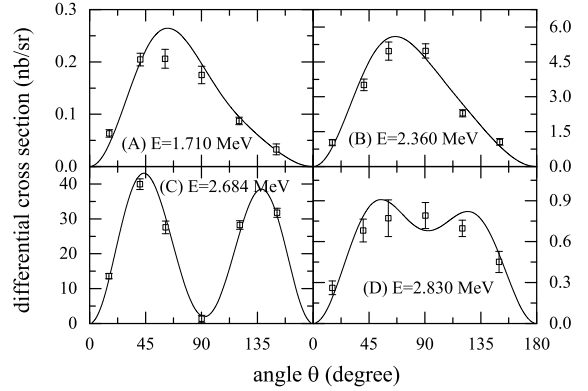}
\caption{ Fits to the $^{12}$C($\alpha,\gamma_0$)$^{16}$O${}_0$ angular distributions of Redder 1987:~\cite{Redd87}
        at $E_{c.m.}$ = 1.710 (A), 2.360 (B), 2.684 (C), and 2.830 (D) MeV.}
\label{fig:Redd}
\end{figure}

\begin{figure}
\center
\includegraphics[scale=0.28]{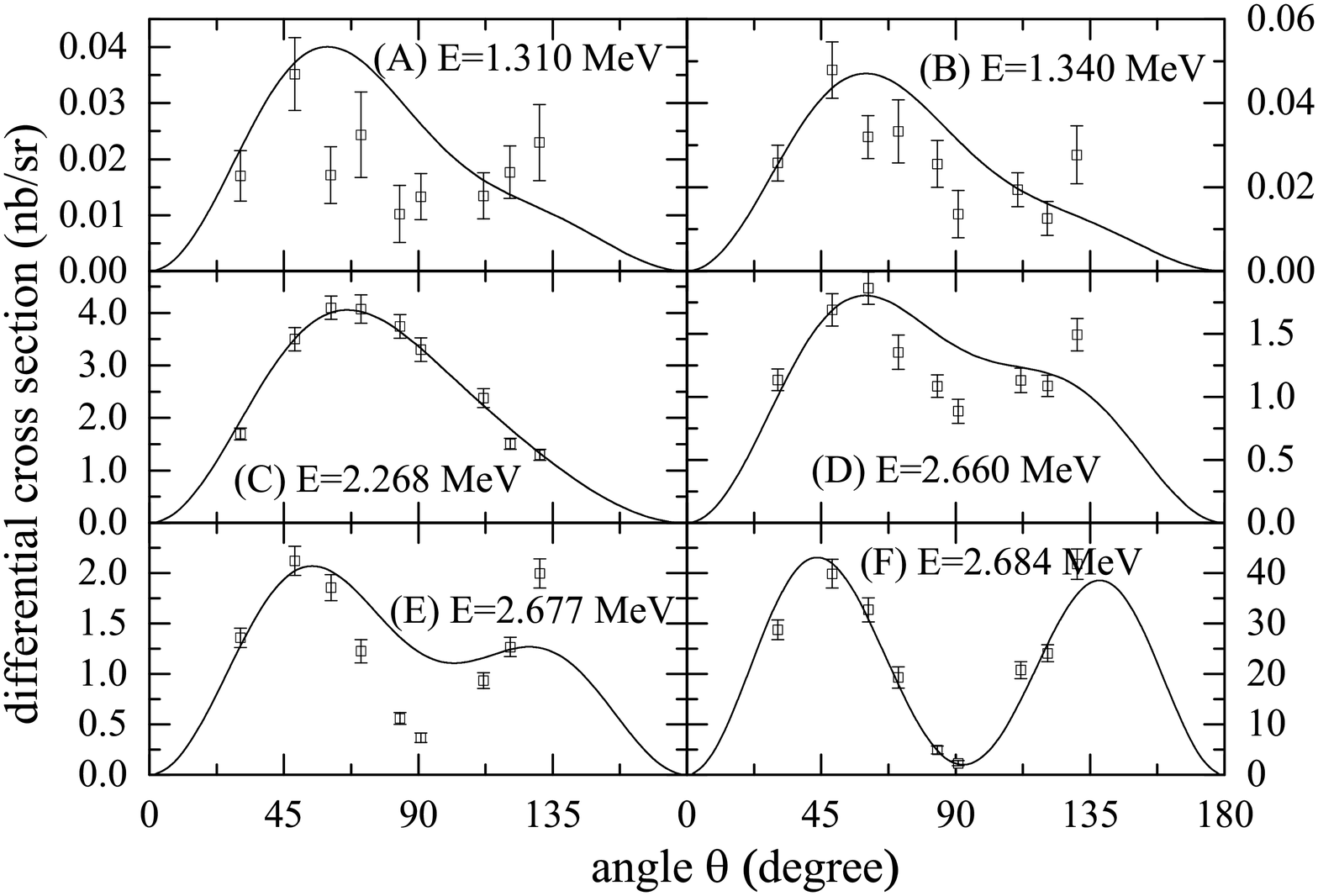}
\caption{
 Fits to the $^{12}$C($\alpha,\gamma_0$)$^{16}$O${}_0$ angular distributions of Assun\c c\~{a}o 2006:~\cite{Assu06}
 at $E_{c.m.}$ = 1.310 (A), 1.340 (B), 2.268 (C), 2.660 (D), 2.677 (E) and 2.684 (F) MeV.}
\label{fig:Assu}
\end{figure}

\begin{figure}
\center
\includegraphics[scale=0.28]{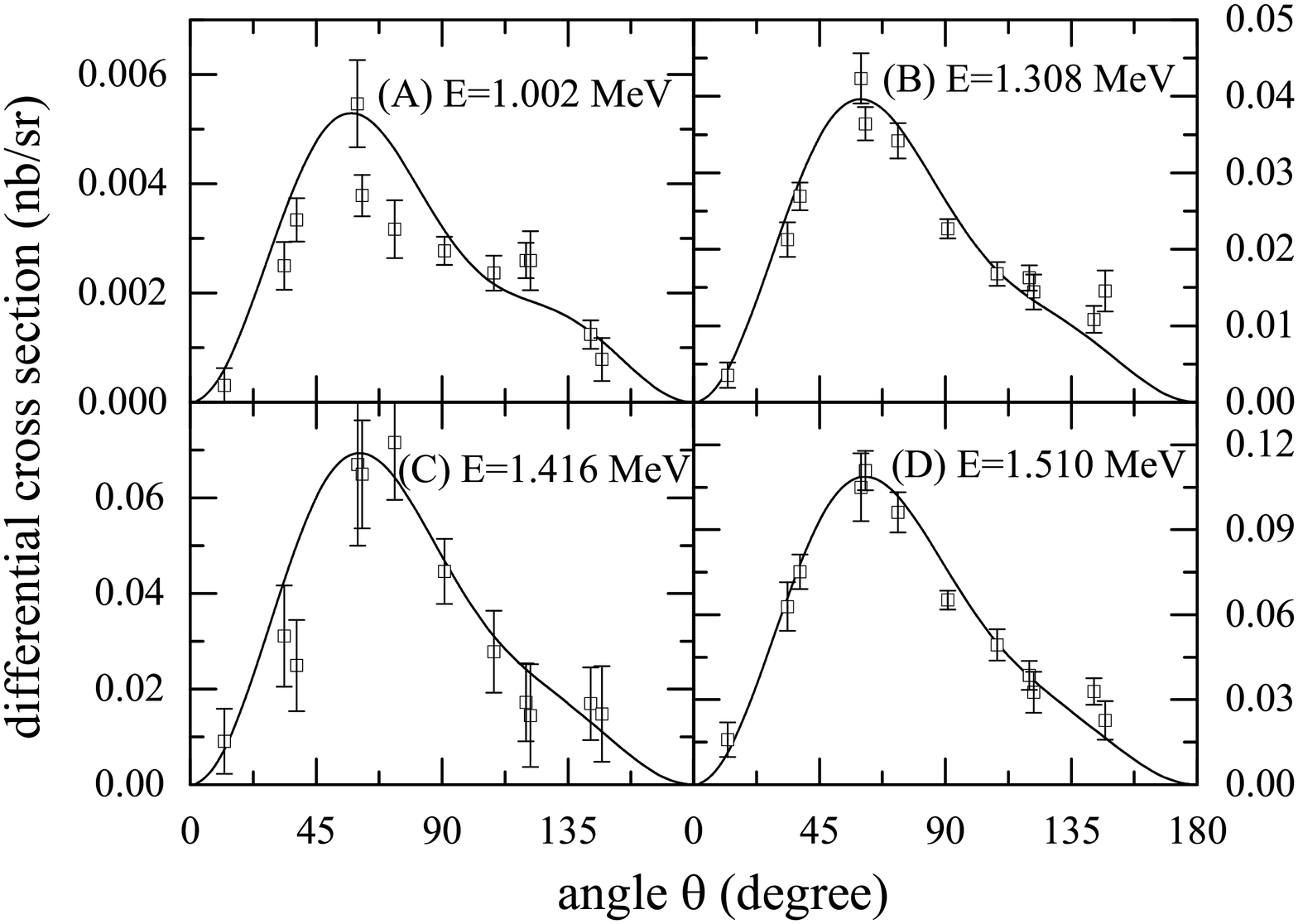}
\caption{
 Fits to the $^{12}$C($\alpha,\gamma_0$)$^{16}$O${}_0$ angular distributions of Plag 2012:~\cite{Plag12}
 at $E_{c.m.}$ = 1.002 (A), 1.308 (B), 1.416 (C) and 1.510 (D) MeV.}
\label{fig:Plag}
\end{figure}

\begin{figure}
\center
\includegraphics[scale=0.28]{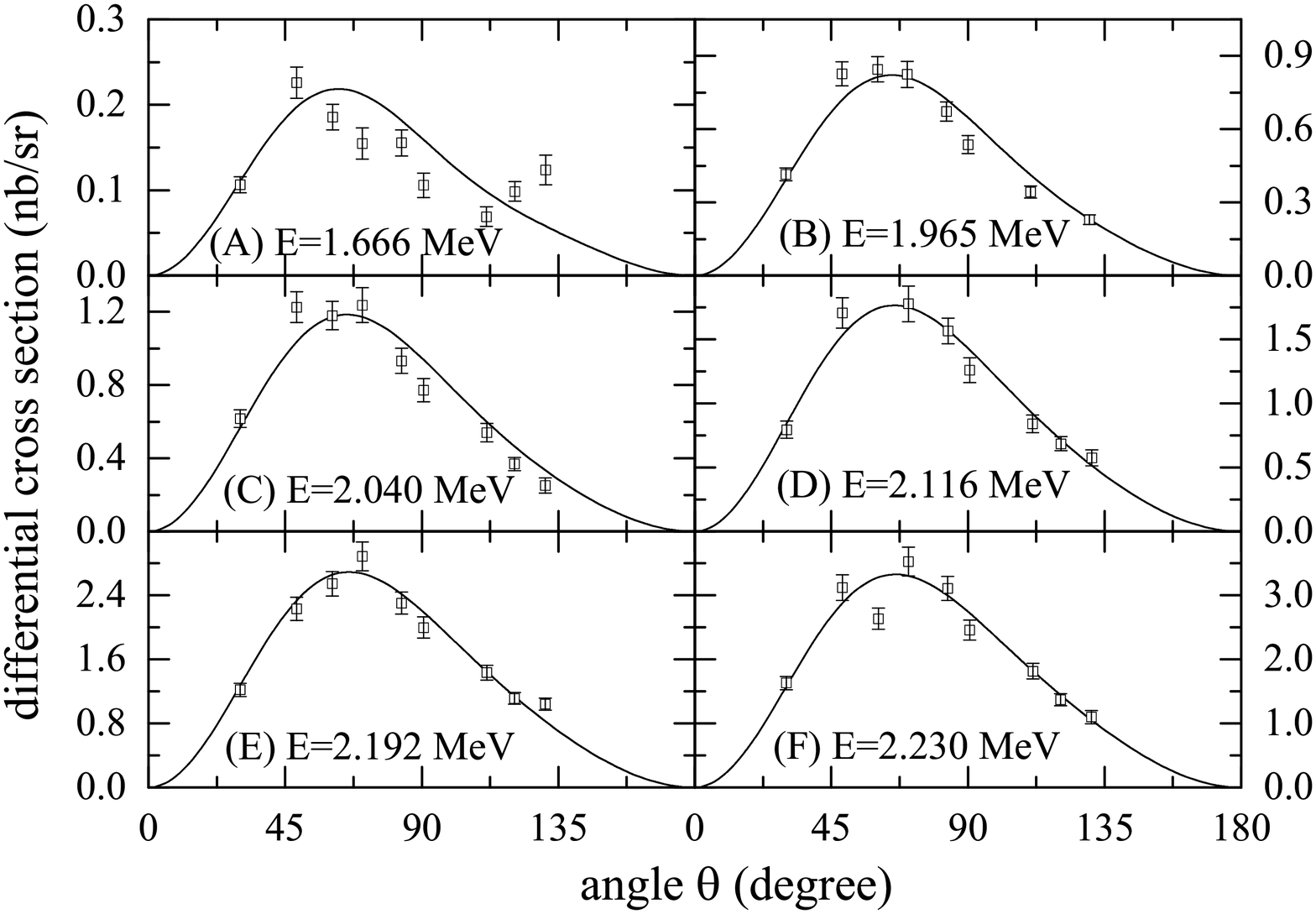}
\caption{
 Fits to the $^{12}$C($\alpha,\gamma_0$)$^{16}$O${}_0$ angular distributions of Fey 2004:~\cite{Fey04}
 at $E_{c.m.}$ = 1.666 (A), 1.965 (B), 2.040 (C), 2.116 (D), 2.192 (E) and 2.230 (F) MeV.}
\label{fig:Fey1}
\end{figure}

\begin{figure}
\center
\includegraphics[scale=0.28]{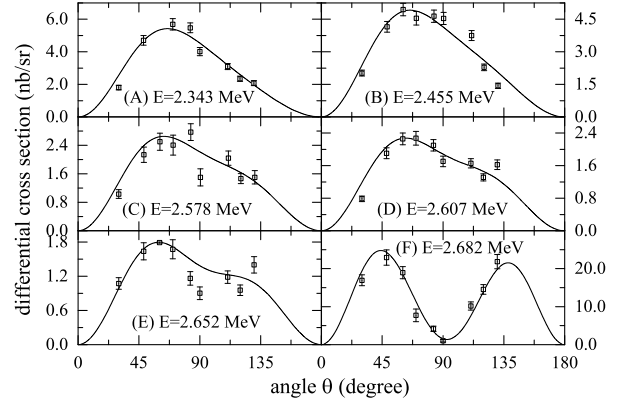}
\caption{
 Fits to the $^{12}$C($\alpha,\gamma_0$)$^{16}$O${}_0$ angular distributions of Fey 2004:~\cite{Fey04}
 at $E_{c.m.}$ = 2.343 (A), 2.455 (B), 2.578 (C), 2.607 (D), 2.652 (E) and 2.682 (F) MeV.}
\label{fig:Fey2}
\end{figure}

\begin{figure}
\center
\includegraphics[scale=0.28]{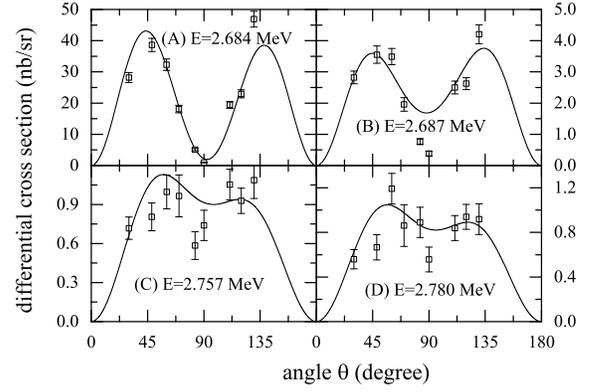}
\caption{
 Fits to the $^{12}$C($\alpha,\gamma_0$)$^{16}$O${}_0$ angular distributions of Fey 2004:~\cite{Fey04}
 at $E_{c.m.}$ = 2.684 (A), 2.687 (B), 2.757 (C) and 2.780 (D) MeV.}
\label{fig:Fey3}
\end{figure}

\begin{figure}
\center
\includegraphics[scale=0.28]{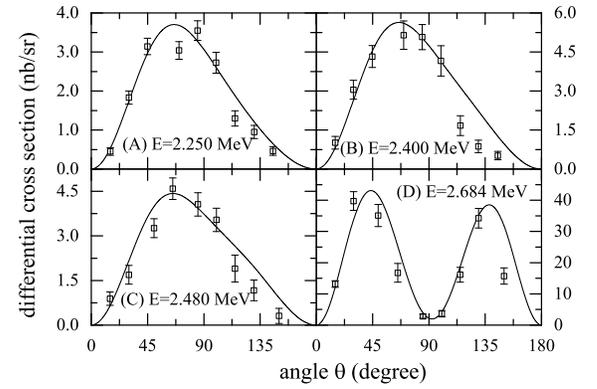}
\caption{
 Fits to the $^{12}$C($\alpha,\gamma_0$)$^{16}$O${}_0$ angular distributions of Kunz 1997:~\cite{Kunz97}
 at $E_{c.m.}$ = 2.250 (A), 2.400 (B), 2.480 (C) and 2.684 (D) MeV.}
\label{fig:Kunz}
\end{figure}

\begin{figure}
\center
\includegraphics[scale=0.28]{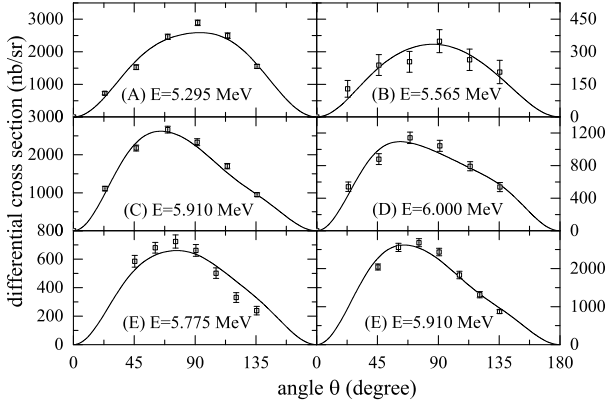}
\caption{
 Fits to the $^{12}$C($\alpha,\gamma_0$)$^{16}$O${}_0$ angular distributions of Larson 1964:~\cite{Lars64} and
 Kernel 1971:~\cite{Kern71} at $E_{c.m.}$ = 5.295 (A), 5.565 (B), 5.910 (C), 6.000 (D), 5.775 (E) and 5.910 (F) MeV.}
\label{fig:Lars}
\end{figure}

\begin{figure}
\center
\includegraphics[scale=0.30]{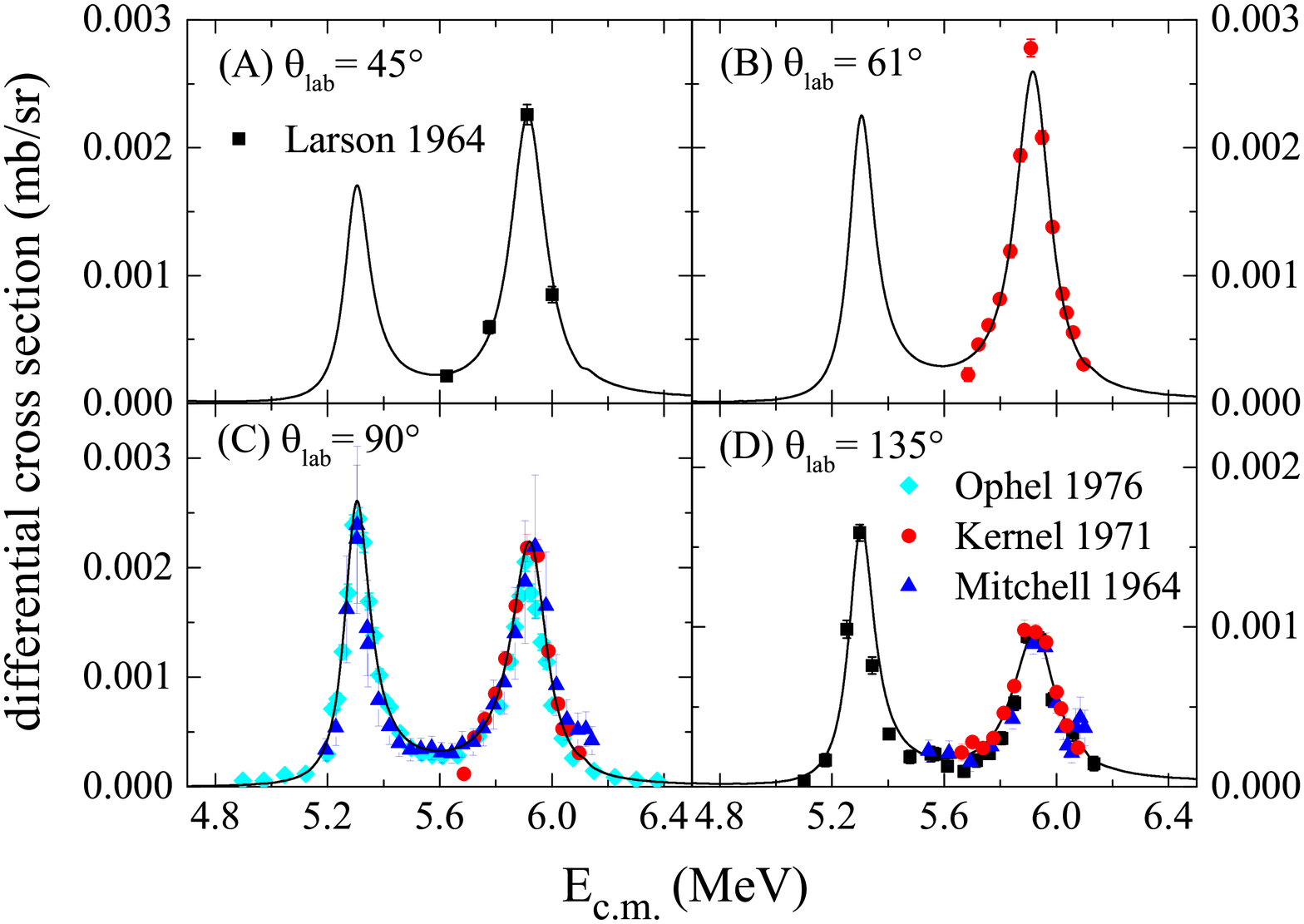}
\caption{
 (Color online) Fits to the $^{12}$C($\alpha,\gamma_0$)$^{16}$O${}_0$ differential cross-section data at $45^{\circ}$ (A),
 $61^{\circ}$ (B), $90^{\circ}$ (C), and $135^{\circ}$ (D) from Larson 1964:~\cite{Lars64}, Kernel 1971:~\cite{Kern71},
 Ophel 1976:~\cite{Ophe76} and Mitchell 1964:~\cite{Mitc64}.}
\label{fig:AGEF}
\end{figure}

The corresponding R-matrix calculation of angular distributions for the reaction $^{12}$C($\alpha,\gamma_0$)$^{16}$O${}_0$
are illustrated in Fig.~\ref{fig:Ouellet_1}- Fig.~\ref{fig:AGEF} at representative $\alpha $ energies from
$E_{c.m.}$ = 1.002 to $E_{c.m.}$ = 6.075 MeV. The angular distribution measured at the $2^+_2$ ($E_{x}$ = 9.84 MeV),
has the familiar E20 pattern, is symmetric with respect to 90${^\circ}$, while the distributions obtained at other energies
are asymmetric about 90${^\circ}$, clearly indicating the presence of both E10 and E20 amplitudes in the capture mechanism.
With the much improved $\gamma $ ray angular distributions in our R-matrix calculation, it can now be possible to derive
more accurate values for the cross sections of the E$_{10}$ and E$_{20}$ transitions to the ground state of $^{16}$O.
Fig.~\ref{fig:SE10} and Fig.~\ref{fig:SE20} show the calculations of the S$_{E10}$ and S$_{E20}$ together with all the available experimental data.
Although the data of S$_{E10}$ and S$_{E20}$ are not used in the fits, these data lie in two sides of our calculation uniformly,
which in turn illustrates the rationality and self-consistency in our R-matrix fit of the angular distributions and S$_{g.s}$,
ground state transitions.

In Refs.~\cite{Ouel96,Gial01} the $1^-_1$ subthreshold state and the $1^-_2$ resonance may interfere destructively and
result in a significantly lower S$_{E10}$. However, the constructive solution is strongly favored
and the destructive interference pattern has been eliminated in our calculation of angular distribution,
resulting in a value of S$_{E10}$(0.3 MeV) = 98.0 $\pm$ 7.0 keVb.

The cross section around the $E_{c.m.}$ = 2.5--3.0 MeV region is a rapidly changing function of energy, which strongly depends upon
the interference scheme between the resonance $2^+_2$ and other E$_{20}$ amplitudes. But the relative
E$_{20}$-E$_{20}$ interference sign is not well determined by the integral capture data, i.e., the best result of Ref.~\cite{Schu12}
in an interference pattern determined by the high-energy data of Ref.~\cite{Schu11} is different from most previous analyses
Refs.~\cite{Dufo08,Kunz02}. The interference scheme has been commendably constrained by the angular distributions calculation
in our R-matrix fitting near this resonance, which are in accordance with the new measurement result of Ref.~\cite{Sayr12},
and the extrapolation values of S$_{E20}$(0.3 MeV) = 56.0 $\pm$ 4.1 keVb in our calculation.

\subsection{$^{12}$C($\alpha,\gamma_1$)$^{16}$O${}_1$}
Radiative $\alpha $-particle capture into the first excited, J$^\pi$ = $0^+$ state at 6.05 MeV excitation energy have been
investigated recently in two independent experiments~\cite{Schu11,Mate06}, however, there exists big differences of S${}_{6.05}$(0.3 MeV)
each other. So it is necessary to give a detailed research and discussion. In the work of C. Matei et.al.~\cite{Mate06},
S${}_{6.05}$(0.3 MeV) = $25_{-16}^{+25}$ keV b is obtained by fitting the experimental results therein. It mainly comes
from S$_{E1}$ and partially from S$_{E2}$ and is with large error. In contrast to the analysis of Ref.~\cite{Mate06},
the extrapolation of Ref.~\cite{Schu11} suggests a negligible contribution from this amplitude, S${}_{6.05}$(0.3 MeV) $<$ 1 keV b
by analyzing their data, which is mainly contributed by S$_{E2}$ while little by S$_{E1}$. Ref.~\cite{Mate06} and Ref.~\cite{Schu11}
use the same experimental method, and both their original data show the $\gamma $ contribution of the first excited state
(0$^{+}_{1}$, 6.05 MeV). But Ref.~\cite{Schu11} concludes that the S${}_{6.05}$ is negligible in the energy region less than 3.3 MeV,
so it only gives the experimental data above the energy.
\begin{figure}
\center
\includegraphics[scale=0.29]{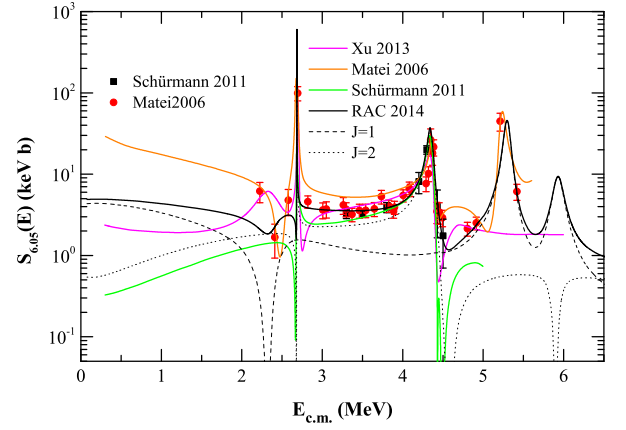}
\caption{
 (Color online) Results of best R-matrix fit for the cascade transitions $S_{6.05}$ from Sch\"urmann 2011:~\cite{Schu11}
 and Matei 2006:~\cite{Mate06}, together with the decomposition into different energy level contributions. For comparison,
 the results of Sch\"urmann 2011:~\cite{Schu11}, Matei 2006:~\cite{Mate06}, and Xu 2013:~\cite{Xu13} are shown in this figure.}
\label{fig:SCG1}
\end{figure}

In our fit, the data of Ref.~\cite{Schu11} is regarded as standard data and the normalization coefficient is 1.03.
The energy regions of the data in Ref.~\cite{Mate06} and Ref.~\cite{Schu11} have overlap around 3.5 MeV. The data of Ref.~\cite{Mate06}
can be normalized by that of Ref.~\cite{Schu11}, and the normalization factor is 0.88. This forms a dataset of S${}_{6.05}$
which covers full energy region with complete energy points and continuous values. This transition can therefore
be estimated to be S${}_{6.05}$(0.3MeV) = 4.9 $\pm$ 1.2 keV b, where S$_{E1}$ is the most important contribution in S${}_{6.05}$(0.3 MeV).
The S${}_{6.05}$ obtained by this work is from the systematic analysis of the whole O$^{16}$ system. Hence compared with previous analysis,
our result is much firmly based on the experiments and is reliable.

\subsection{$^{12}$C($\alpha,\gamma_2$)$^{16}$O${}_2$}

Very little data exists about the transition into the $E_{x}$ = 6.13 MeV state (J$^\pi$ = 3$^-$) except for the 2$^{+}_{3}$ resonance
at $E_{x}$ = 11.60 MeV and 3$^{-}_{2}$ resonance at $E_{x}$ = 11.52 MeV of Ref.~\cite{Schu11}. The parameters of these resonance
can be sufficient to describe this data and the fit result in S${}_{6.13}$(0.3 MeV) = 0.2 $\pm$ 0.1 keVb.
\begin{figure}
\center
\includegraphics[scale=0.29]{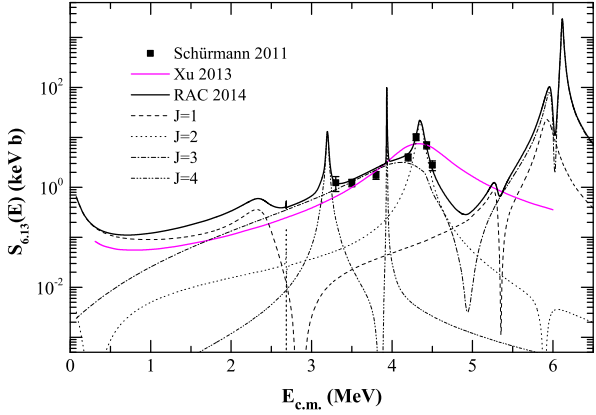}
\caption{
 (Color online) Results of the best R-matrix fit for the cascade transitions $S_{6.13}$ from Sch\"urmann 2011:~\cite{Schu11}
 together with the decomposition into different energy level contributions. For comparison, the result of Xu 2013:~\cite{Xu13}
 is shown in this figure.}
\label{fig:SCG2}
\end{figure}

\subsection{$^{12}$C($\alpha,\gamma_3$)$^{16}$O${}_3$}

\begin{figure}
\center
\includegraphics[scale=0.29]{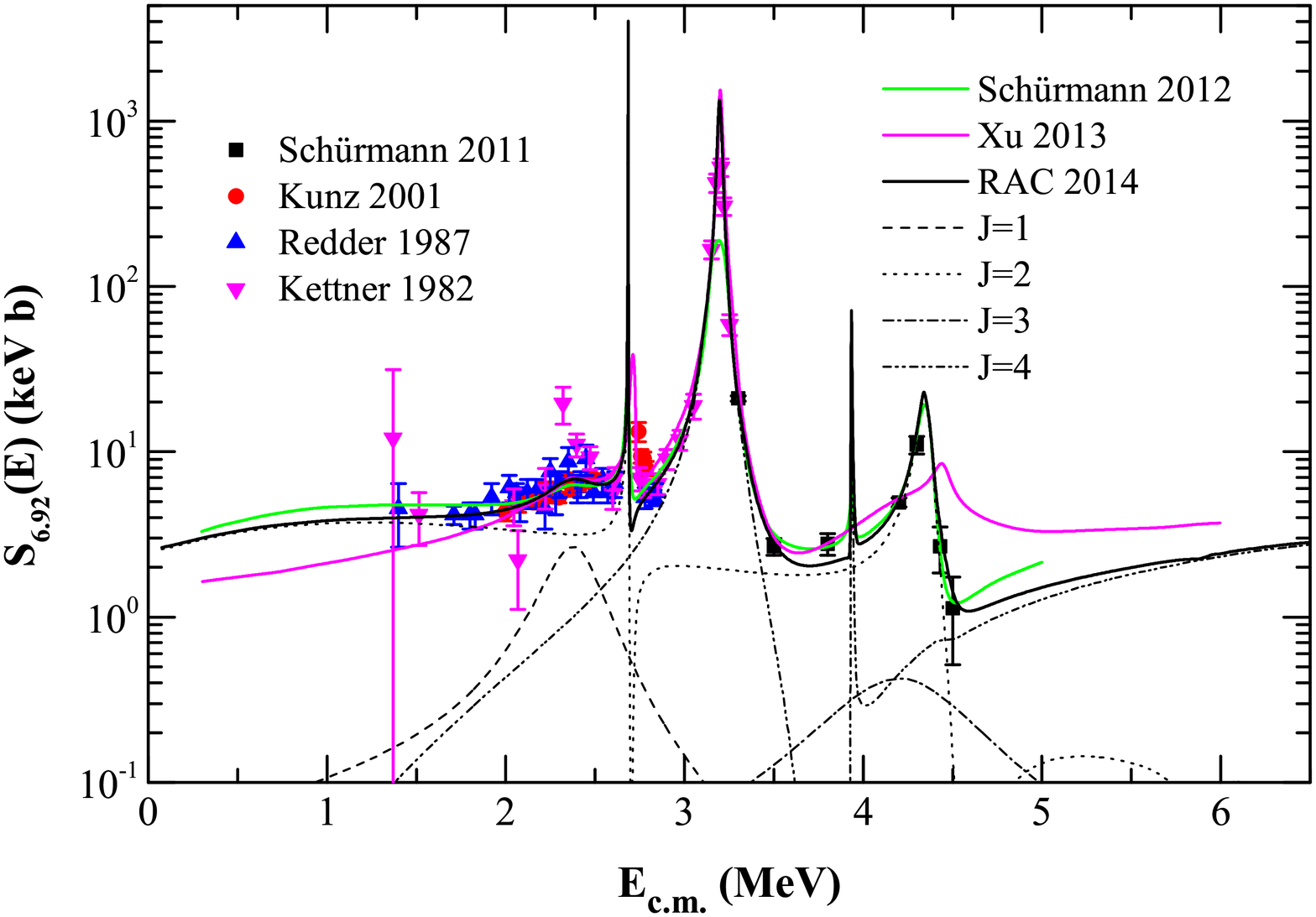}
\caption{
 (Color online) Results of the best R-matrix fit for the cascade transitions $S_{6.92}$ from Redder 1987:~\cite{Redd87},
 Kettner 1982:~\cite{Kett82}, Kunz 2001:~\cite{Kunz01}, and Sch\"urmann 2011:~\cite{Schu11},
 together with the decomposition into different energy level contributions. For comparison, the results of
 Sch\"urmann 2012:~\cite{Schu12} and Xu 2013:~\cite{Xu13} are shown in this figure.}
\label{fig:SCG3}
\end{figure}

Four cascade data of S${}_{6.92}$~\cite{Redd87,Kett82,Kunz01,Schu11} cover a range from $E_{c.m.}$ = 1.4 to 5.5 MeV,
and the cross section at astrophysical energy is largely governed by the direct capture process, from s-, d- and g-wave captures
in Refs.~\cite{Redd87,Kett82,Mate08,Schu11}. With the reasonable normalization of Refs.~\cite{Redd87,Kett82},
a good fit to these experimental data are achieved from resonance parameters and a direct capture parameter for
$J^{\pi }{\rm =}{{\rm 2}}^{{\rm +}}_1$ (see parameter table), resulting in S${}_{6.92}$(0.3 MeV) = 3.0 $\pm$ 0.4 keV b,
which is consistent with the result of Ref.~\cite{Schu12}. Fig.~\ref{fig:SCG3} shows the results of S${}_{6.92}$
as well as its decomposition into different level contributions. Also, the normalization factors are given in Table~\ref{tab:Normalization}.

\subsection{$^{12}$C($\alpha,\gamma_4$)$^{16}$O${}_4$}
\begin{figure}
\center
\includegraphics[scale=0.29]{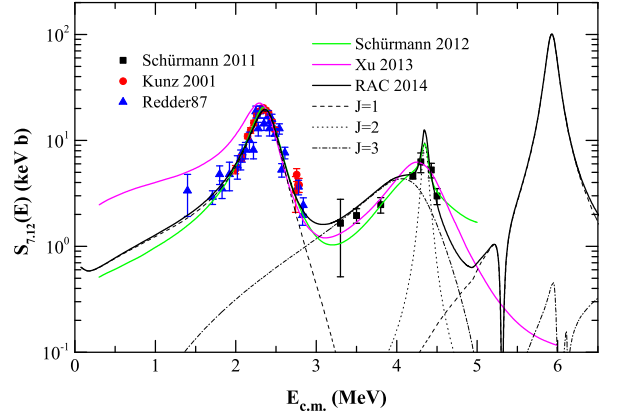}
\caption{
 (Color online) Results of the best R-matrix fit for the cascade transitions $S_{7.12}$ from Redder 1987:~\cite{Redd87},
 Kunz 2001:~\cite{Kunz01}, and Sch\"urmann 2011:~\cite{Schu11}, together with the decomposition into different energy level contributions.
 For comparison, the results of Sch\"urmann 2012:~\cite{Schu12} and Xu 2013:~\cite{Xu13} are shown in this figure.}
\label{fig:SCG4}
\end{figure}

The capture of S${}_{7.12}$ would be expected to proceed mainly via p- and f-wave direct process and resonance transition
at low energies~\cite{Redd87,Schu11}. With resonance parameters and a direct capture parameter for $J^{\pi }{\rm =}{{\rm 1}}^{{\rm -}}_1$,
a good fit to the experimental data~\cite{Redd87,Kunz01,Schu11} is obtained (see parameter table). And the extrapolated S factor
for this transition is also small, S${}_{7.12}$(0.3 MeV) = 0.6 $\pm$ 0.2 keVb. The normalization factors of these applied data are given
in Table~\ref{tab:Normalization}. Fig.~\ref{fig:SCG4} shows the results of S${}_{7.12}$ as well as its decomposition into
different level contributions.

\subsection{{$^{12}$C($\alpha,\alpha$)$^{12}$C}}

\begin{figure}
\center
\includegraphics[scale=0.28]{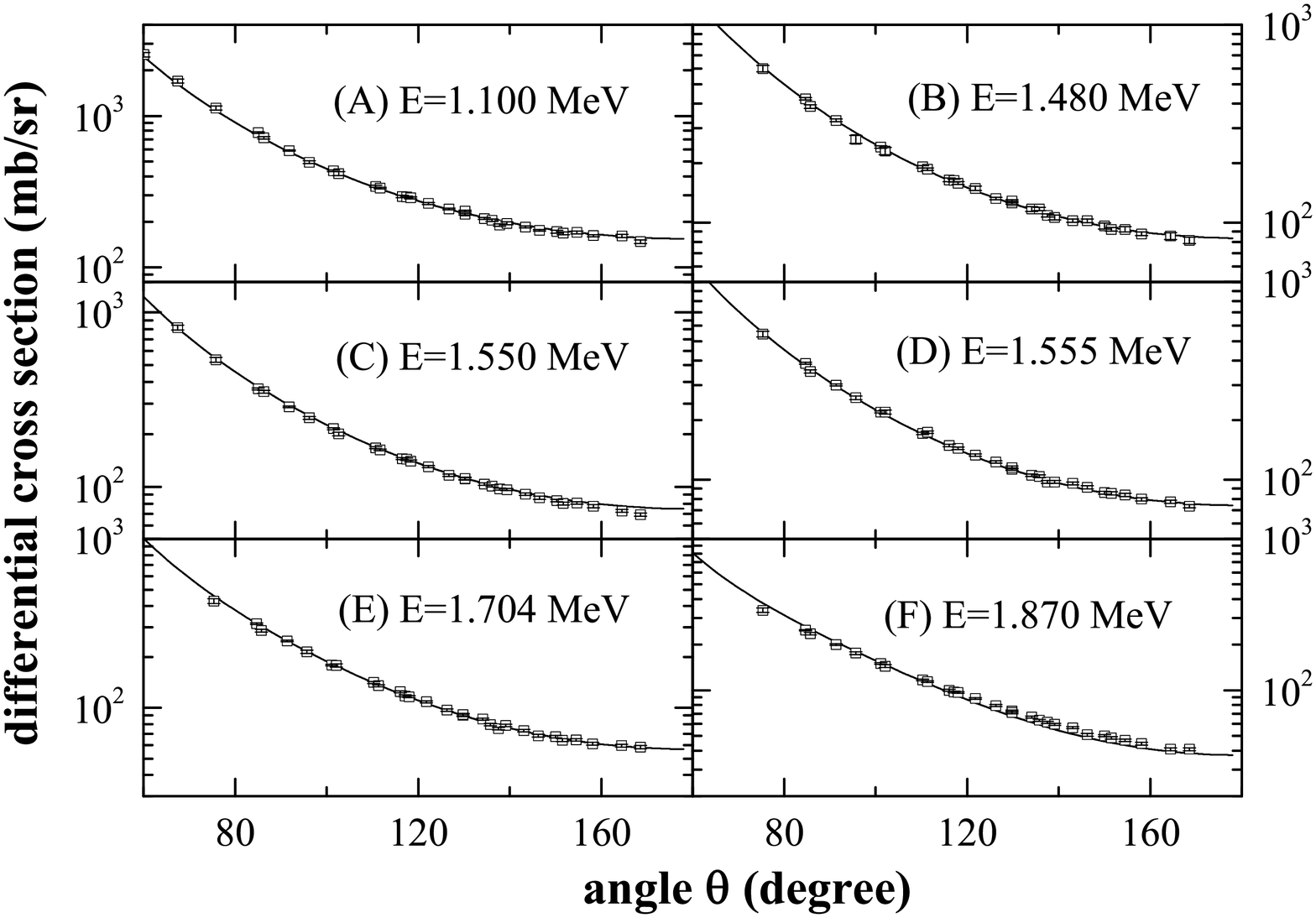}
\caption{
 Fits to the {$^{12}$C($\alpha,\alpha$)$^{12}$C} angular distributions of Plaga 1987:~\cite{Plag87}
 at $E_{c.m.}$ = 1.100 (A), 1.480 (B), 1.550 (C), 1.555 (D), 1.704 (E), and 1.870 (F) MeV.}
\label{fig:Plaga_1}
\end{figure}

\begin{figure}
\center
\includegraphics[scale=0.28]{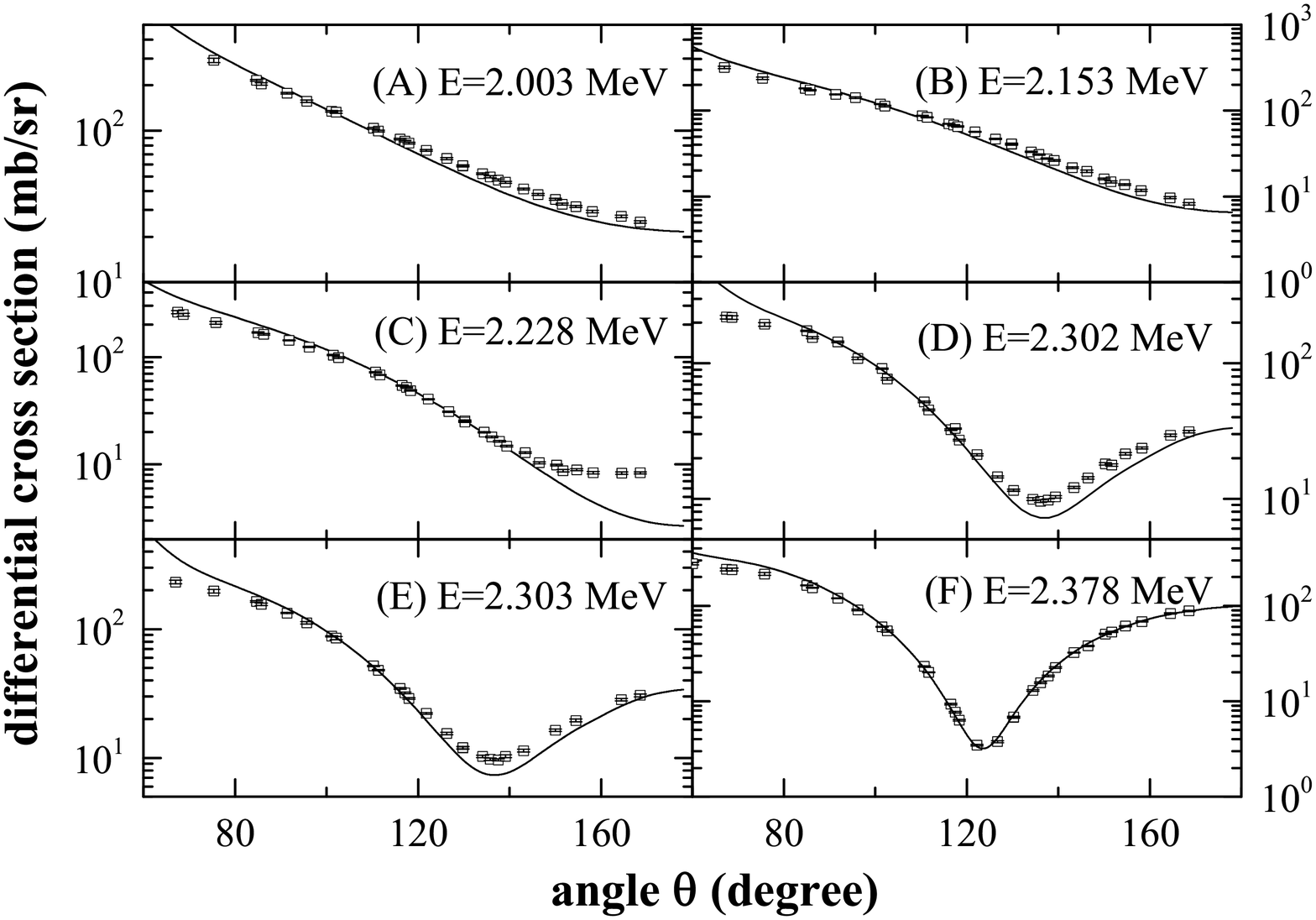}
\caption{
 Fits to the {$^{12}$C($\alpha,\alpha$)$^{12}$C} angular distributions of Plaga 1987:~\cite{Plag87}
 at $E_{c.m.}$ = 2.003 (A), 2.153 (B), 2.228 (C), 2.302 (D), 2.303 (E), and 2.378 (F) MeV.}
\label{fig:Plaga_2}
\end{figure}

\begin{figure}
\center
\includegraphics[scale=0.28]{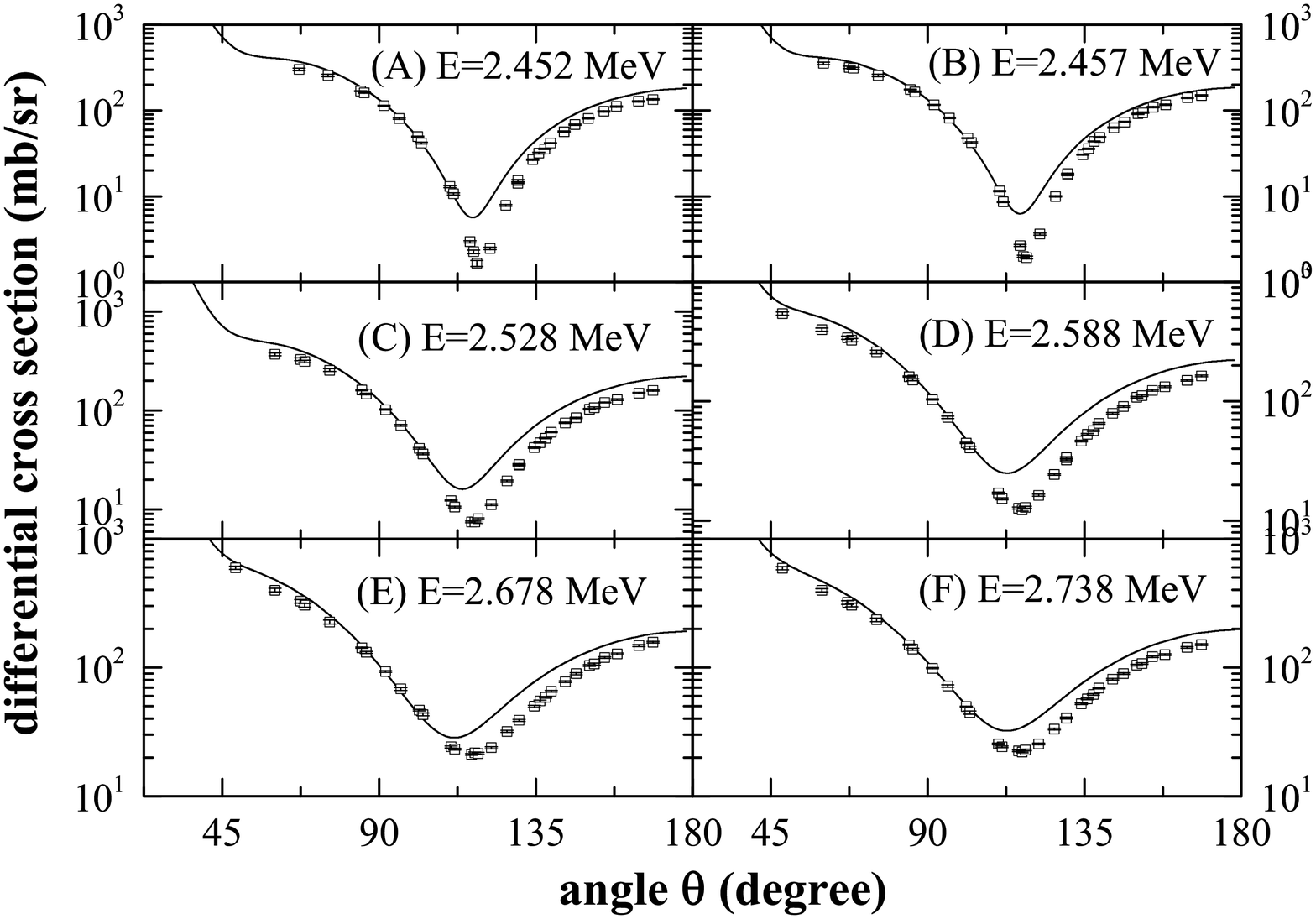}
\caption{
 Fits to the {$^{12}$C($\alpha,\alpha$)$^{12}$C} angular distributions of Plaga 1987:~\cite{Plag87}
 at $E_{c.m.}$ = 2.452 (A), 2.457 (B), 2.528 (C), 2.588 (D), 2.678 (E), and 2.738 (F) MeV.}
\label{fig:Plaga_3}
\end{figure}

\begin{figure}
\center
\includegraphics[scale=0.28]{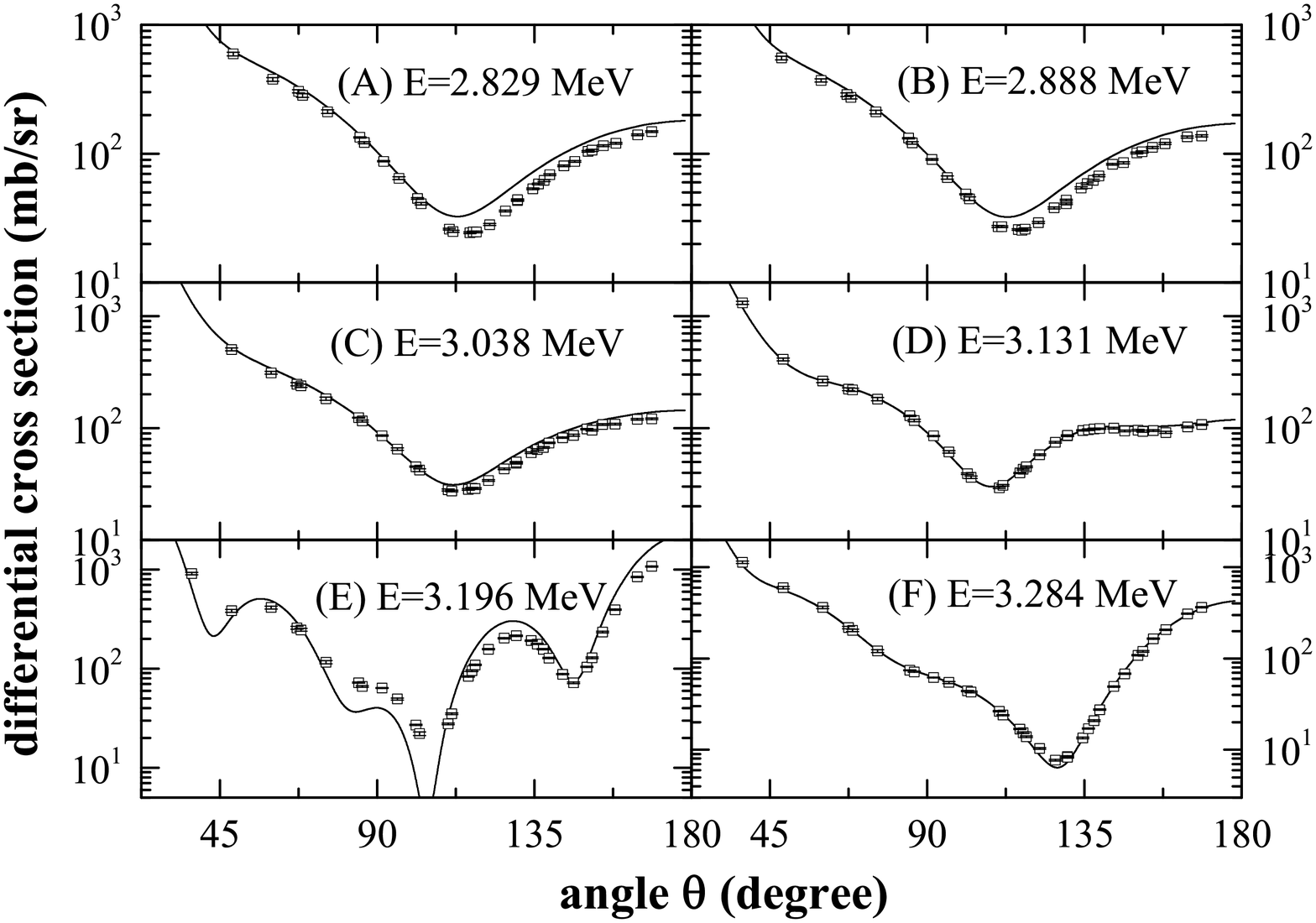}
\caption{
 Fits to the {$^{12}$C($\alpha,\alpha$)$^{12}$C} angular distributions of Plaga 1987:~\cite{Plag87}
 at $E_{c.m.}$ = 2.829 (A), 2.888 (B), 3.038 (C), 3.131 (D), 3.196 (E), and 3.284 (F) MeV.}
\label{fig:Plaga_4}
\end{figure}

\begin{figure}
\center
\includegraphics[scale=0.28]{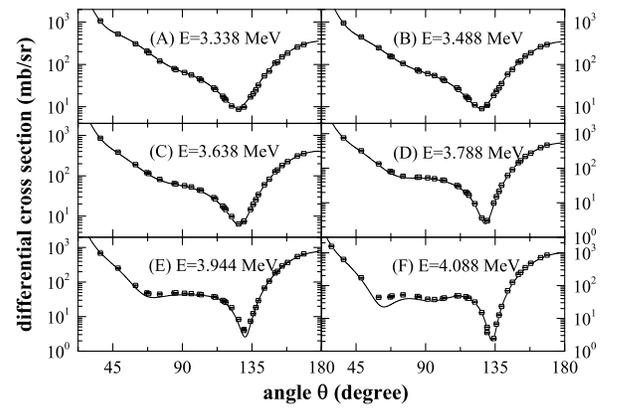}
\caption{
 Fits to the {$^{12}$C($\alpha,\alpha$)$^{12}$C} angular distributions of Plaga 1987:~\cite{Plag87}
 at $E_{c.m.}$ = 3.338 (A), 3.488 (B), 3.638 (C), 3.788 (D), 3.944 (E), and 4.088 (F) MeV.}
\label{fig:Plaga_5}
\end{figure}

\begin{figure}
\center
\includegraphics[scale=0.28]{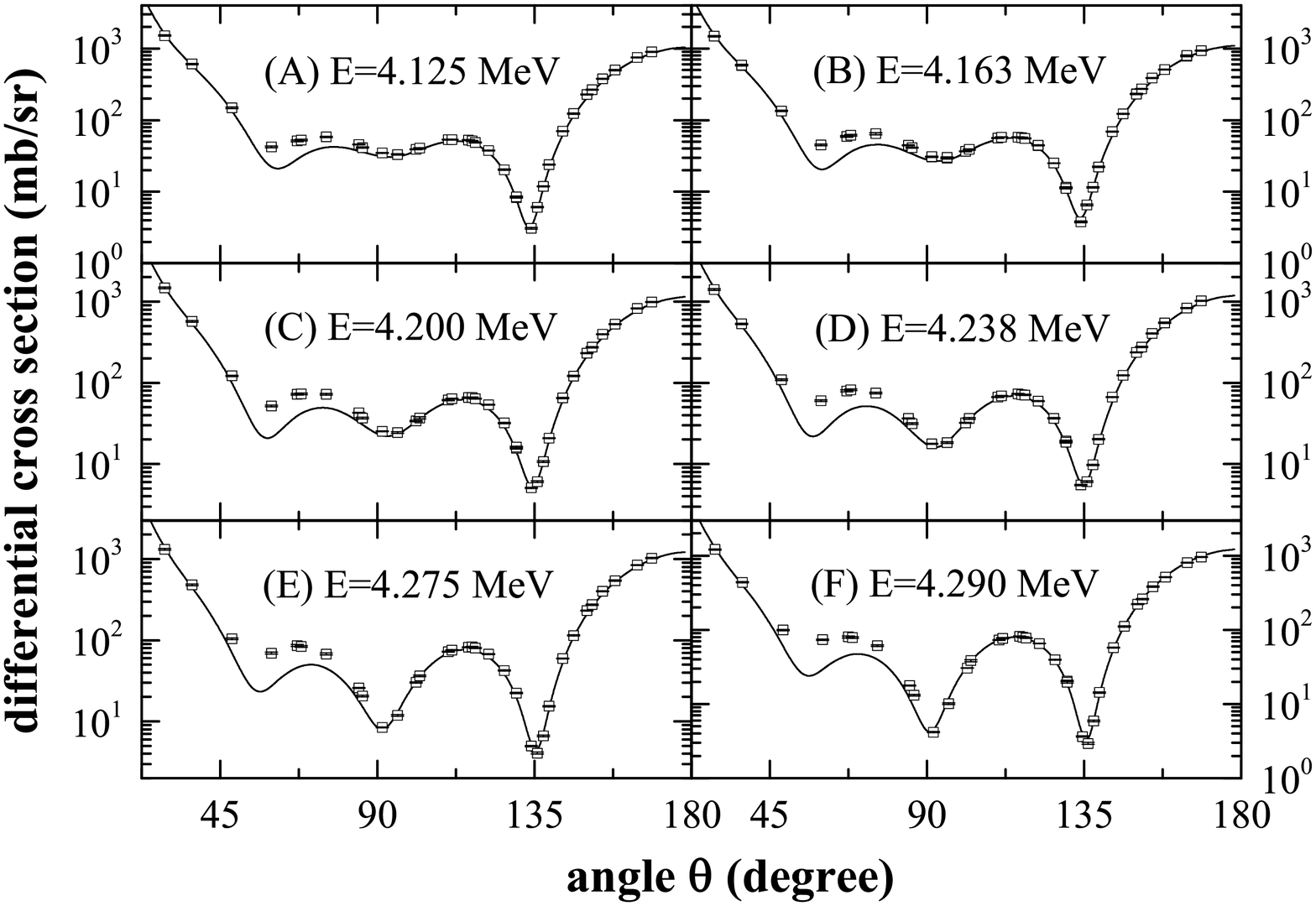}
\caption{
 Fits to the {$^{12}$C($\alpha,\alpha$)$^{12}$C} angular distributions of Plaga 1987:~\cite{Plag87}
 at $E_{c.m.}$ = 4.125 (A), 4.163 (B), 4.200 (C), 4.238 (D), 4.275 (E), and 4.290 (F) MeV.}
\label{fig:Plaga_6}
\end{figure}

\begin{figure}
\center
\includegraphics[scale=0.28]{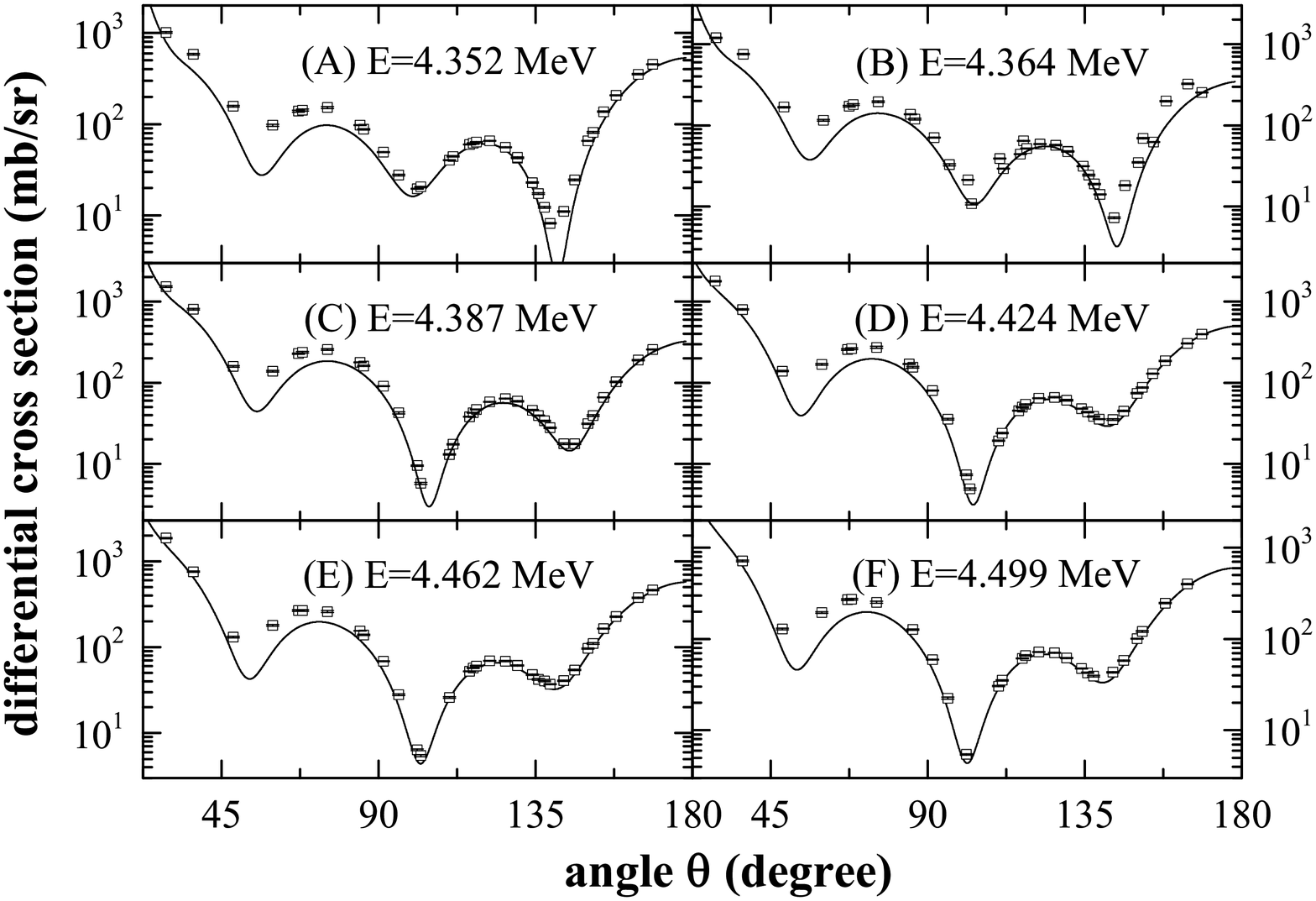}
\caption{
 Fits to the {$^{12}$C($\alpha,\alpha$)$^{12}$C} angular distributions of Plaga 1987:~\cite{Plag87}
 at $E_{c.m.}$ = 4.352 (A), 4.364 (B), 4.387 (C), 4.424 (D), 4.462 (E), and 4.499 (F) MeV.}
\label{fig:Plaga_7}
\end{figure}

\begin{figure}
\center
\includegraphics[scale=0.28]{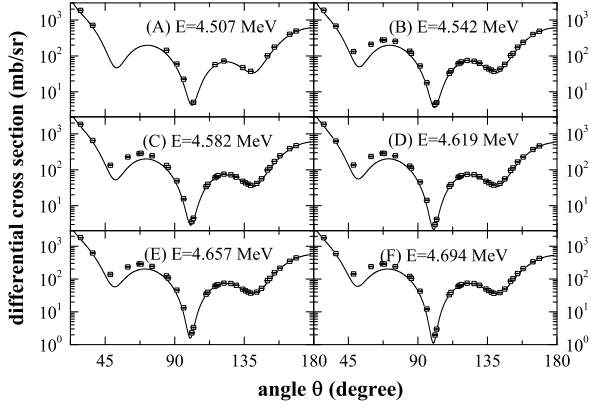}
\caption{
 Fits to the {$^{12}$C($\alpha,\alpha$)$^{12}$C} angular distributions of Plaga 1987:~\cite{Plag87}
 at $E_{c.m.}$ = 4.507 (A), 4.542 (B), 4.582 (C), 4.619 (D), 4.657 (E), and 4.694 (F) MeV.}
\label{fig:Plaga_8}
\end{figure}

\begin{figure}
\center
\includegraphics[scale=0.28]{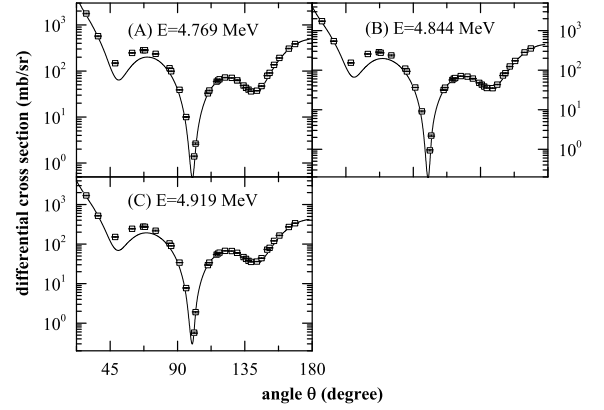}
\caption{
 Fits to the {$^{12}$C($\alpha,\alpha$)$^{12}$C} angular distributions of Plaga 1987:~\cite{Plag87}
 at $E_{c.m.}$ = 4.769 (A), 4.844 (B) and 4.919 (C) MeV.}
\label{fig:Plaga_9}
\end{figure}

\begin{figure}
\center
\includegraphics[scale=0.28]{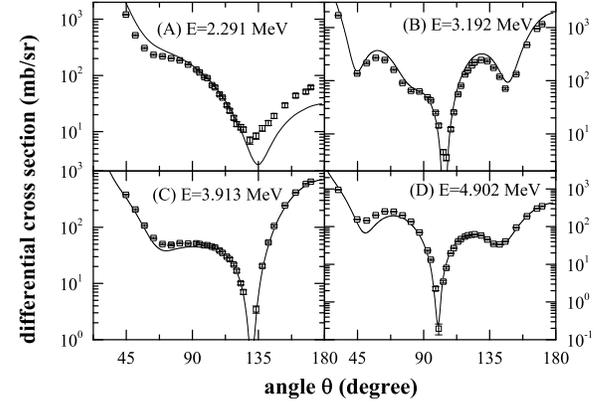}
\caption{
 Fits to the {$^{12}$C($\alpha,\alpha$)$^{12}$C} angular distributions of Tischhauser 2009:~\cite{Tisc09}
 at $E_{c.m.}$ = 2.291 (A), 3.192 (B), 3.913 (C) and 4.902 (D) MeV.}
\label{fig:Tisc}
\end{figure}

\begin{figure}
\center
\includegraphics[scale=0.28]{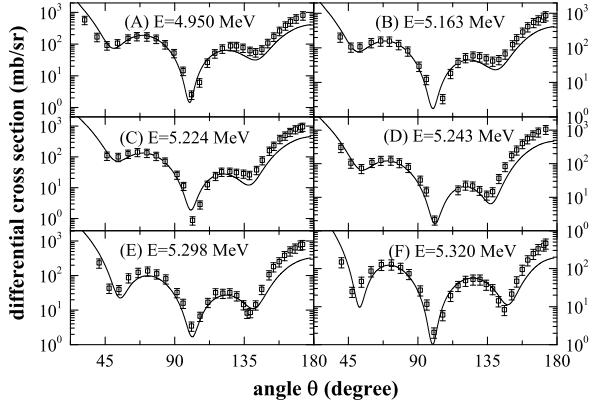}
\caption{
 Fits to the {$^{12}$C($\alpha,\alpha$)$^{12}$C} angular distributions of Morris 1968:~\cite{Morr68}
 at $E_{c.m.}$ = 4.950 (A), 5.163 (B), 5.224 (C), 5.243 (D), 5.298 (E) and 5.320 (F) MeV.}
\label{fig:Morris_6.1}
\end{figure}

\begin{figure}
\center
\includegraphics[scale=0.28]{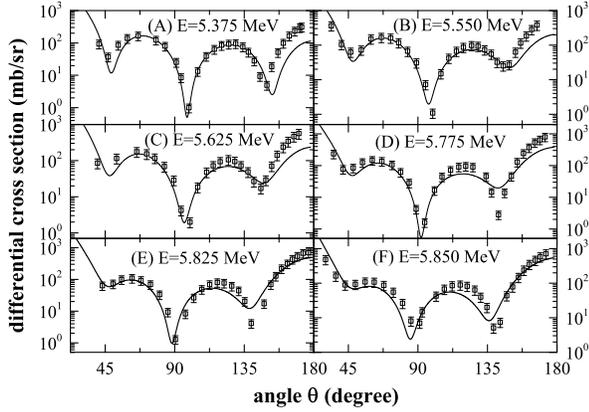}
\caption{
 Fits to the {$^{12}$C($\alpha,\alpha$)$^{12}$C} angular distributions of Morris 1968~\cite{Morr68}
 at $E_{c.m.}$ = 5.375 (A), 5.550 (B), 5.625 (C), 5.775 (D), 5.825 (E) and 5.850 (F) MeV.}
\label{fig:Morris_6.2}
\end{figure}

The most widely used elastic scattering data of $\alpha$-particles on $^{12}$C contain the $\alpha $ particle information
for all relevant states in $^{16}$O, which can be obtained with rather high accuracy. Previous elastic-scattering data
have been used to determine the scattering phase shifts for individual angular momenta Refs.~\cite{Schu12,Tang10} and so on.
Such a procedure is necessary in cases when the analysis of ${}^{12}$C($\alpha $, $\gamma $)${}^{16}$O is restricted to only
one particular angular momentum, but the interference structures in all data, associated with all the resonance states,
have been neglected~\cite{Schu12,Buch96}. Taking all angular momenta into account simultaneously, R-matrix fits of four
group angular distributions and the associated data~\cite{Plag87,Tisc02,Tisc09,Morr68,Brun75} are presented from $E_{c.m.}$ =1.1
to $E_{c.m.}$ = 5.85 MeV in Fig.~\ref{fig:Plaga_1} to Fig.~\ref{fig:Morris_6.2} with the strictly theoretical formulae of
Eq.~\ref{EQ_Diff}. In order to reduce the space of the paper, the figures from $E_{c.m.}$ = 5.85 MeV to $E_{c.m.}$ = 7.5 MeV of
Refs.~\cite{Morr68,Brun75} are not shown in this paper.

The scattering data by Plaga \emph{et al}.~\cite{Plag87} were obtained in the considerably better energy range in comparing
with the other studies. Differential cross-section data for all 35 angles in the range ${\theta }_{lab}$ = 22${}^\circ $
to 163${}^\circ $ and for 51 energies from $E_{c.m.}$ = 1.0 to 4.9 MeV are included in the fit. Data points in the vicinity of
narrow resonances are also contained for this analysis. Level parameters from the fit are in an excellent agreement with
those reported in Ref.~\cite{Till93}. The interference structures in the data, associated with resonance states
in the energy range covered by this data, are well reproduced by the R-matrix fits. All the results are shown in
Figs.\ref{fig:Plaga_1} to Figs.\ref{fig:Plaga_9}.

Recently the angular distributions of ${}^{{\rm 12}}{{\rm C(}\alpha {\rm ,}\alpha {\rm )}{}^{{\rm 12}}{{\rm C}}}$ in the
$\alpha$-energy range of 2.6--8.2 MeV, at angles from 24${}^\circ $ to 166${}^\circ $ have been measured at the University
of Notre Dame using an array of 32 silicon detectors~\cite{Tisc02,Tisc09}. The relative differential cross-section excitation curves
for eight selected detector angles and the four angular distributions for energies near the $E_{c.m.}$ = 2.291 $(1^{-}_{2})$,
3.192 $(4^{+}_{2})$, 3.913 $(2^{+}_{3})$, and 4.902 $(0^{+}_{2})$ MeV resonances are available. To reduce the amount of computations,
only the four angular distributions are employed in this fit, which the angular distributions are found to be in a good agreement with
those data. Fits are shown in Fig.\ref{fig:Tisc}.

The best quality $\alpha$-scattering cross-section data above proton separation energies are shown in Fig.\ref{fig:Morris_6.1}
and Fig.\ref{fig:Morris_6.2} of Ref.~\cite{Morr68}, which have good coherence to experimental data.

\subsection{{$^{12}$C($\alpha,\alpha${$_{1}$})$^{12}$C} and {$^{12}$C($\alpha,p$)$^{15}$N}}

\begin{figure}
\center
\includegraphics[scale=0.28]{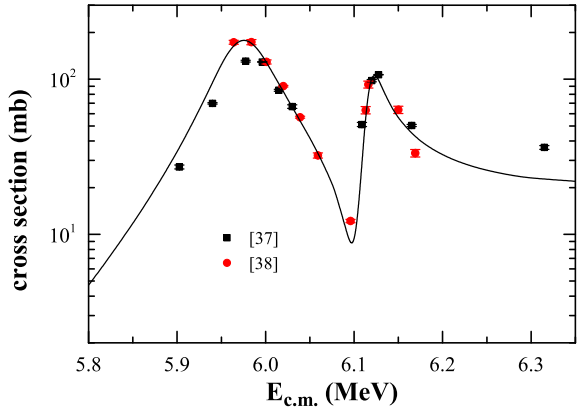}
\caption{
 (Color online) Fits to the {$^{12}$C($\alpha,\alpha${$_{1}$})$^{12}$C} cross section of Mitchell 1965:~\cite{Mitc65}
 and deBoer 2012:~\cite{Debo12a} .}
\label{fig:CSAA1}
\end{figure}

\begin{figure}
\center
\includegraphics[scale=0.28]{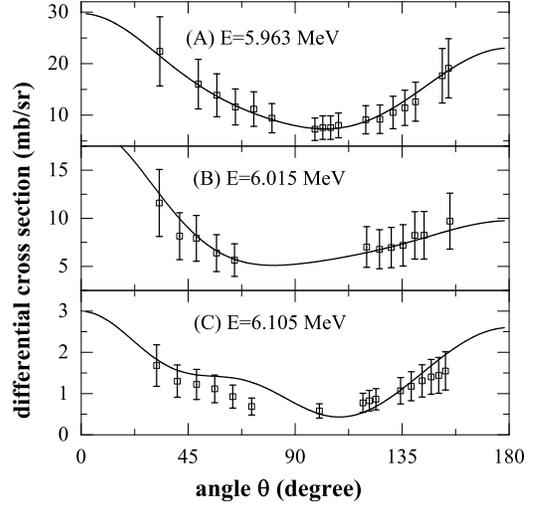}
\caption{
 Fits to the {$^{12}$C($\alpha,\alpha${$_{1}$})$^{12}$C} angular distributions of Mitchell 1965:~\cite{Mitc65}
 at $E_{c.m.}$ = 5.963 (A), 6.015 (B) and 6.105 (C) MeV.}
\label{fig:MA1D}
\end{figure}

\begin{figure}
\center
\includegraphics[scale=0.28]{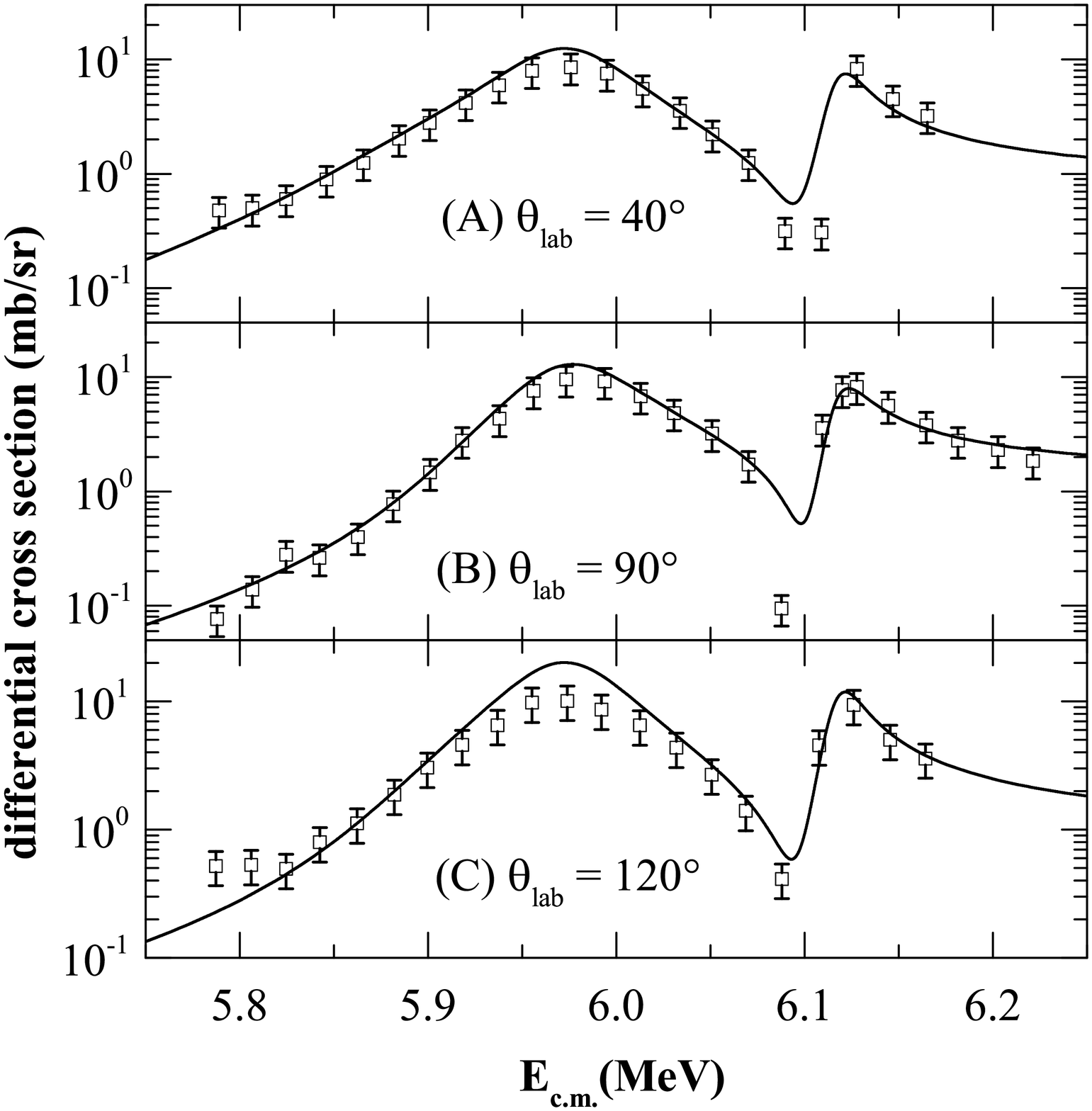}
\caption{
 Fits to the {$^{12}$C($\alpha,\alpha${$_{1}$})$^{12}$C} differential cross section data of Mitchell 1965:~\cite{Mitc65}.}
\label{fig:MA1E}
\end{figure}

\begin{figure}
\center
\includegraphics[scale=0.28]{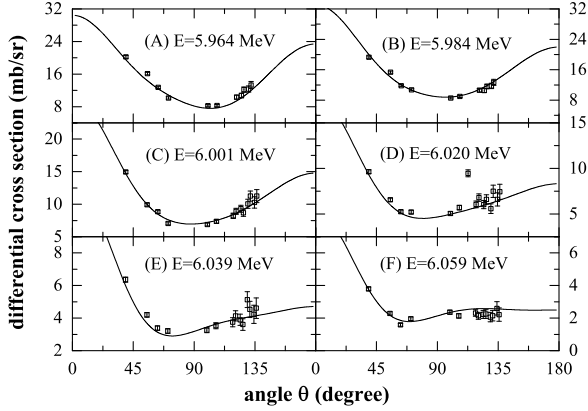}
\caption{
 Fits to the {$^{12}$C($\alpha,\alpha${$_{1}$})$^{12}$C} angular distributions of deBoer 2012:~\cite{Debo12a}
 at $E_{c.m.}$ = 5.964 (A), 5.984 (B), 6.001 (C), 6.020 (D), 6.039 (E) and 6.059 (F) MeV.}
\label{fig:DeBo1}
\end{figure}

\begin{figure}
\center
\includegraphics[scale=0.28]{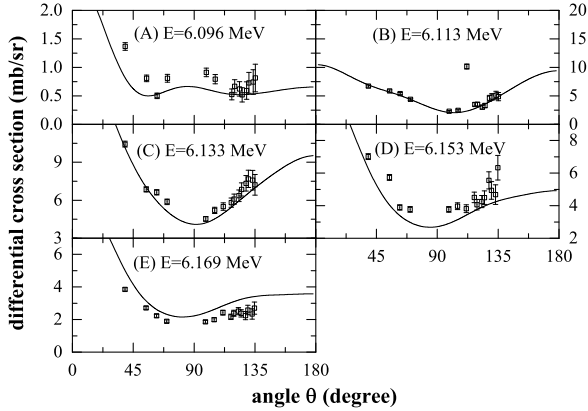}
\caption{
 Fits to the {$^{12}$C($\alpha,\alpha${$_{1}$})$^{12}$C} angular distributions of deBoer 2012:~\cite{Debo12a}
 at $E_{c.m.}$ = 6.096 MeV (A), 6.113 (B), 6.133 (C), 6.153 (D) and 6.169 (E) MeV.}
\label{fig:DeBo2}
\end{figure}

\begin{figure}
\center
\includegraphics[scale=0.28]{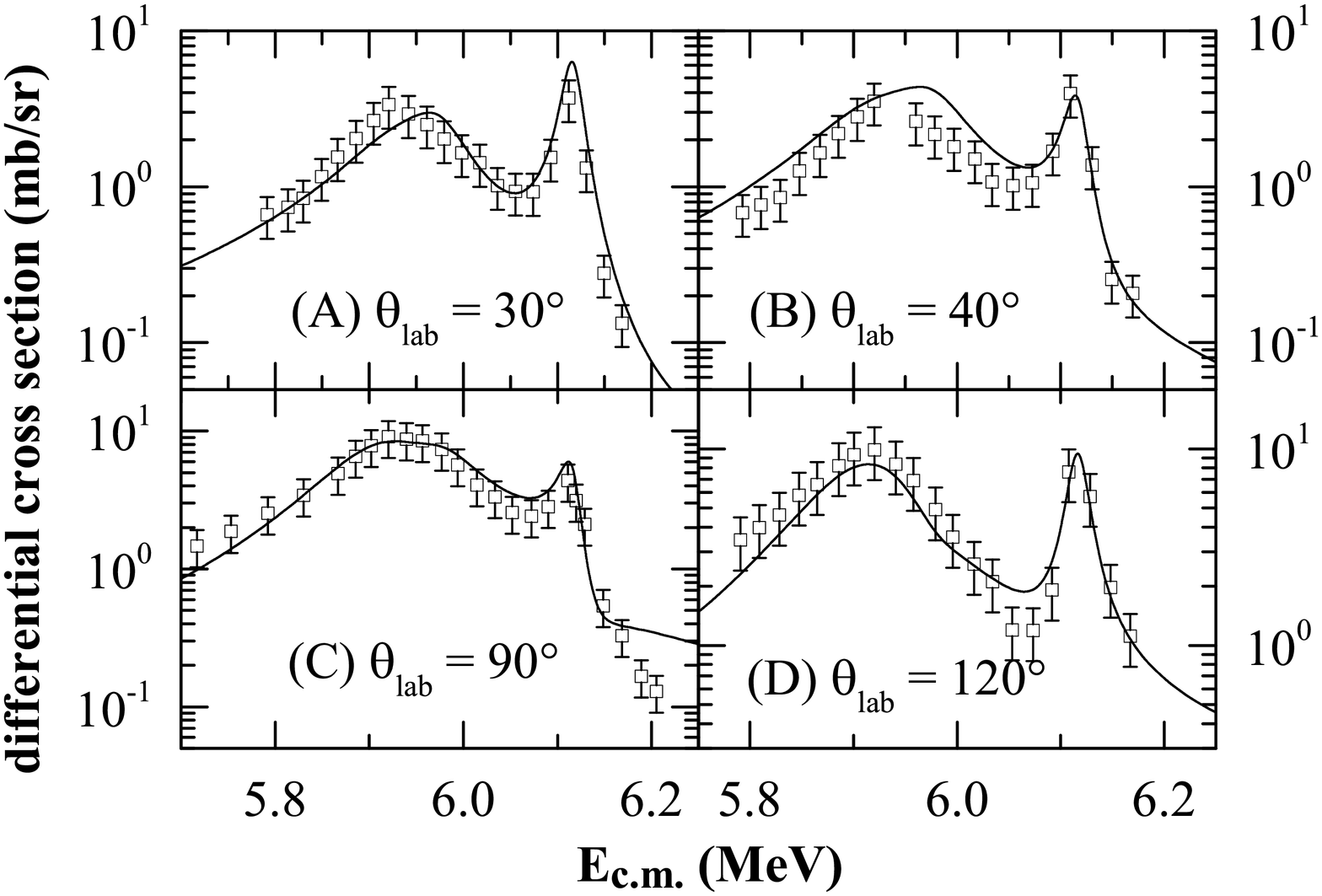}
\caption{
 Fits to the {$^{12}$C($\alpha,p$)$^{15}$N} differential cross section data of Mitchell 1965:~\cite{Mitc65}.}
\label{fig:MAPE}
\end{figure}

Good fits of $^{12}$C($\alpha,{\alpha}_{1}$)$^{12}$C and $^{12}$C($\alpha,p$)$^{15}$N are helpful to reduce the uncertainty
produced by the  distant levels, and then to improve the fit precision of {$^{12}$C($\alpha,\gamma$)$^{16}$O} S factor subsequently.
Previously, the only available angular distribution data of the $^{12}$C($\alpha,{\alpha}_{1}$)$^{12}$C reaction has been obtained
at incident energies E${}_{c.m.}$ = 5.963, 6.015 and 6.105 MeV in Ref.~\cite{Mitc65}. And the excitation curves of
$^{12}$C($\alpha,{\alpha}_{1}$)$^{12}$C and $^{12}$C($\alpha,p$)$^{15}$N for four selected detector angles have been measured
in the same energy range in the paper. But an absolute scaling was not reported in these measurements. In
Ref.~\cite{Debo12a,Debo12b}, new yield-ratio data for the reactions $^{12}$C($\alpha,{\alpha}_{1}$)$^{12}$C and $^{12}$C($\alpha,p$)$^{15}$N
were performed at the University of Notre Dame in order to provide additional data for a comprehensive
\textit{R}-matrix analysis of compound-nucleus reactions populating ${}^{16}$O. The data are in the form of yield ratios where
the $^{12}$C($\alpha,{\alpha}_{0}$)$^{12}$C yields measured at an angle ${\theta }_{{\rm lab}}$ = 58.9${}^\circ $ are used
as the reference data. And the transformational angular distribution data and cross sections are obtained by a private communication
with Dr. R. J. deBoer.

The R-matrix fits of $^{12}$C($\alpha,{\alpha}_{1}$)$^{12}$C angle-integrated cross section are shown by the solid lines
in Fig.\ref{fig:CSAA1}, illustrating the consistency level of the simultaneous fit to the data of Ref.~\cite{Debo12a}
together with the relative data in Ref.~\cite{Mitc65}. Fits of the available angular distribution data are shown in
Fig.\ref{fig:MA1D} to Fig.\ref{fig:DeBo2}, in which the significant contributions originate from states at
E$_{\textrm{x}}$ = 12.95 $(2^{+}_{4})$, 13.13 $(3^{-}_{3})$, and 13.27 $(3^{-}_{4})$ MeV. Fits for the angular distribution data of
Ref.~\cite{Mitc65} for the reaction $^{12}$C($\alpha,p$)$^{15}$N are shown in Fig.\ref{fig:MAPE}, and the normalization factors are given
in Table~\ref{tab:Normalization}.

\begin{table*}
\caption{\label{tab:Normalization} Scaling factors for datasets which have no reported absolute scale,
        and the normalization coefficient for the absolute data.}
\begin{ruledtabular}
\begin{tabular}{cccccccccc}
Figure No. &  Ref. & Normalization & $\chi^{2}_{C}$ &  $ndp$  & Figure No.  & Ref.  & Normalization  & $\chi^{2}_{C}$ &  $ndp$ \\ \hline
$4		$ & $[7] $ &	$1.03\times10^{+00}$ &	$1.437$ &	$91 $ & $19(B)$ &	$[17]$ &	$1.29\times10^{-08}$ &	$3.993$ &	$9  $   \\
$4		$ & $[8] $ &	$1.03\times10^{+00}$ &	$0.959$ &	$7  $ & $19(C)$ &	$[17]$ &	$1.08\times10^{-08}$ &	$1.712$ &	$9  $	\\
$4		$ & $[10]$ &	$1.00\times10^{+00}$ &	$1.056$ &	$2  $ & $19(D)$ &	$[17]$ &	$1.25\times10^{-08}$ &	$1.060$ &	$9  $	\\
$4		$ & $[9] $ &	$1.03\times10^{+00}$ &	$4.196$ &	$4  $ & $16(A)$ &	$[9] $ &	$2.21\times10^{-10}$ &	$2.472$ &	$12 $	\\
$5		$ & $[26]$ &	$1.00\times10^{+00}$ &	$1.406$ &	$93 $ & $16(B)$ &	$[9] $ &	$3.13\times10^{-10}$ &	$2.357$ &	$12 $	\\
$5 		$ & $[27]$ &	$1.00\times10^{+00}$ &	$1.406$ &	$91 $ & $16(C)$ &	$[9] $ &	$2.20\times10^{-09}$ &	$0.384$ &	$12 $	\\
$5      $ & $[28]$ &	$1.00\times10^{+00}$ &	$1.406$ &	$75 $ &	$16(D)$ &	$[9] $ &	$4.33\times10^{-10}$ &	$1.526$ &	$12 $	\\
$27-31  $ &	$[32]$ &	$1.00\times10^{+00}$ &	$1.763$ &	$823$ &	$\textrm{Not shown}$ &	$[9] $ &	$1.74\times10^{-08}$ &	$3.088$ &	$12 $	\\
$32-35  $ &	$[32]$ &	$1.00\times10^{+00}$ &	$1.185$ &	$794$ &	$14(A)$ &	$[12]$ &	$7.10\times10^{-08}$ &	$2.800$ &	$6  $	\\
$36     $ &	$[34]$ &	$1.00\times10^{+00}$ &	$1.652$ &	$128$ &	$14(B)$ &	$[12]$ &	$1.50\times10^{-06}$ &	$2.806$ &	$6  $	\\
$37,38  $ &	$[35]$ &	$1.00\times10^{+00}$ &	$1.034$ &	$613$ &	$14(C)$ &	$[12]$ &	$1.18\times10^{-05}$ &	$2.145$ &	$6  $	\\
$\textrm{Not shown}$ &	$[36]$ &	$1.00\times10^{+00}$ &	$1.060$ &	$244$ &	$14(D)$ &	$[12]$ &	$3.04\times10^{-07}$ &	$1.744$ &	$6  $	\\
$6      $ &	$[12]$ &	$1.00\times10^{+00}$ &	$2.690$ &	$24 $ &	$21(A)$ &	$[19]$ &	$1.93\times10^{-06}$ &	$3.672$ &	$7  $	\\
$6      $ &	$[8] $ &	$1.03\times10^{+00}$ &	$2.327$ &	$7  $ &	$21(B)$ &	$[19]$ &	$2.21\times10^{-06}$ &	$1.006$ &	$7  $	\\
$6      $ &	$[13]$ &	$9.70\times10^{-01}$ &	$1.753$ &	$9  $ &	$21(C)$ &	$[19]$ &	$1.96\times10^{-06}$ &	$2.862$ &	$7  $	\\
$6      $ &	$[16]$ &	$1.00\times10^{+00}$ &	$0.464$ &	$20 $ &	$21(D)$ &	$[19]$ &	$2.12\times10^{-06}$ &	$1.536$ &	$7  $	\\
$6      $ &	$[15]$ &	$1.00\times10^{+00}$ &	$0.788$ &	$20 $ &	$21(E)$ &	$[21]$ &	$1.28\times10^{-06}$ &	$2.867$ &	$7  $	\\
$6      $ &	$[9] $ &	$1.03\times10^{+00}$ &	$0.784$ &	$4  $ &	$21(F)$ &	$[21]$ &	$1.37\times10^{-06}$ &	$2.415$ &	$7  $	\\
$6      $ &	$[18]$ &	$1.03\times10^{+00}$ &	$2.467$ &	$4  $ &	$\textrm{\textrm{Not shown}}$ &	$[21]$ &	$1.90\times10^{-06}$ &	$1.264$ &	$7  $	\\
$6      $ &	$[23]$ &	$8.70\times10^{-01}$ &	$1.196$ &	$48 $ &	$21(C)$ &	$[20]$ &	$1.10\times10^{+00}$ &	$2.980$ &	$40 $	\\
$6      $ &	$[67]$ &	$8.00\times10^{-01}$ &	$1.936$ &	$24 $ &	$22(A)$ &	$[19]$ &	$2.02\times10^{-06}$ &	$1.261$ &	$4  $	\\
$6      $ &	$[20]$ &	$1.00\times10^{+00}$ &	$0.001$ &	$1  $ &	$22(D)$ &	$[19]$ &	$1.99\times10^{-06}$ &	$2.819$ &	$20 $	\\
$9-11   $ &	$[13]$ &	$9.70\times10^{-01}$ &	$1.897$ &	$96 $ &	$22(B)$ &	$[21]$ &	$1.53\times10^{-06}$ &	$2.176$ &	$13 $	\\
$12     $ &	$[18]$ &	$1.03\times10^{+00}$ &	$2.346$ &	$14 $ &	$22(C)$ &	$[21]$ &	$1.44\times10^{-06}$ &	$1.659$ &	$13 $	\\
$15(A)  $ &	$[16]$ &	$5.30\times10^{-10}$ &	$1.675$ &	$9  $ &	$22(D)$ &	$[21]$ &	$2.12\times10^{-06}$ &	$1.267$ &	$13 $	\\
$15(B)  $ &	$[16]$ &	$4.14\times10^{-10}$ &	$1.683$ &	$9  $ &	$22(C)$ &	$[22]$ &	$6.38\times10^{-05}$ &	$0.785$ &	$29 $	\\
$15(C)  $ &	$[16]$ &	$4.68\times10^{-09}$ &	$1.089$ &	$9  $ &	$22(D)$ &	$[22]$ &	$1.02\times10^{-03}$ &	$1.531$ &	$14 $	\\
$15(D)  $ &	$[16]$ &	$6.15\times10^{-09}$ &	$4.002$ &	$9  $ &	$ 23  $ &	$[24]$ &	$8.00\times10^{-01}$ &	$2.276$ &	$32 $	\\
$15(E)  $ &	$[16]$ &	$5.41\times10^{-09}$ &	$4.872$ &	$9  $ &	$ 23  $ &	$[8] $ &	$1.03\times10^{+00}$ &	$0.888$ &	$7  $	\\
$15(F)  $ &	$[16]$ &	$1.31\times10^{-07}$ &	$2.576$ &	$9  $ &	$ 24  $ &	$[8] $ &	$1.03\times10^{+00}$ &	$1.379$ &	$7  $	\\
$20(A)  $ &	$[14]$ &	$2.36\times10^{-06}$ &	$1.730$ &	$9  $ &	$ 25  $ &	$[15]$ &	$1.00\times10^{+00}$ &	$1.076$ &	$16 $	\\
$20(B)  $ &	$[14]$ &	$2.01\times10^{-06}$ &	$1.784$ &	$9  $ &	$ 25  $ &	$[12]$ &	$3.62\times10^{-01}$ &	$0.745$ &	$25 $	\\
$20(C)  $ & $[14]$ &	$1.48\times10^{-06}$ &	$1.328$ &	$9  $ &	$ 25  $ &	$[23]$ &	$3.53\times10^{-01}$ &	$2.155$ &	$21 $	\\
$20(D)  $ &	$[14]$ &	$3.39\times10^{-06}$ &	$2.857$ &	$9  $ &	$ 25  $ &	$[8] $ &	$1.03\times10^{+00}$ &	$1.268$ &	$7  $	\\
$13(A)  $ &	$[11]$ &	$9.23\times10^{-08}$ &	$1.884$ &	$10 $ &	$ 26  $ &	$[15]$ &	$1.00\times10^{+00}$ &	$1.579$ &	$16 $	\\
$13(B)  $ &	$[11]$ &	$9.16\times10^{-08}$ &	$2.213$ &	$10 $ &	$ 26  $ &	$[12]$ &	$4.42\times10^{-01}$ &	$2.199$ &	$25 $	\\
$13(C)  $ &	$[11]$ &	$8.16\times10^{-08}$ &	$1.543$ &	$10 $ &	$ 26  $ &	$[8] $ &	$1.03\times10^{+00}$ &	$0.492$ &	$7  $	\\
$13(D)  $ &	$[11]$ &	$6.98\times10^{-08}$ &	$1.643$ &	$10 $ &	$ 39  $ &	$[37]$ &	$1.70\times10^{-02}$ &	$1.560$ &	$11 $	\\
$17(A)  $ &	$[17]$ &	$5.08\times10^{-10}$ &	$4.156$ &	$9  $ &	$ 39  $ &	$[38]$ &	$1.00\times10^{+00}$ &	$0.380$ &	$11 $	\\
$17(B)  $ &	$[17]$  &	$6.49\times10^{-10}$ &	$2.460$ &	$8  $ &	$ 40  $ &	$[37]$ &	$3.66\times10^{+00}$ &	$0.037$ &	$17 $	\\
$17(C)  $ &	$[17]$  &	$2.01\times10^{-09}$ &	$2.533$ &	$9  $ &	$ 40  $ &	$[37]$ &	$3.60\times10^{+00}$ &	$0.068$ &	$12 $	\\
$17(D)  $ &	$[17]$  &	$3.42\times10^{-09}$ &	$0.942$ &	$8  $ &	$ 40  $ &	$[37]$ &	$1.52\times10^{-01}$ &	$0.658$ &	$16 $	\\
$17(E)  $ &	$[17]$  &	$2.53\times10^{-09}$ &	$1.599$ &	$9  $ &	$ 41  $	&	$[37]$ &	$5.54\times10^{+00}$ &	$0.829$ &	$21 $	\\
$17(F)  $ &	$[17]$  &	$3.93\times10^{-09}$ &	$1.708$ &	$9  $ &	$ 41  $	&	$[37]$ &	$5.32\times10^{+00}$ &	$0.834$ &	$25 $	\\
$18(A)  $ &	$[17]$  &	$4.67\times10^{-09}$ &	$2.490$ &	$8  $ &	$ 41  $	&	$[37]$ &	$5.58\times10^{-01}$ &	$1.079$ &	$21 $	\\
$18(B)  $ &	$[17]$  &	$5.66\times10^{-09}$ &	$3.480$ &	$9  $ &	$ 42  $	&	$[38]$ &	$1.00\times10^{+00}$ &	$2.078$ &	$153$	\\
$18(C)  $ &	$[17]$  &	$1.67\times10^{-08}$ &	$1.960$ &	$9  $ &	$ 44  $	&	$[37]$ &	$4.20\times10^{+00}$ &	$0.804$ &	$21 $	\\
$18(D)  $ &	$[17]$  &	$3.83\times10^{-09}$ &	$2.645$ &	$9  $ &	$ 44  $	&	$[37]$ &	$2.12\times10^{+00}$ &	$0.778$ &	$20 $	\\
$18(E)  $ &	$[17]$  &	$9.17\times10^{-09}$ &	$2.877$ &	$9  $ &	$ 44  $	&	$[37]$ &	$1.48\times10^{+00}$ &	$0.877$ &	$24 $	\\
$18(F)  $ &	$[17]$  &	$1.19\times10^{-07}$ &	$1.970$ &	$9  $ & $ 44  $	&	$[37]$ &	$1.10\times10^{+00}$ &	$0.696$ &	$21 $	\\
$19(A)  $ &	$[17]$  &	$2.53\times10^{-08}$ &	$4.693$ &	$9 $ & & & & &\\
\end{tabular}
\end{ruledtabular}
\end{table*}

\begin{table*}
\caption{\label{tab:tablelevel01} Parameters obtained from the R-matrix fits, particle partial widths for the levels,
    and the radiation widths for the $\gamma$ -ray transitions considered in this work compared with those from literatures. }
\begin{ruledtabular}
\begin{tabular}{ccccccccc}
Res. no. & $E_x$(MeV) & $J^{\pi}_{n}$ & $E_{\lambda}$(MeV) & $\alpha(I_{1}+I_{2})$  & $(s,l)$ & $\gamma_{\alpha sl}$(MeV)$^{1/2}$ &  $\Gamma$(keV)  & Ref.~\cite{Till93}  \\ \hline
$01$ & $6.049 $ & $0^+_1$ & $-0.1113 $ & $\alpha+^{12}$C         & $(0,0)$ & $-1.37(18)\times10^{-1} $ & $Fixed                 $ & $19.7(55)               $\\
&    &          &         &              $\gamma_{0}+^{16}$O$_{0}$ & $(1,1)$ & $ 2.14(41)\times10^{-5} $ & $6.88(50)\times10^{-12}$ & $6.86(50)\times10^{-12} $\\

$02$ & $12.049$ & $0^+_2$ & $4.865   $ & $\alpha+^{12}$C         & $(0,0)$ & $ 4.99(151)\times10^{-3}$ & $1.02(25)\times10^{0}  $ & $1.5(5)\times10^{0}     $\\
&    &          &         &              $\gamma_{0}+^{16}$O$_{0}$ & $(1,1)$ & $ 6.25(247)\times10^{-4}$ &                          &                          \\

$03$ & $14.032$ & $0^+_3$ & $6.557   $ & $\alpha+^{12}$C         & $(0,0)$ & $ 1.65(5)\times10^{-1}  $ & $Fixed                 $ & $\Gamma_{\alpha}/\Gamma_{TOT}=0.9$\\

$04$ & $15.066$ & $0^+_4$ & $7.904   $ & $\alpha+^{12}$C         & $(0,0)$ & $ 4.33\times10^{-1}     $ & $Fixed                 $ & $\Gamma_{\alpha}/\Gamma_{TOT}=0.35$\\
&    &          &         &              $\alpha_{1}+^{12}$C     & $(2,2)$ & $ 5.91\times10^{-2}     $ & $Fixed                 $ & $\Gamma_{TOT}=166(30)   $\\
&    &          &         &              $p+^{15}$N              & $(1,1)$ & $-2.85\times10^{-2}     $ & $Fixed                 $ &                          \\

$05$ & $BG$\footnotemark[1]     & $0^+_5$ & $30.740  $ & $\alpha+^{12}$C         & $(0,0)$ & $ 2.23(9)\times10^{ 0}  $ &                          &                          \\
&    &          &         &              $\gamma_{0}+^{16}$O$_{0}$ & $(1,1)$ & $-3.01(20)\times10^{-4} $ &                          &                          \\

$06$ & $7.119 $ & $1^-_1$ & $ -0.4515$ & $\alpha+^{12}$C         & $(0,1)$ & $-8.08    \times10^{-2} $ & $Fixed                 $ & $62.0(170)$              \\
&    &          &         &              $\gamma_{0}+^{16}$O$_{0}$ & $(1,0)$ & $ 2.20(107)\times10^{-4}$ & $5.7(3)\times10^{-5}   $ & $5.5(3)\times10^{-5}$    \\
&    &          &         &              $\gamma_{0}+^{16}$O$_{0}$ & $(1,2)$ & $ 1.61(46)\times10^{-2} $ &                          &                          \\
&    &          &         &              $\gamma_{1}+^{16}$O$_{1}$ & $(1,0)$ & $-2.06(898)\times10^{-6}$ & $3.0(1)\times10^{-10}  $ & $<3.0\times10^{-10}$     \\
&    &          &         &              $\gamma_{2}+^{16}$O$_{2}$ & $(2,1)$ & $-7.32(83)\times10^{-4} $ & $4.5(1)\times10^{-8}   $ & $4.6(10)\times10^{-8}$   \\
&    &          &         &              $\gamma_{3}+^{16}$O$_{3}$ & $(1,0)$ & $5.54(1300)\times10^{-6}$ & $5.5(2)\times10^{-10}  $ & $<1.0\times10^{-9}$      \\
&    &          &         &              $\gamma_{4}+^{16}$O$_{4}$\footnotemark[2] & $(0,1)$ & $-1.00(67)\times10^{-2} $ &                          &                          \\

$07$ & $9.585 $ & $1^-_2$ & $2.295   $ & $\alpha+^{12}$C         & $(0,1)$ & $ 3.37(3)\times10^{-1}  $ & $322(5)                $ & $420(20)$                \\
&    &          &         &              $\gamma_{0}+^{16}$O$_{0}$ & $(1,0)$ & $-7.37(25)\times10^{-5} $ & $7.25(23)\times10^{-6} $ & $1.56(12)\times10^{-5}$      \\
&    &          &         &              $\gamma_{0}+^{16}$O$_{0}$ & $(1,2)$ & $-2.99(9)\times10^{-3}  $ &                          &                          \\
&    &          &         &              $\gamma_{1}+^{16}$O$_{1}$ & $(1,0)$ & $-1.39(15)\times10^{-4} $ &                          &                          \\
&    &          &         &              $\gamma_{3}+^{16}$O$_{3}$ & $(1,0)$ & $-6.46(45)\times10^{-5} $ & $5.8(8)\times10^{-7}   $ & $1.4(14)\times10^{-6}$       \\
&    &          &         &              $\gamma_{4}+^{16}$O$_{4}$ & $(0,1)$ & $ 1.36(68)\times10^{-3} $ & $1.5(15)\times10^{-6}  $ & $7.8(16)\times10^{-6}$       \\

$08$ & $12.442$ & $1^-_3$ & $5.289  $ &  $\alpha+^{12}$C         & $(0,1)$ & $ 1.13(2)\times10^{-1}  $ & $162(7)                $ & $102(4)$                 \\
&    &          &         &              $\gamma_{0}+^{16}$O$_{0}$ & $(1,0)$ & $-8.15(137)\times10^{-4}$ & $2.3(3)\times10^{-3}   $ & $1.2(2)\times10^{-2}$    \\
&    &          &         &              $\gamma_{0}+^{16}$O$_{0}$ & $(1,2)$ & $-2.73(13)\times10^{-2} $ &                          &                          \\
&    &          &         &              $\gamma_{1}+^{16}$O$_{1}$ & $(1,0)$ & $-5.25(59)\times10^{-4} $ & $1.2(3)\times10^{-4}   $ & $1.2(6)\times10^{-4}$    \\
&    &          &         &              $\gamma_{2}+^{16}$O$_{2}$ & $(2,1)$ & $ 1.39(51)\times10^{-3} $ & $3.4(25)\times10^{-5}  $ & $7.0(30)\times10^{-5}$\footnotemark[3]       \\
&    &          &         &              $\gamma_{4}+^{16}$O$_{4}$ & $(0,1)$ & $ 2.89(55)\times10^{-3} $ & $9.0(34)\times10^{-5}  $ & $1.3(5)\times10^{-4}$\footnotemark[3]       \\
&    &          &         &              $\alpha_{1}+^{12}$C     & $(2,1)$ & $ 1.12(4)\times10^{-1}  $ & $2.9(2)\times10^{-2}   $ & $2.5\times10^{-2}$       \\
&    &          &         &              $p+^{15}$N              & $(1,0)$ & $ 1.16(10)\times10^{-1} $ & $1.8(3)\times10^{-1}   $ & $9.0(1)\times10^{-1}$    \\

$09$ & $13.088$ & $1^-_4$ & $ 5.859 $  & $\alpha+^{12}$C         & $(0,1)$ & $-4.71(49)\times10^{-2} $ & $27(4)                 $ & $45(18)$                 \\
&    &          &         &              $\gamma_{0}+^{16}$O$_{0}$ & $(1,0)$ & $-3.56(52)\times10^{-3} $ & $1.20(4)\times10^{-2}  $ & $3.2(5)\times10^{-2}$    \\
&    &          &         &              $\gamma_{0}+^{16}$O$_{0}$ & $(1,2)$ & $-2.42(14)\times10^{-2} $ &                          &                          \\
&    &          &         &              $\gamma_{1}+^{16}$O$_{1}$ & $(1,0)$ & $-7.77(118)\times10^{-4}$ & $2.7(8)\times10^{-4}   $ & $2.4(5)\times10^{-4}$    \\
&    &          &         &              $\gamma_{2}+^{16}$O$_{2}$ & $(2,1)$ & $-3.54(86)\times10^{-3} $ & $2.8(13)\times10^{-4}  $ & $4.0(20)\times10^{-4}$\footnotemark[3]       \\
&    &          &         &              $\gamma_{4}+^{16}$O$_{4}$ & $(0,1)$ & $-9.53(136)\times10^{-3}$ & $1.3(4)\times10^{-3}   $ & $1.35(40)\times10^{-3}$  \\
&    &          &         &              $\alpha_{1}+^{12}$C     & $(2,1)$ & $ 8.30(52)\times10^{-2} $ & $8.5(9)\times10^{-1}   $ & $1.0\times10^{ 0}$       \\
&    &          &         &              $p+^{15}$N              & $(1,0)$ & $ 2.35(13)\times10^{-1} $ & $3.4(4)\times10^{ 1}   $ & $1.1(2)\times10^{ 2}$    \\

$10$ & $17.510$ & $1^-_5$ & $ 9.710 $  & $\alpha+^{12}$C         & $(0,1)$ & $-8.99(387)\times10^{-2}$ & $Fixed                 $ & $29(9)$                  \\

$11$ & $BG$     & $1^-_6$ & $ 11.434$  & $\alpha+^{12}$C         & $(0,1)$ & $ 1.23(1)\times10^{-2}  $ &                          &                          \\
&    &          &         &              $\gamma_{0}+^{16}$O$_{0}$ & $(1,0)$ & $-5.75(65)\times10^{-4} $ &                          &                          \\
&    &          &         &              $\gamma_{0}+^{16}$O$_{0}$ & $(1,2)$ & $ 7.40(237)\times10^{-3}$ &                          &                          \\
&    &          &         &              $\gamma_{1}+^{16}$O$_{1}$ & $(1,0)$ & $-9.71(133)\times10^{-4}$ &                          &                          \\
&    &          &         &              $p+^{15}$N              & $(1,0)$ & $-8.95(290)\times10^{-2}$ &                          &                          \\

$12$ & $6.917 $ & $2^+_1$ & $ -0.2449$ & $\alpha+^{12}$C         & $(0,2)$ & $ 1.68    \times10^{-1} $ & $Fixed                 $ & $26.7(103)$              \\
&    &          &         &              $\gamma_{0}+^{16}$O$_{0}$ & $(1,1)$ & $ 1.99(8)\times10^{-3}  $ & $9.9(3)\times10^{-5}   $ & $9.7(3)\times10^{-5}$    \\
&    &          &         &              $\gamma_{0}+^{16}$O$_{0}$ & $(1,3)$ & $ 2.81(37)\times10^{-1} $ &                          &                          \\
&    &          &         &              $\gamma_{1}+^{16}$O$_{1}$ & $(1,1)$ & $-6.75(5)\times10^{-4}  $ & $2.7(3)\times10^{-8}   $ & $2.7(3)\times10^{-8}$    \\
&    &          &         &              $\gamma_{2}+^{16}$O$_{2}$ & $(2,0)$ & $ 1.30(29)\times10^{-5} $ & $9.0(30)\times10^{-9}  $ & $9.0\times10^{-9}$       \\
&    &          &         &              $\gamma_{3}+^{16}$O$_{3}$\footnotemark[4] & $(1,1)$ & $ 1.96(23)\times10^{-2} $ &                          &                          \\

\end{tabular}
\end{ruledtabular}
\footnotetext[1]{BG is the abbreviation of background level.}
\footnotetext[2]{Direct capture for the $S_{7.12}$.}
\footnotetext[3]{Ref.~\cite{Debo13}.}
\footnotetext[4]{Direct capture for the $S_{6.92}$.}
\end{table*}

\begin{table*}
\caption{\label{tab:tablelevel23} Parameters obtained from the R-matrix fits, particle partial widths for the levels,
        and the radiation widths for the $\gamma$-ray transitions considered in this work compared with those from literatures. }
\begin{ruledtabular}
\begin{tabular}{ccccccccc}
Res. no. & $E_x$(MeV) & $J^{\pi}_{n}$ & $E_{\lambda}$(MeV) & $\alpha(I_{1}+I_{2})$  & $(s,l)$ & $\gamma_{\alpha sl}$(MeV)$^{1/2}$ &  $\Gamma$(keV)  & Ref.~\cite{Till93}  \\ \hline
$13$ & $9.844 $ & $2^+_2$ & $ 2.684  $ & $\alpha+^{12}$C         & $(0,2)$ & $ 1.34(2)\times10^{-2}  $ & $0.71(19)              $ & $0.625(100)$             \\
&    &          &         &              $\gamma_{0}+^{16}$O$_{0}$ & $(1,1)$ & $-1.62(3)\times10^{-4}  $ & $2.0(1)\times10^{-6}   $ & $5.7(6)\times10^{-6}$    \\
&    &          &         &              $\gamma_{0}+^{16}$O$_{0}$ & $(1,3)$ & $-1.03(3)\times10^{-2}  $ &                          &                          \\
&    &          &         &              $\gamma_{1}+^{16}$O$_{1}$ & $(1,1)$ & $-3.27(43)\times10^{-4} $ & $4.4(12)\times10^{-7}  $ & $1.9(4)\times10^{-6}$    \\
&    &          &         &              $\gamma_{3}+^{16}$O$_{3}$ & $(1,1)$ & $-1.03(9)\times10^{-3  }$ & $2.1(4)\times10^{-6}   $ & $2.2(4)\times10^{-6}$    \\

$14$ & $11.520 $ & $2^+_3$ & $ 4.314 $ & $\alpha+^{12}$C         & $(0,2)$ & $ 6.86(3)\times10^{-2}  $ & $74(1)                 $ & $71(5)$                  \\
&    &          &         &              $\gamma_{0}+^{16}$O$_{0}$ & $(1,1)$ & $ 1.77(17)\times10^{-3} $ & $6.8(2)\times10^{-4}   $ & $6.1(2)\times10^{-4}$    \\
&    &          &         &              $\gamma_{0}+^{16}$O$_{0}$ & $(1,3)$ & $-2.02(15)\times10^{-2} $ &                          &                          \\
&    &          &         &              $\gamma_{1}+^{16}$O$_{1}$ & $(1,1)$ & $ 1.64(13)\times10^{-3} $ & $3.1(5)\times10^{-5}   $ & $3.0(5)\times10^{-5}$       \\
&    &          &         &              $\gamma_{2}+^{16}$O$_{2}$ & $(2,0)$ & $ 2.46(16)\times10^{-4} $ & $2.2(3)\times10^{-5}   $ & $2.0\times10^{-5}$\footnotemark[1]       \\
&    &          &         &              $\gamma_{3}+^{16}$O$_{3}$ & $(1,1)$ & $-1.66(14)\times10^{-3} $ & $2.0(3)\times10^{-5}   $ & $2.9(7)\times10^{-5}$       \\
&    &          &         &              $\gamma_{4}+^{16}$O$_{4}$ & $(2,0)$ & $-1.90(32)\times10^{-4} $ & $1.1(4)\times10^{-5}   $ & $ <5.0 \times10^{-6}$       \\

$15$ & $13.020 $ & $2^+_4$ & $ 5.833 $ & $\alpha+^{12}$C         & $(0,2)$ & $ 8.15(20)\times10^{-2} $ & $112(5)                $ & $150(10)$                \\
&    &          &         &              $\gamma_{0}+^{16}$O$_{0}$ & $(1,1)$ & $-8.51(66)\times10^{-4} $ & $2.2(2)\times10^{-4}   $ & $7.0(20)\times10^{-4}$   \\
&    &          &         &              $\gamma_{0}+^{16}$O$_{0}$ & $(1,3)$ & $-7.54(74)\times10^{-2} $ &                          &                          \\
&    &          &         &              $\alpha_{1}+^{12}$C     & $(2,0)$ & $-4.04(37)\times10^{-2} $ & $4.1(8)\times10^{-1}   $ & $5.0(20)\times10^{-1}$\footnotemark[1]   \\
&    &          &         &              $p+^{15}$N              & $(1,1)$ & $-8.23(45)\times10^{-2} $ & $1.6(2)\times10^{ 0}   $ & $1.5(2)\times10^{ 0}$\footnotemark[1]    \\

$16$ & $15.90 $ & $2^+_5$ & $ 8.300  $ & $\alpha+^{12}$C         & $(0,2)$ & $ 2.22(9)\times10^{-1}  $ & $Fixed                 $ & $\Gamma_{TOT}=600$       \\
&    &          &         &              $\gamma_{0}+^{16}$O$_{0}$ & $(1,1)$ & $-1.34(8)\times10^{-3}  $ & $Fixed                 $ & $\Gamma_{\alpha}\Gamma_{\gamma}/\Gamma_{TOT}=0.4eV$\\

$17$ & $16.443$ & $2^+_6$ & $ 9.281  $ & $\alpha+^{12}$C         & $(0,2)$ & $-3.37(26)\times10^{-2} $ & $Fixed                 $ & $\Gamma_{\alpha}/\Gamma_{TOT}=0.28$\\
&    &          &         &              $\gamma_{0}+^{16}$O$_{0}$ & $(1,1)$ & $ 1.38(7)\times10^{-3}  $ & $Fixed                 $ & $\Gamma_{\alpha}\Gamma_{\gamma}/\Gamma_{TOT}=0.45eV$\\
&    &          &         &              $\alpha_{1}+^{12}$C     & $(2,0)$ & $-9.50\times10^{-3}     $ & $Fixed                 $ & $\Gamma_{TOT}=22(3)$     \\
&    &          &         &              $p+^{15}$N              & $(1,1)$ & $ 8.75\times10^{-3}     $ & $Fixed                 $ &                          \\

$18$ & $17.129$ & $2^+_7$ & $ 9.967  $ & $\alpha+^{12}$C         & $(0,2)$ & $-7.68\times10^{-2}     $ & $Fixed                 $ & $\Gamma_{\alpha}/\Gamma_{TOT}=0.37$\\
&    &          &         &              $\alpha_{1}+^{12}$C     & $(2,0)$ & $ 6.62\times10^{-3}     $ & $Fixed                 $ & $\Gamma_{TOT}=107(14)$   \\
&    &          &         &              $p+^{15}$N              & $(1,1)$ & $ 1.88\times10^{-2}     $ & $Fixed                 $ &                          \\

$19$ & $BG    $ & $2^+_8$ & $ 22.618 $ & $\alpha+^{12}$C         & $(0,2)$ & $ 2.22(2)\times10^{ 0}  $ &                          &                          \\
&    &          &         &              $\gamma_{0}+^{16}$O$_{0}$ & $(1,1)$ & $ 1.15(14)\times10^{-3} $ &                          &                          \\
&    &          &         &              $\gamma_{0}+^{16}$O$_{0}$ & $(1,3)$ & $ 1.67(48)\times10^{-1} $ &                          &                          \\
&    &          &         &              $\gamma_{1}+^{16}$O$_{1}$ & $(1,1)$ & $ 1.01(12)\times10^{-2} $ &                          &                          \\
&    &          &         &              $\gamma_{3}+^{16}$O$_{3}$ & $(1,1)$ & $ 3.18(21)\times10^{-3} $ &                          &                          \\
&    &          &         &              $\alpha_{1}+^{12}$C     & $(2,0)$ & $-4.90(42)\times10^{ 0} $ &                          &                          \\
&    &          &         &              $p+^{15}$N              & $(1,1)$ & $-1.35(58)\times10^{ 0} $ &                          &                          \\

$20$ & $6.1299$ & $3^-_1$ & $-0.1032$  & $\alpha+^{12}$C         & $(0,3)$ & $-7.39(17)\times10^{-2} $ & $5.44(25)              $ & $2.35(80)$               \\
&    &          &         &              $\gamma_{0}+^{16}$O$_{0}$ & $(1,2)$ & $ 5.83(31)\times10^{-4} $ & $2.47(7)\times10^{-8}  $ & $2.60(13)\times10^{-8}$  \\

$21$ & $11.600$ & $3^-_2$ & $ 3.969$   & $\alpha+^{12}$C         & $(0,3)$ & $ 3.14(2)\times10^{-1}  $ & $718(10)               $ & $800(100)$               \\
&    &          &         &              $\gamma_{2}+^{16}$O$_{2}$ & $(2,1)$ & $5.38(3930)\times10^{-5}$ & $2.4(346)\times10^{-8} $ & $1.0\times10^{-5}$\footnotemark[1]       \\
&    &          &         &              $\gamma_{3}+^{16}$O$_{3}$ & $(3,0)$ & $-1.05(59)\times10^{-4} $ & $2.8(31)\times10^{-6}  $ & $1.0\times10^{-5}$\footnotemark[1]       \\
&    &          &         &              $\gamma_{4}+^{16}$O$_{4}$ & $(2,1)$ & $ 6.13(540)\times10^{-4}$ & $1.6(29)\times10^{-6}  $ & $2.0\times10^{-5}$\footnotemark[1]       \\

$22$ & $13.129$ & $3^-_3$ & $ 6.082$   & $\alpha+^{12}$C         & $(0,3)$ & $ 1.02(3)\times10^{-1}  $ & $69(5)                 $ & $90(14)$                 \\
&    &          &         &              $\gamma_{0}+^{16}$O$_{0}$ & $(1,2)$ & $-1.63(79)\times10^{-3} $ & $9.1(30)\times10^{-6}  $ & $1.0\times10^{-5}$       \\
&    &          &         &              $\gamma_{2}+^{16}$O$_{2}$ & $(2,1)$ & $-1.83(30)\times10^{-2} $ & $7.5(25)\times10^{-3}  $ & $8.0\times10^{ 0}$\footnotemark[1]       \\
&    &          &         &              $\alpha_{1}+^{12}$C     & $(2,1)$ & $-2.88(30)\times10^{-1} $ & $4.26(32)\times10^{ 1} $ & $2.09(6)\times10^{ 1}$   \\
&    &          &         &              $p+^{15}$N              & $(1,2)$ & $-1.12(119)\times10^{-3}$ & $9.8(6)\times10^{-1}   $ & $1.0\times10^{ 0}$       \\

$23$ & $13.259$ & $3^-_4$ & $ 6.063$   & $\alpha+^{12}$C         & $(0,3)$ & $-3.77(21)\times10^{-2} $ & $13(4)                 $ & $9(4)$                    \\
&    &          &         &              $\gamma_{2}+^{16}$O$_{2}$ & $(2,1)$ & $-1.79(15)\times10^{-2} $ & $8.0(13)\times10^{-3}  $ & $9.2(15)\times10^{-3}$\footnotemark[1]    \\
&    &          &         &              $\alpha_{1}+^{12}$C     & $(2,1)$ & $ 3.44(15)\times10^{-2} $ & $2.96(36)\times10^{ 2} $ & $8.2(11)\times10^{ 0}$    \\
&    &          &         &              $p+^{15}$N              & $(1,2)$ & $ 4.43(12)\times10^{-1} $ & $1.6(1)\times10^{ 1}   $ & $4.1\times10^{ 0}$        \\

$24$ & $14.100$ & $3^-_5$ & $ 7.149$   & $\alpha+^{12}$C         & $(0,3)$ & $-5.154\times10^{-6}    $ & $Fixed                 $ & $150(75)$                 \\
&    &          &         &              $\alpha_{1}+^{12}$C     & $(2,1)$ & $-1.03(63)\times10^{-2 }$ & $Fixed                 $ & $\Gamma_{TOT}=750(200)$   \\

$25$ & $15.408$ & $3^-_6$ & $ 8.246$   & $\alpha+^{12}$C         & $(0,3)$ & $-2.57\times10^{-7}     $ & $Fixed                 $ & $\Gamma_{\alpha}/\Gamma_{TOT}=0.58$\\
&    &          &         &              $\alpha_{1}+^{12}$C     & $(2,1)$ & $ 2.24\times10^{-1}     $ & $Fixed                 $ & $\Gamma_{TOT}=133(7)$   \\
&    &          &         &              $p+^{15}$N              & $(1,2)$ & $ 9.83\times10^{-6}     $ &                          &                          \\

\end{tabular}
\end{ruledtabular}
\footnotetext[1]{Ref~\cite{Debo13}.}
\end{table*}

\begin{table*}
\caption{\label{tab:tablelevel345} Parameters obtained from the R-matrix fits, particle partial widths for the levels,
        and the radiation widths for the $\gamma$ -ray transitions considered in this work compared with those from the literature. }
\begin{ruledtabular}
\begin{tabular}{ccccccccc}
Res. no. & $E_x$(MeV) & $J^{\pi}_{n}$ & $E_{\lambda}$(MeV) & $\alpha(I_{1}+I_{2})$  & $(s,l)$ & $\gamma_{\alpha sl}$(MeV)$^{1/2}$ &  $\Gamma$(keV)  & Ref.~\cite{Till93}  \\ \hline

$26$ & $15.828$ & $3^-_7$ & $ 8.266$   & $\alpha+^{12}$C         & $(0,3)$ & $-2.24\times10^{-1}     $ & $Fixed                 $ & $\Gamma_{\alpha}/\Gamma_{TOT}=0.21$\\
&    &          &         &              $\alpha_{1}+^{12}$C     & $(2,1)$ & $ 1.03\times10^{-2}     $ & $Fixed                 $ & $\Gamma_{TOT}=703(113)$   \\

$27$ & $BG    $ & $3^-_8$ & $ 22.618 $ & $\alpha+^{12}$C         & $(0,3)$ & $-1.64(1)\times10^{ 0}  $ &                          &                          \\
&    &          &         &              $\gamma_{4}+^{16}$O$_{4}$ & $(2,1)$ & $-1.58(63)\times10^{-2} $ &                          &                          \\

$28$ & $10.361 $ & $4^+_1$ & $ 3.196 $ & $\alpha+^{12}$C         & $(0,4)$ & $ 2.17(3)\times10^{-1}  $ & $28.4(6)               $ & $26(3)$                  \\
&    &          &         &              $\gamma_{0}+^{16}$O$_{0}$ & $(1,3)$ & $1.15(2650)\times10^{-4}$ & $5.6(20)\times10^{-11} $ & $5.6(20)\times10^{-11}$  \\
&    &          &         &              $\gamma_{2}+^{16}$O$_{2}$ & $(4,0)$ & $ 4.27(47)\times10^{-5} $ & $4.7(10)\times10^{-7}  $ & $<1.0\times10^{-6}$      \\
&    &          &         &              $\gamma_{3}+^{16}$O$_{3}$ & $(3,1)$ & $-4.25(33)\times10^{-3} $ & $5.2(8)\times10^{-5}   $ & $6.2((6)\times10^{-5}$   \\

$29$ & $11.094 $ & $4^+_2$ & $ 3.934 $ & $\alpha+^{12}$C         & $(0,4)$ & $ 9.78(108)\times10^{-3}$ & $0.296(32)             $ & $0.28(5)$                \\
&    &          &         &              $\gamma_{2}+^{16}$O$_{2}$ & $(4,0)$ & $ 1.05(19)\times10^{-4} $ & $3.7(1.3)\times10^{-6} $ & $3.1(13)\times10^{-6}$   \\
&    &          &         &              $\gamma_{3}+^{16}$O$_{3}$ & $(3,1)$ & $ 7.04(82)\times10^{-4} $ & $2.7(6)\times10^{-6}   $ & $2.5(6)\times10^{-6}$     \\

$30$ & $13.879 $ & $4^+_3$ & $ 6.835 $ & $\alpha+^{12}$C         & $(0,4)$ & $-5.00(56)\times10^{-2} $ & $50.3(49)              $ & $\Gamma_{\alpha}/\Gamma_{TOT}=0.65(5)$\\
&    &          &         &              $\alpha_{1}+^{12}$C     & $(2,1)$ & $-2.01(14)\times10^{-1} $ & $\Gamma_{TOT}=74(7)    $ & $\Gamma_{TOT}=77(7)$     \\

$31$ & $16.844 $ & $4^+_4$ & $ 9.682 $ & $\alpha+^{12}$C         & $(0,4)$ & $ 1.06(21)\times10^{-1} $ & $Fixed$ & $\Gamma_{TOT}=567(60),\Gamma_{\alpha}/\Gamma_{TOT}=0.28$\\

$32$ & $BG     $ & $4^+_5$ & $ 30.230$ & $\alpha+^{12}$C         & $(0,4)$ & $ 2.10(3)\times10^{ 0}  $ &                          &                          \\
&    &          &         &              $\gamma_{3}+^{16}$O$_{3}$ & $(3,1)$ & $-3.44(30)\times10^{-2} $ &                          &                          \\
&    &          &         &              $\alpha_{1}+^{12}$C     & $(2,2)$ & $-8.50(118)\times10^{-1}$ &                          &                          \\

$33$ & $14.660 $ & $5^-_1$ & $ 7.498 $ & $\alpha+^{12}$C         & $(0,5)$ & $ 3.47\times10^{-1}     $ & $Fixed$ & $\Gamma_{TOT}=672(11),\Gamma_{\alpha}/\Gamma_{TOT}=0.94$\\

$34$ & $16.900 $ & $5^-_2$ & $ 9.748 $ & $\alpha+^{12}$C         & $(0,5)$ & $ 2.9\times10^{-1}      $ & $Fixed$ & $\Gamma_{\alpha}=700$\\

$35$ & $BG     $ & $5^-_3$ & $ 23.514$ & $\alpha+^{12}$C         & $(0,5)$ & $ 1.21\times10^{ 0}     $ &                          &                          \\

$36$ & $14.805 $ & $6^+_1$ & $ 7.657 $ & $\alpha+^{12}$C         & $(0,6)$ & $-2.25\times10^{-1}     $ & $Fixed$ & $\Gamma_{TOT}=70(8),\Gamma_{\alpha}/\Gamma_{TOT}=0.28$\\

$37$ & $16.275 $ & $6^+_2$ & $ 9.133 $ & $\alpha+^{12}$C         & $(0,6)$ & $ 2.83\times10^{-1}     $ & $Fixed$ & $\Gamma_{TOT}=422(14)$\\

\end{tabular}
\end{ruledtabular}
\end{table*}

\section{Summary}

This study presents a new R-matrix theory for the {$^{12}$C($\alpha,\gamma$)$^{16}$O} S factor at helium burning temperature,
and a number of applications to demonstrate the applicability and versatility of this theory. The final result of S(0.3 MeV) = 162.7 $\pm$ 7.3
represents the most precise extrapolation of the {$^{12}$C($\alpha,\gamma$)$^{16}$O} S factor at helium burning temperature based on
a set of complementary data including all available information of $^{16}$O system so far. This is to our knowledge the first published analysis
meeting the precision requirements on {$^{12}$C($\alpha,\gamma$)$^{16}$O}. The whole S factor from 0.3 MeV to 10 MeV provides astrophysical
reaction rate of {$^{12}$C($\alpha,\gamma$)$^{16}$O} with a sound basis for researches of nucleosynthesis and evolution of stars.

\begin{acknowledgments}

The authors would like to thank Prof F. Strieder and Dr. R. J. deBoer for many helps in data of cascade captures and
{$^{12}$C($\alpha,\alpha${$_{1}$})$^{12}$C}, respectively, and are also indebted to Prof. Carl R. Brune and Prof. P. Descouvemont,
for helps about their results. In addition,  we appreciate Dr. Xiaodong Tang for reading the manuscript  and giving some comments. This work is supported partially by the National Science Foundation of China under Grant No. 11421505, 91126017,
No.11175233 and by the Tsinghua University Initiative Scientific Research Program, China. No.20111081104.
\end{acknowledgments}

{}

\end{document}